\documentstyle[12pt,subeqn,psfig]{article}

\advance\textwidth by 100pt
\advance\oddsidemargin by -50pt
\advance\evensidemargin by -50pt
\newcommand{\eq}[1]{(\ref{#1})}
\renewcommand{\thesection}{\arabic{section}}
\renewcommand{\theequation}{\thesection.\arabic{equation}}
\newcommand{\cleqn}{\setcounter{equation}{0}}

\newcommand {\bdd} {{\stackrel{\leftrightarrow}{\partial}}}

\newcommand {\barssq}   {\bar{s}^2}

\newcommand {\hate}     {\hat{e}}
\newcommand {\hatg}     {\hat{g}}
\newcommand {\hats}     {\hat{s}}
\newcommand {\hatc}     {\hat{c}}
\newcommand {\hatgz}    {\hat{g}_Z}

\newcommand {\hatesq}   {\hat{e}^2}
\newcommand {\hatgsq}   {\hat{g}^2}
\newcommand {\hatssq}   {\hat{s}^2}
\newcommand {\hatcsq}   {\hat{c}^2}
\newcommand {\hatgzsq}  {\hat{g}_Z^2}

\newcommand {\msbar} {\overline{\rm MS}}
\newcommand {\mwsq}  {m_W^2}
\newcommand {\mzsq}  {m_Z^2}

\newcommand{\scfersumude}{{\sum_{\scfer = \scup, \scdown, \sclepton}}}

\newcommand{\scfersumi}{{\sum_{\stackrel{\scfer = \scup, \scdown, 
         \sclepton}{i = 1, 2}} }}

\newcommand{\etal}{{\em et al.}}
\newcommand{\scfer}{{\tilde{f}}}

\newcommand{\scferi}{{\tilde{f}_i}}
\newcommand{\scferl}{{\tilde{f}_L}}
\newcommand{\scferr}{{\tilde{f}_R}}
\newcommand{\scferone}{{\tilde{f}_1}}
\newcommand{\scfertwo}{{\tilde{f}_2}}
\newcommand{\scup}{{\tilde{u}}}

\newcommand{\scq}{{\tilde{Q}}}
\newcommand{\scl}{{\tilde{L}}}
\newcommand{\scu}{{\tilde{U}}}
\newcommand{\scd}{{\tilde{D}}}
\newcommand{\sce}{{\tilde{E}}}
\newcommand{\scupl}{{\tilde{u}_L}}
\newcommand{\scupr}{{\tilde{u}_R}}
\newcommand{\scupone}{{\tilde{u}_1}}
\newcommand{\scuptwo}{{\tilde{u}_2}}
\newcommand{\scdown}{{\tilde{d}}}

\newcommand{\scdownl}{{\tilde{d}_L}}
\newcommand{\scdownr}{{\tilde{d}_R}}
\newcommand{\scdownone}{{\tilde{d}_1}}
\newcommand{\scdowntwo}{{\tilde{d}_2}}
\newcommand{\sclepton}{{\tilde{e}}}

\newcommand{\scleptonl}{{\tilde{e}_L}}
\newcommand{\scleptonr}{{\tilde{e}_R}}
\newcommand{\scleptonone}{{\tilde{e}_1}}
\newcommand{\scleptontwo}{{\tilde{e}_2}}
\newcommand{\scnu}{{\tilde{\nu}}}

\newcommand{\scnul}{{\tilde{\nu}_L}}

\newcommand{\cosf}{{\cos\theta_{\tilde{f}}}}
\newcommand{\cosu}{{\cos\theta_{\tilde{u}}}}
\newcommand{\cosd}{{\cos\theta_{\tilde{d}}}}
\newcommand{\cosl}{{\cos\theta_{\tilde{e}}}}
\newcommand{\sinf}{{\sin\theta_{\tilde{f}}}}
\newcommand{\sinu}{{\sin\theta_{\tilde{u}}}}
\newcommand{\sind}{{\sin\theta_{\tilde{d}}}}
\newcommand{\sinl}{{\sin\theta_{\tilde{e}}}}
\newcommand{\cosfsq}{{\cos^2\theta_{\tilde{f}}}}
\newcommand{\cosusq}{{\cos^2\theta_{\tilde{u}}}}
\newcommand{\cosdsq}{{\cos^2\theta_{\tilde{d}}}}
\newcommand{\coslsq}{{\cos^2\theta_{\tilde{e}}}}
\newcommand{\sinfsq}{{\sin^2\theta_{\tilde{f}}}}
\newcommand{\sinusq}{{\sin^2\theta_{\tilde{u}}}}
\newcommand{\sindsq}{{\sin^2\theta_{\tilde{d}}}}
\newcommand{\sinlsq}{{\sin^2\theta_{\tilde{e}}}}
\newcommand{\phasepossf}{{e^{i\phi_{\tilde{f}}}}}
\newcommand{\phasenegsf}{{e^{-i\phi_{\tilde{f}}}}}
\newcommand{\phasepossu}{{e^{i\phi_{\tilde{u}}}}}
\newcommand{\phasenegsu}{{e^{-i\phi_{\tilde{u}}}}}
\newcommand{\phasepossd}{{e^{i\phi_{\tilde{d}}}}}
\newcommand{\phasenegsd}{{e^{-i\phi_{\tilde{d}}}}}
\newcommand{\phasepossel}{{e^{i\phi_{\tilde{e}}}}}
\newcommand{\phasenegsel}{{e^{-i\phi_{\tilde{e}}}}}

\newcommand{\phasenegdiff}{{e^{-i({\phi_{\tilde{u}}}-{\phi_{\tilde{d}}})}}}
\newcommand {\pitgg}    {\Pi_T^{\gamma\gamma}}
\newcommand {\pitgz}    {\Pi_T^{\gamma Z}}
\newcommand {\pitzz}    {\Pi_T^{ZZ}}
\newcommand {\pitww}    {\Pi_T^{WW}}

\newcommand {\pitggg} {{\Pi}_{T,\gamma}^{\gamma\gamma}}

\newcommand {\pitgzg} {{\Pi}_{T,\gamma}^{\gamma Z}}
\newcommand {\pitzzz} {{\Pi}_{T,Z}^{ZZ}}

\newcommand {\qsq} {{q^2}}

\newcommand {\HHbar}
{{\stackrel{{\scriptscriptstyle (}-{\scriptscriptstyle )}}{H}}}
\newcommand {\GammaGammabar}
{{\stackrel{{\scriptscriptstyle (}-{\scriptscriptstyle )}}{\Gamma}}}
\newcommand {\hhbar}
{{\stackrel{{\scriptscriptstyle (}-{\scriptscriptstyle )}}{h}}}
\newcommand {\lambdalambdabar}
{{\stackrel{{\scriptscriptstyle (}-{\scriptscriptstyle )}}{\lambda}}}

\newcommand {\boxes}  {{\rm [Box]}}

\def\new{\rm new}

\def\to{\rightarrow}

\def\ov{\overline}

\def\eeww{e^- e^+ \to W^- W^+}

\def\etal{{\it et al.~}}

\def\mt{m_t^{}}
\def\mh{m_H^{}}
\def\gev{{\rm GeV}}
\def\alphas{\alpha_s}

\def\mw{m_W^{}}
\def\mz{m_Z^{}}
\def\mzsq{m_Z^2}

\newcommand{\beq}{\begin{equation}}
\newcommand{\eeq}{\end{equation}}
\newcommand{\bea}{\begin{eqnarray}}
\newcommand{\eea}{\end{eqnarray}}
\newcommand{\bsub}{\begin{subequations}}
\newcommand{\esub}{\end{subequations}}
\renewcommand{\theequation}{\thesection.\arabic{equation}}

\def\PRD#1#2#3{Phys. Rev. {\bf D#1} (19#2) #3}
\def\NPB#1#2#3{Nucl. Phys. {\bf B#1} (19#2) #3}

\def\ZPC#1#2#3{Z. Phys. {\bf C#1} (19#2) #3}
\def\EPJC#1#2#3{Eur. Phys. J. {\bf C#1} (19#2) #3}
\def\PLB#1#2#3{Phys. Lett. {\bf B#1} (19#2) #3}
\def\PRL#1#2#3{Phys. Rev. Lett. {\bf #1} (19#2) #3}

\makeatletter
\newtoks\@stequation

\def\subequations{\refstepcounter{equation}%
  \edef\@savedequation{\the\c@equation}%
  \@stequation=\expandafter{\theequation}
  \edef\@savedtheequation{\the\@stequation}
  \edef\oldtheequation{\theequation}%
  \setcounter{equation}{0}%
  \def\theequation{\oldtheequation\alph{equation}}}

\def\endsubequations{%
  \ifnum\c@equation < 2 \@warning{Only \the\c@equation\space subequation
    used in equation \@savedequation}\fi
  \setcounter{equation}{\@savedequation}%
  \@stequation=\expandafter{\@savedtheequation}%
  \edef\theequation{\the\@stequation}%
  \global\@ignoretrue}

\def\eqnarray{\stepcounter{equation}\let\@currentlabel\theequation
\global\@eqnswtrue\m@th
\global\@eqcnt\z@\tabskip\@centering\let\\\@eqncr
$$\halign to\displaywidth\bgroup\@eqnsel\hskip\@centering
     $\displaystyle\tabskip\z@{##}$&\global\@eqcnt\@ne
      \hfil$\;{##}\;$\hfil
     &\global\@eqcnt\tw@ $\displaystyle\tabskip\z@{##}$\hfil
   \tabskip\@centering&\llap{##}\tabskip\z@\cr}

\makeatother

\begin{document}
\thispagestyle{empty}
\vspace*{-15mm}
\baselineskip 10pt
\begin{flushright}
\begin{tabular}{l}
{\bf KEK-TH-670}\\
{\bf KA-TP-27-1999}\\
{\bf UR-1601}\\
{\bf DESY 99-201}\\
{\bf hep-ph/0002066}
\end{tabular}
\end{flushright}
\baselineskip 24pt 
\vglue 15mm 
\begin{center}
{\Large\bf One-loop sfermion corrections to 
    $\eeww$ in the MSSM\\
}
\vspace{5mm}

\baselineskip 18pt 
\def\thefootnote{\fnsymbol{footnote}}
\setcounter{footnote}{0}

{\bf 
Sher Alam$^{1}$, 
Kaoru Hagiwara$^{1}$,
Shinya Kanemura$^2$, \\
Robert Szalapski$^3$ and 
Yoshiaki Umeda$^{1,4}$}

\vspace{5mm}
{\it 
$^1$Theory Group, KEK, Tsukuba, Ibaraki 305-0801, Japan\\
$^2$Institut f\"ur Theoretische Physik, Universit\"at Karlsruhe, 
D-76128 Karlsruhe, Germany\\
$^3$Department of Physics and Astronomy, University of Rochester, 
        Rochester, NY 14627-0171, USA\\
$^4$ II Institut f\"{u}r Theoretische Physik, Universit\"{a}t Hamburg,
 D-22761 Hamburg, Germany}
\end{center}


\vspace*{15mm}

\begin{center}
\begin{abstract}
\vspace*{4mm}
\begin{minipage}{15cm}

We study one-loop effects of sfermions on helicity amplitudes for $\eeww$  
in the Minimal Supersymmetric Standard Model.  
The one-loop contributions are calculated in the $\overline{\rm MS}$ 
renormalization scheme. 
In order to verify the validity of the analytic calculation and the 
numerical program, the following tests are performed.        
(i) The BRS sum rules hold exactly among the analytic expressions of the 
    form factors of the $\eeww$ amplitude and those of the amplitudes where
    the external $W^\pm$ bosons are replaced by the corresponding Goldstone 
    bosons $\chi^\pm$, hence they hold within the expected accuracy 
    of the numerical program.  
(ii) The one-loop sfermion contribution to the amplitudes  
     decouple in the heavy mass limit. This property is used to test the 
     overall normalization of the amplitudes.
     In order to observe the analytically exact decoupling, the amplitudes 
     are expanded by the $\overline{\rm MS}$ couplings of the Standard Model.
(iii) The high-energy analytic formulas of the helicity amplitudes, 
      which are verified by using the equivalence theorem analytically, 
      are used for the numerical test of the high energy behavior of the 
      amplitudes.
We then investigate the magnitude of the one-loop effects on each helicity 
amplitude which may be measured by experiments at future linear colliders.    
Under the constraint from available precision data, 
the one-loop corrections to a few helicity amplitudes 
(for example, the longitudinal-$W$-pair production) can be at most  
from $-0.8$\% to $+0.6$\% in magnitude in the observables.
The corrections in the helicity-summed cross sections are smaller, 
typically a few times $\pm 0.1$\% at large scattering angles. 

\end{minipage}
\end{abstract}
\end{center}

\newpage
\baselineskip 16pt 
\def\thefootnote{\arabic{footnote}}
\setcounter{footnote}{0}
\section{Introduction}
\cleqn

\hspace*{12pt}
As collider experiments move to higher energy and higher luminosity 
we are able to probe previously untested aspects of the Standard Model 
(SM) and to search for new physics beyond the SM.  
A leading candidate for the new physics is the Supersymmetric Standard 
Model (SUSY SM).  
If nature has indeed the supersymmetry broken at the weak-scale, 
we should expect to observe the loop level corrections due to
superpartner particles as well as those from the SM particles. 
On the other hand, through these loop-level predictions of the SUSY the 
non-observation of new-physics effects may be used to place constraints 
on the SUSY Lagrangian.  The energy upgrades of the LEP facility at
CERN, LEP~2, and the possibility of a future linear collider such as
JLC, NLC and TESLA motivate us to study $W$-boson pair production through 
electron-positron annihilation\cite{hpzh87,wwcross96}.  
%

In this paper, reflecting the above prospects, the one-loop contributions 
of sfermions to helicity amplitudes for $\eeww$ are investigated in the 
Minimal Supersymmetric Standard Model (MSSM).  
The new physics effects on these amplitudes have been investigated  
in a generic framework in Refs.~\cite{holdom91,hhis96,ads}.
In the SM, the earliest works for the radiative 
corrections to $\eeww$\cite{lemoine80,philippe} have been followed 
by authors of Refs.~\cite{eeww-smrc,ann,jeger,beenakker} for the process 
with the on-shell $W$ bosons. 
The study of radiative corrections to the off-shell $W$-pair production 
has been developed in Refs.~\cite{off-shell,beenakker-rev}.    
The complete SUSY corrections to the differential cross section of 
$\eeww$ have been calculated in Ref.~\cite{alam} in the model with 
spontaneously broken supersymmetry. The sfermion corrections to the 
process $\eeww$ have been discussed in Ref.~\cite{hempfling}.
The trilinear gauge-boson vertices, $\gamma WW$ and $Z WW$, are the 
important ingredients of this process\cite{gg79,hpzh87,hisz93,tgbc96}.  
Several authors have calculated one-loop SUSY contributions to the trilinear  
$\gamma WW$ and $ZWW$ vertices\cite{triple-mssm}.  

In the calculation of the one-loop effects to the $\eeww$ helicity 
amplitudes, a form-factor decomposition of helicity amplitudes is 
invaluable\cite{gg79,hpzh87,jeger}. 
For this reason we present our results by extending the formalism of 
Ref.~\cite{hhis96}. 
In Sec.~2, the essential aspects of the form-factor formalism and 
the helicity amplitudes for the process $\eeww$ 
are reviewed. The formalism is also extended such that the unphysical scalar 
polarization of the final-state $W$ bosons may be also 
studied\cite{hisz93,brs}.  
This will be important when we employ the BRS sum rules 
later for the test of the one-loop form factor calculation.  
A form-factor decomposition for the processes including the
Nambu-Goldstone bosons 
($e^-e^+ \rightarrow \chi^\mp\,W^\pm$ and 
 $e^-e^+ \rightarrow  \chi^-\,\chi^+$) is then presented along with the
 BRS sum rules among the form factors of $W^-W^+$ production and
 those of $W^\mp\chi^\pm$ or $\chi^-\chi^+$ production processes. 
In Sec.~3, we discuss the tree-level results of the helicity amplitudes.
In Sec.~4, we calculate the one-loop sfermion effects on the form factors
of each process in the $\overline{\rm MS}$ scheme\cite{msbar}. 
One of the difficulties of performing loop-level calculations 
is determining the reliability of the results.
This is especially so in the process $\eeww$ where subtle cancellation 
among diagrams which individually grow with energy takes place.
Violation of the gauge-theory cancellation due to incomplete higher-order
terms can hence lead to artificially large corrections.  
Therefore, Sec.~5 is devoted to test our calculation by using 
the following three instruments:   
\begin{description}
\item[(i)] 
      $\;$
      From the global BRS invariance of the electroweak theory\cite{brs-org}, 
      we obtain sum rules among the form factors of 
      $\eeww$ and those of the processes in which one or two 
      external $W$ bosons are replaced by the corresponding 
      Nambu-Goldstone bosons\cite{brs,gkn}. 
\item[(ii)] 
      $\,$
      After properly renormalizing the $\overline{\rm MS}$ couplings 
      to ensure the observed values of the low-energy electroweak observables 
      ($\alpha$, $\mz$ and $G_F$), the full one-loop amplitudes  
      reduce to those of the SM in the large mass limit of the SUSY
      particles\cite{decoupling,dobado}. 
      By expanding the one-loop amplitudes in terms of
      $\msbar$ couplings of the SM, the decoupling property can be
      observed exactly. 
\item[(iii)]
     The analytic expressions of the amplitudes in the high-energy limit 
     are useful to test the numerical program for the of the one-loop 
     $\eeww$ amplitudes in this limit. 
     Such analytic expressions are confirmed by using the equivalence 
     between the longitudinally polarized $W$ bosons and their 
     associated Goldstone bosons, so call the equivalence 
     theorem\cite{et,et2}. 
\end{description}
\noindent
The test (i) ensures the gauge-theory cancellation among the one-loop 
corrected amplitudes, and hence shows the correctness of the loop 
calculation.  
The test (ii) ensures the validity of the renormalization scheme and 
shows the correctness of the overall normalization factors such as 
the external wave-function corrections which cannot be tested by the BRS 
sum rules.  
The test (iii) demonstrates the stability of our numerical program at 
high energies.   
A presentaion of these tests is one main part in this paper in addition 
to evaluation of the magnitude of the sfermion contribution. 
In Sec.~6, we will show our numerical results of the 
$\eeww$ helicity amplitudes, 
and we will examine in which case the sfermion effects become substantial.  
The sfermion effects on the $\eeww$ cross section 
are then discussed under the constraint from the direct search experiments 
and the electroweak precision measurements\cite{pdg99,ch99}. 
In Sec.~\ref{sec-conclusions}, we discuss the results and 
present our conclusion.
To establish our notation and conventions, we present the relevant portions 
of the SUSY Lagrangian in Appendix~A.  This includes the  
sfermion--gauge boson and sfermion--Nambu-Goldstone boson 
interactions.
Appendix~B contains all the explicit formulas for the one-loop sfermion 
contributions to the $\eeww$ form factors.   
In Appendix~C, we calculate sfermion effects on the form factors of 
the processes where the Nambu-Goldstone bosons appear in external lines.
Useful formulas of the loop integral functions for the heavy-mass limit 
and the high-energy limit are given in Appendix~D.

\section{The helicity amplitudes}
\cleqn

\subsection{$\eeww$}
\label{sec-eeww-amp}

\begin{figure}[t]
\begin{center}
\leavevmode\psfig{file=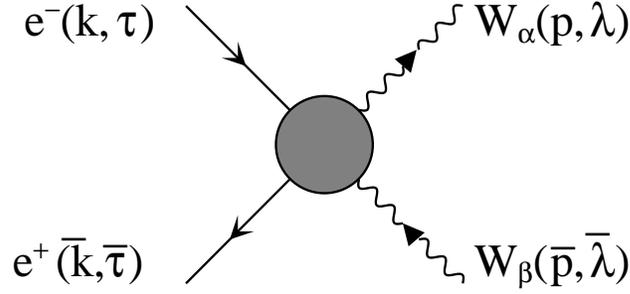,angle=0,height=4cm,silent=0}
\end{center}
\caption{The process $\eeww$\/ with momentum and helicity assignments. 
The momenta $k$ and $\overline{k}$ are incoming, but $p$ and 
$\overline{p}$ are outgoing. The arrows in the $W$ boson lines indicate 
the flow of a negative electric charge.}
\label{fig-eeww-blob} 
\end{figure}

\hspace*{12pt}
The process,  
$e^-(k,\tau) + e^+(\overline{k},\overline{\tau}) \rightarrow
W^-(p,\lambda) + W^+(\overline{p},\overline{\lambda})$, 
is depicted in Fig.~\ref{fig-eeww-blob}.  
The momenta of the $e^-$, $e^+$, $W^-$ and $W^+$ are $k$, 
$\overline{k}$, $p$ and $\overline{p}$, respectively.  
The helicity of the incoming $e^-$ ($e^+$) is given by $\frac{1}{2}\tau$ 
($\frac{1}{2}\overline{\tau}$), and $\lambda$ ($\overline{\lambda}$) is 
the helicity of the outgoing $W^-$ ($W^+$).  
In the limit of massless electrons only $\overline{\tau} = -\tau$ 
helicity amplitudes survive, and they may be written 
for each set of $\{\tau, \lambda, \ov{\lambda}\}$ as\cite{hhis96,brs} 
\begin{equation}
\label{amp-eeww}
{\cal M}^{\lambda \ov{\lambda}}_\tau
  (\eeww)
= \sum_{i=1}^{16} F_{i,\tau}(s,t)\, j_\mu(k,\overline{k},\tau) 
T_i^{\mu\alpha\beta} \epsilon_\alpha(p,\lambda)^\ast 
\epsilon_\beta(\overline{p},\overline{\lambda})^\ast  \;,
\end{equation}
where all dynamical information is contained in the form-factors 
$F_{i,\tau}(s,t)$ with $s = (k + \overline{k})^2 \equiv q^2$ 
and $t = (k - p)^2$.  
The other factors in Eq.~(\ref{amp-eeww}) are of a purely kinematical 
nature; 
$\epsilon_\alpha(p,\lambda)^\ast$ and 
$\epsilon_\beta(\overline{p},\overline{\lambda})^\ast$ are the polarization 
vectors for $W^-$ and $W^+$, respectively, 
and 
$j_\mu(k,\overline{k},\tau)$ is the fermion current of massless electrons. 
The 16 independent basis tensors, $T^{\mu\alpha\beta}_i$, 
are defined by Eqs.~(2.6) in Ref.~\cite{brs}. 
For the process with physically polarized $W$ bosons 
($\lambda, \ov{\lambda} = -, +$ or $0$) are decomposed 
by the first 9 of $T_i^{\mu\alpha\beta}$. 
The rest of $T_i^{\mu\alpha\beta}$ are needed for treating the process 
including unphysical scalar $W$ bosons ($\lambda, \ov{\lambda} = S$).
The properties of $F_{i,\tau}(s,t)$ under the discrete transformations of 
charge conjugation ($C$), parity inversion ($P$) and the combined 
transformation $CP$ are summarized in Table~\ref{table-c-p-cp}.  
\begin{table}[t]
\begin{center}
\begin{tabular}{|c||c|c|c|c|c|c|c|c|c|}\hline
&\hspace*{0.3cm}$F_1$\hspace*{0.3cm} &\hspace*{0.3cm}$F_2$\hspace*{0.3cm} 
&\hspace*{0.3cm}$F_3$\hspace*{0.3cm} &\hspace*{0.3cm}$F_4$\hspace*{0.3cm} 
&\hspace*{0.3cm}$F_5$\hspace*{0.3cm} &\hspace*{0.3cm}$F_6$\hspace*{0.3cm} 
&\hspace*{0.3cm}$F_7$\hspace*{0.3cm} &\hspace*{0.3cm}$F_8$\hspace*{0.3cm} 
&\hspace*{0.3cm}$F_9$\hspace*{0.3cm} \\ 
\hline \hline
$C$  & $+$ & $+$ & $+$ & $-$ & $-$ & $+$ & $+$ & $+$ & $-$ \\ \hline
$P$  & $+$ & $+$ & $+$ & $+$ & $-$ & $-$ & $-$ & $+$ & $-$ \\ \hline
$CP$ & $+$ & $+$ & $+$ & $-$ & $+$ & $-$ & $-$ & $+$ & $+$ \\ \hline
\end{tabular}
\end{center}
\caption{The properties of the form factors $F_{i,\tau}(s,t)$ under the 
discrete transformations $C$, $P$ and $CP$. Only those which contribute to 
physical processes are listed.}
\label{table-c-p-cp}
\end{table}

Finally, the 18 physical helicity amplitudes are given 
in terms of the form factors ($F_{1,\tau}$ - $F_{9,\tau}$) by 
\begin{subequations}
\begin{eqnarray}
\!\!\!\!\!\!\!\!\!\!\!\!
M^{00}_{\tau} &=& -s \left[ -\gamma^2\beta (1+\beta^2) F_{1,\tau} 
+ 4\beta^3\gamma^4 F_{2,\tau} +2\beta\gamma^2 F_{3,\tau}
 - 2 \gamma^2 \cos\theta F_{8,\tau}
\right] \sin \theta , \label{m00}\\ 
\!\!\!\!\!\!\!\!\!\!\!\!
M^{\pm 0}_{\tau} &=& s \gamma \left[ \beta (
F_{3,\tau} 
- i  F_{4,\tau} 
\pm        \beta F_{5,\tau}) 
\pm i  F_{6,\tau}  
\pm (\tau \mp 2\cos\theta) F_{8,\tau} 
\mp 4  \gamma^2 \beta \cos \theta F_{9,\tau}
\right] \frac{(\tau \pm \cos\theta)}{\sqrt{2}} , \label{m+0}\\ 
\!\!\!\!\!\!\!\!\!\!\!\!
M^{0\pm}_{\tau} &=& s\gamma\left[ \beta (
F_{3,\tau} + i F_{4,\tau} \mp \beta F_{5,\tau} ) \pm i F_{6,\tau}
\mp (\tau \pm 2\cos\theta)
F_{8,\tau}
\pm 4  \gamma^2 \cos \theta F_{9,\tau}
\right]\frac{(\tau \mp \cos\theta)}{\sqrt{2}} , \label{m0+}\\ 
\!\!\!\!\!\!\!\!\!\!\!\!
M^{\pm\pm}_{\tau}  &=& s\left[
- \beta F_{1,\tau} 
\mp i F_{6,\tau} 
\mp 4 i \beta^2 \gamma^2 F_{7,\tau} 
+ \cos\theta F_{8,\tau}
+ 4 \beta \gamma^2 \tau F_{9,\tau}
 \right] \sin\theta ,\label{m++} \\
\!\!\!\!\!\!\!\!\!\!\!\!
M^{\pm\mp}_{\tau} &=& 
\mp s ( F_{8,\tau} \pm 4 \beta \gamma^2 F_{9,\tau}) 
                      (\tau \pm \cos\theta ) \sin\theta ,\label{m+-}
\end{eqnarray}
\end{subequations}
where the scattering angle $\theta$ is measured between the momentum 
vectors of the $e^-$ and $W^-$,  and  
\begin{equation}
\beta = \sqrt{1-m_W^2/E_W^2}\;,\makebox[.5cm]{} \gamma = E_W/m^{}_W
\;,\makebox[.5cm]{}E_W = \sqrt{s}/2  \;.
\end{equation}

\subsection{$e^-e^+ \rightarrow     W^\mp \chi^\pm$ 
             and $e^-e^+\rightarrow \chi^- \chi^+$}
\label{subsec-amp-eewx}

\begin{figure}[t]
\begin{center}
\leavevmode\psfig{file=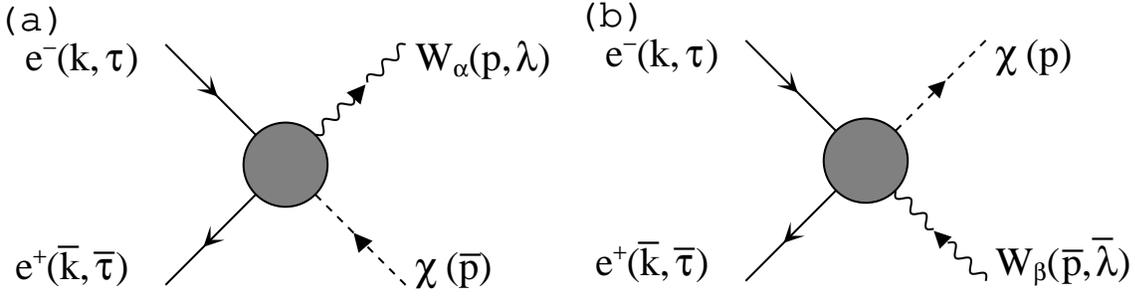,angle=0,height=4cm,silent=0}
\end{center}
\caption{The processes (a) $e^-e^+\rightarrow W^-\chi^+$ and (b)
$e^-e^+\rightarrow \chi^-W^+$ with the momentum and helicity assignments 
chosen to agree with those in 
Fig.~\protect{\ref{fig-eeww-blob}}. The momenta $k$ and $\overline{k}$ are 
incoming, but $p$ and $\overline{p}$ are outgoing. The arrows in the $W$
and $\chi$ lines indicate the flow of a negative electric charge.}
\label{fig-eewx-eexw-blob}
\end{figure}

\hspace*{12pt}
The process for $e^-e^+ \rightarrow W^\mp \chi^\pm$ are 
shown in Figs.~\ref{fig-eewx-eexw-blob}(a) and \ref{fig-eewx-eexw-blob}(b). 
Our phase convention for the Nambu-Goldstone bosons $\chi^\pm$ 
is that of Ref.~\cite{hisz93}.  
We decompose the amplitudes as\cite{brs} 
\begin{subequations}
\begin{eqnarray}
 {\cal M}^\lambda_\tau (e^-e^+\rightarrow W^-\chi^+)
& = &   i \sum_{j=1}^{4} H_{j,\tau}(s,t)\, j_\mu(k,\overline{k},\tau) 
S_j^{\mu\alpha} \epsilon_\alpha(p,\lambda)^\ast \;,\label{amp-eewx}\\
 {\cal M}^{\ov{\lambda}}_\tau (e^-e^+ \rightarrow \chi^- W^+)
& = &  i \sum_{j=1}^{4} \overline{H}_{j,\tau}(s,t) 
 \, j_\mu(k,\overline{k},\tau) \overline{S}_j^{\mu\beta} 
\epsilon_\beta (\overline{p},\overline{\lambda})^\ast \;.\label{amp-eexw}
\end{eqnarray}
\end{subequations}
In \eq{amp-eewx}, there are four independent basis tensors, 
$S_i^{\mu\alpha}$ ($i=1$-$4$), corresponding to the four, three physical 
plus one scalar, polarizations of the $W^-$ boson.
The corresponding form factors are given by $H_{i,\tau}(s,t)$.
A second set of four tensors, $\ov{S}_i^{\mu\beta}$, is introduced to the 
$\chi^- W^+$ production. The corresponding form factors are written by 
$\overline{H}_{i,\tau}(s,t)$.
The basis tensors $S_i^{\mu\alpha}$ and 
$\ov{S}_i^{\mu\beta}$ are given in Ref~\cite{brs}.

\begin{figure}[t]
\begin{center}
\leavevmode\psfig{file=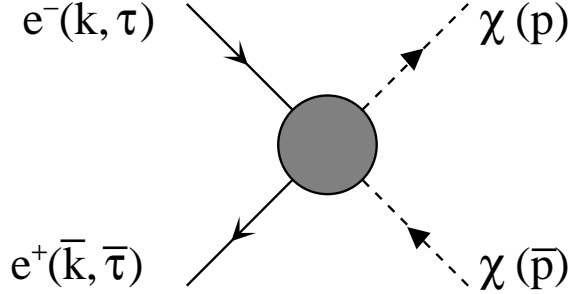,angle=0,height=4cm,silent=0}
\end{center}
\caption{The process $e^-e^+\rightarrow \chi^-\chi^+$ with momentum and 
helicity assignments chosen to coincide with those in 
Fig.~\protect{\ref{fig-eeww-blob}}. The momenta $k$ and $\overline{k}$ are 
incoming, but $p$ and $\overline{p}$ are outgoing.
The arrows in the $\chi^\pm$ boson lines indicate the flow of a negative 
electric charge. 
}
\label{fig-eexx-blob}
\end{figure}

Next, the amplitude for the process $e^-e^+ \rightarrow \chi^-\chi^+$ shown in 
Fig.~\ref{fig-eexx-blob}
may be expressed as 
\begin{equation}
{\cal M}_\tau(e^-e^+ \rightarrow \chi^-\chi^+) 
       =  P^\mu j_\mu(k,\bar{k},\tau) R_\tau(s,t)\;.
\label{amp-wwxx}       
\end{equation}
Notice that there is only one form factor, $R_\tau(s,t)$, which carries an
index for the electron helicity.

\section{The tree-diagram contributions to the helicity amplitudes}
\label{sec-tree}
\cleqn

\begin{figure}[tbp]
\begin{center}
\leavevmode\psfig{file=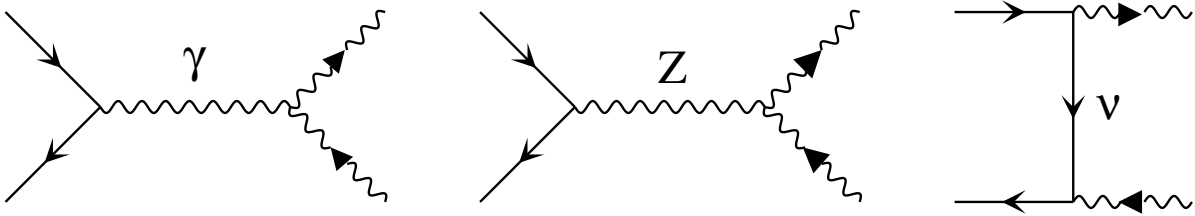,angle=0,height=3cm,silent=0}
\end{center}
\caption{Tree-level Feynman diagrams for $\eeww$.
The arrows on the $W$-boson lines indicate the flow of a negative electric 
charge.}
\label{fig-eeww-tree}
\end{figure}

\hspace*{12pt}
Before going to discuss the one-loop sfermion effects, let us study  
the behavior of the tree-level helicity amplitudes for 
$\eeww$. The Feynman graphs are shown in Fig.~\ref{fig-eeww-tree}.  
The contribution of these diagrams to the form factors introduced in Sec.~2 
is expressed by  
\begin{eqnarray} 
      F_{i,\tau}^{tree}(s,t)  =  
         \frac{\hatesq}{s} Q_e  f_i^{\gamma\,(0)}  
       + \frac{\hatgsq}{s - m_W^2/\hatcsq}
        \big(T^3_e-\hatssq Q_e\big) f_i^{Z\,(0)} 
  + \frac{T^3_e\hatgsq}{2t} 
      f_i^{t\,(0)} \, , \label{big-Ftree}
\end{eqnarray}
where $Q_e = -1$ and $T^3_e$ are the electric charge and the 
third component of weak iso-spin of the electron, respectively.
At tree level,  
we should set the $Z$-boson mass in the propagator as 
\begin{eqnarray}
m_Z^2 = \frac{\mwsq}{\hatcsq}, \;\; ({\rm tree} \; {\rm level}), 
\label{treemz}  
\end{eqnarray}
in order for the tree-level BRS sum-rules to be satisfied\cite{brs}. 
The coefficients $f^{\gamma(0)}_i$, $f^{Z(0)}_i$ and 
$f^{t(0)}_i$ are listed in Table~\ref{table-smf0}.
  
\begin{table}[t]
\begin{center}
\begin{tabular}{|c||cccccccccccccccc|}\hline
i & 1 & 2 & 3 & 4 & 5 & 6 & 7 & 8 & 9 & 10 & 11 & 12 & 13 & 14 & 15 & 16 \\ 
\hline \hline
$f_{i}^{\gamma\,(0)}$ &1&&2&& &&& &&$-1$&&&1&&& \\ 
$f_{i}^{Z\,(0)}$      &1&&2&& &&& &&$-1$&&&1&&& \\ 
$f_{i}^{t\,(0)}$      &1&&2&&1&&&1&&$-2$&&&2&&& \\ \hline
\end{tabular}
\end{center}
\caption{Explicit values for 
the $f_i^{\gamma\,(0)}$, $f_i^{Z\,(0)}$ and $f_i^{t\,(0)}$ in 
Eq.~\eq{big-Ftree}.  
Only nonzero values are shown.}
\label{table-smf0} 
\end{table}

We here take the $W$-boson mass 
$\mwsq$ and the $\msbar$ couplings $\hatesq$ and $\hatssq$  as 
the input parameters. 
In the SM, we can determine the $\overline{\rm MS}$ running coupling
constants from the relations\cite{hagrev}
\begin{subequations}
\label{msbarsm}
\begin{eqnarray}
  \frac{1}{\hat{\alpha}(\mz)} &=& 
         \frac{1}{\overline{\alpha}(\mzsq)} 
         - 0.88 + \frac{8}{9\pi} 
          \left( 1 + \frac{\alpha_s}{\pi} \right) 
         \ln \frac{\mt}{\mz}\,, \\ 
   \frac{\hatssq(\mz)}{\hat{\alpha}(\mz)} &=& 
   \frac{\overline{s}^2(\mzsq)}{\overline{\alpha}(\mzsq)} 
         - 0.11 + \frac{1}{3 \pi} 
          \left( 1 + \frac{\alpha_s}{\pi} \right)        
         \ln \frac{\mt}{\mz}\,,                     
\end{eqnarray}
\end{subequations}
where the effective charges\cite{hhkm94} are estimated as 
$1/{\overline{\alpha}(\mzsq)} = 128.75 \pm 0.09$\cite{alpha_ej} 
and $\barssq (\mzsq) = 0.23035 \pm 0.00023$ for $\mt = 175 \gev$ and 
$\mh = 100 \gev$.
By inserting the mean values into Eqs.~(\ref{msbarsm}), we find
\vspace{-6mm}
\begin{subequations}
\label{msbarinput}
\begin{eqnarray}
  \frac{\hatesq_{\rm SM}(m_Z)}{4 \pi} &=& \frac{1}{128.06}, \\
  \hatssq_{\rm SM}(m_Z) &=& 0.2313,  
\end{eqnarray}
\end{subequations}
for $\mt = 175 \gev$ and $\alpha_s= 0.118$.
These couplings follow the SM renormalization group equation\footnote{
  We note here that both the magnitudes of the 
  $\overline{\rm MS}$ couplings in Eqs.~(\ref{msbarinput}) and the
  renormalization group equations in Eqs.~(\ref{msrge}) are for the SM
  with all 6 quark flavors. We include the effect of the top-quark 
  at the $\mz$-scale in order to avoid introducing the transition 
  between the 5-quark and 6-quark regime in the SM.} 
\begin{subequations}
\label{msrge}
\begin{eqnarray}
\frac{1}{\hat{e}^2_{\rm SM} (\mu)} &=& \frac{1}{\hat{e}^2_{\rm SM} (m^{}_Z)}
-\frac{11}{3}\frac{1}{16\pi^2}
\log{\frac{\mu^2}{m^2_Z}}, \\
\frac{1}{\hat{g}^2_{\rm SM} (\mu)} &=& \frac{1}{\hat{g}^2_{\rm SM} (m^{}_Z)}
+\frac{19}{6}\frac{1}{16\pi^2}
\log{\frac{\mu^2}{m^2_Z}} . 
\end{eqnarray}
\end{subequations}
We take the value of 
$m^{}_W$ as\cite{mwvalue}   
\begin{eqnarray}
  m^{}_W &=& 80.41 {\rm GeV}.  \label{mwvalue}
\end{eqnarray}

\begin{figure}[t]
\begin{center}
$\begin{array}{rl}
\leavevmode\psfig{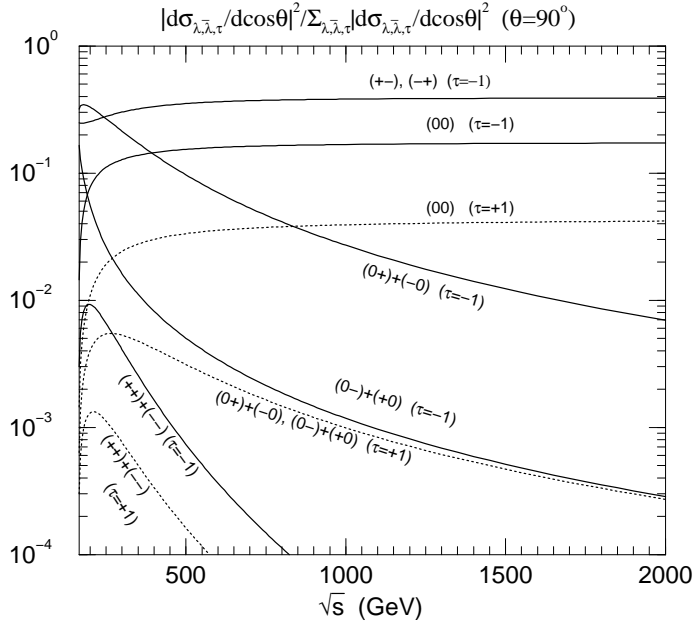} 
\end{array}$
\end{center}
\caption{The ratio of the squared helicity amplitudes to the helicity 
sum of them at the scattering angle $90^{\circ}$ 
for $\sqrt{s}$=161 - 2000 GeV.
} 
\label{fig:amp2vstot}
\end{figure}
The $\sqrt{s}$ dependences of the tree-level helicity amplitudes at 
the scattering angle $\theta$ = $90^{\circ}$ are seen in 
Fig.~\ref{fig:amp2vstot}.
At high energies where the $W$-boson mass is negligible, 
the helicity amplitudes $M^{+-}$, $M^{-+}$ and $M^{00}$ become 
important. The amplitudes $M^{++}$ and $M^{--}$ decrease as 
$1/s$, while $M^{0+}$, $M^{-0}$, $M^{0-}$ and 
$M^{+0}$ decrease as $1/\sqrt{s}$ at high energies. 
At moderately high energies, e.g.\ at $\sqrt{s}\sim 500\gev$, 
the five amplitudes, $M^{+-}$, $M^{-+}$, $M^{00}$, $M^{0+}$ and 
$M^{-0}$ are significant for left-handed electron ($\tau=-$), 
while only $M^{00}$ is significant for right-handed electron 
($\tau=+$).  

\section{One-loop sfermion contributions to the form factors}
\label{sec-ff}
\cleqn

\hspace*{12pt}
In this section, we calculate the one-loop contributions of squarks
and sleptons to the form factors. 
We present the Lagrangian for the sfermion sector in Appendix A. 

\subsection{The scheme of the one-loop calculation}

\hspace*{12pt}
We choose our input electroweak parameters to express the 
tree-level amplitudes that satisfy the BRS sum rules\cite{brs}.  
The $\overline{\rm MS}$ couplings of the MSSM, $\hat{e}^2(\mu)$ and
$\hat{g}^2(\mu)$ are used as the expansion parameters.  
They are related to the SM $\overline{\rm MS}$ couplings in
Eqs.~(\ref{msbarinput}) by the matching condition
\begin{subequations}
\label{eq-mssm}
\begin{eqnarray}
\frac{1}{\hat{e}^2 (\mu)} &=& \frac{1}{\hat{e}^2_{\rm SM} (\mu)}
         - \frac{1}{16\pi^2}
\left[
\frac{16}{3}
\log{\mu^2}-  \sum_{f} N_c^f \frac{Q^2_{{f}}}{3}
\left( \log{m_{\tilde{f}_1}^2} + \log{m_{\tilde{f}_2}^2} \right)
\right] , \label{eq-emssm}\\
\frac{1}{\hat{g}^2 (\mu)} &=& \frac{1}{\hat{g}^2_{\rm SM} (\mu)}
- \frac{1}{16\pi^2}
\left[2\log{\mu^2}-  \sum_{{f}} 
       N_c^f \frac{T^3_{{f}_L^{}} Q_{{f}}}{3} 
\left(\cos^2 \theta_{\tilde{f}} \log{m^2_{\tilde{f}_1}}
+ \sin^2 \theta_{\tilde{f}} \log{m^2_{\tilde{f}_2}}
\right)\right] ,  \label{eq-gmssm} 
\end{eqnarray}
\end{subequations}
where the color factor $N_c^f$ is $3$ for squarks and $1$ for
sleptons, and the other notation of the sfermion sector is defined 
in Appendix A. 
The above conditions ensure that physical observables at low energies 
remain the same when all the sfermion masses are large. In this paper, 
we do not consider contributions of the fermionic supersymmetric 
particles (charginos, neutralinos, and gluinos) nor those from the 
extra Higgs bosons.  These particles are assumed to be even 
heavier, and we work within the effective MSSM with squarks and 
sleptons only. The effects of -ino particles and the extra Higgs 
bosons will be studied elsewhere\cite{hku00}.
The other $\msbar$ couplings are obtained from $\hatesq$ and $\hatgsq$ as 
\begin{eqnarray}
  \hatssq(\mu)  = \frac{\hatesq(\mu)}{\hatgsq(\mu)}\,,  \;\;  
  \hatcsq(\mu)  = 1 - \hatssq(\mu)\,,                   \;\;
  \hatgzsq(\mu) = \frac{\hatgsq(\mu)}{\hatcsq(\mu)}\,. 
\end{eqnarray}
These three input parameters $\{ \mw, \hatesq(\mu), \hatssq(\mu) \}$
are consistently employed in the evaluation of all loop integrals and 
form factors. 
All the relevant diagrams are Taylor-expanded by the $\hatgsq$ 
(or $\hatesq$) and only the terms up to 
${\cal O}(\hatg^4)$ are taken into account.

The ${\overline{\rm MS}}$ masses of the vector bosons are defined 
in terms of $\hatesq (= \hatesq(\mu))$, $\hatgsq (= \hatssq(\mu))$ 
and $m_W$ by 
\begin{eqnarray}
  \hat{m}_W^2 &=& m_W^2 + \Pi_T^{WW}(m_W^2)      \, , \\
  \hat{m}_Z^2 &=& \frac{\hat{m}_W^2}{\hat{c}^2} 
   = \frac{1}{\hat{c}^2} 
     \left\{ m_W^2 + \Pi_T^{WW} (m_W^2) \right\} \, ,
\end{eqnarray}
where $\Pi_T^{WW}(q^2)$ is the $W$-boson 
two-point function in the $\msbar$ scheme\cite{hhkm94}. 
The physical mass of the $Z$ boson is then obtained as 
\begin{eqnarray}
  m_Z^2 = \frac{m_W^2}{\hat{c}^2} 
   + \frac{1}{\hat{c}^2} \Pi_T^{WW} (m_W^2) - \Pi_T^{ZZ} 
 \left(\frac{m_W^2}{\hat{c}^2}\right) \equiv 
  \frac{m_W^2}{\hat{c}^2} + \Delta\; , \label{defmz}
\end{eqnarray}
where deviation from the tree-level expression Eq.~(\ref{treemz}) is 
denoted by $\Delta$. The $Z$-boson propagator should then be 
expanded and truncated as 
\begin{eqnarray}
  \frac{1}{s - m_Z^2} = \frac{1}{s - (m_W^2/\hat{c}^2)} 
     \left\{ 1 + \frac{\Delta}{s - m_W^2/\hat{c}^2}\right\}\;. \label{expmz}
\end{eqnarray}
In Ref.~\cite{brs}, we have demonstrated that by this scheme  
the BRS sum rules hold exactly. 

\subsection{$\eeww$}
\label{subsec-ff-eeww}

\hspace*{12pt}
In the one-loop level, the form factors $F_{i,\tau}(s,t)$ defined 
in Eq.~(\ref{amp-eeww}) may be written as 
\begin{eqnarray}
  F_{i,\tau} = F_{i,\tau}^{(0)} + F_{i,\tau}^{(1)},  \label{ff01}
\end{eqnarray}
where $F_{i,\tau}^{\rm (0)}$ and $F_{i,\tau}^{\rm (1)}$ are the 
${\cal O}(\hatgsq)$ and ${\cal O}(\hatg^4)$ contributions, respectively. 
Although we are interested in the $\eeww$ amplitudes with the physically 
polarized external $W$ bosons ($\lambda, \overline{\lambda} = 0, \pm$),  
in order to test the form factors by the BRS sum rules,   
we have to consider the cases in which one or two external $W$ bosons 
are unphysical too; {\it i.e.} $\lambda$ and/or $\overline{\lambda} = S$.  
Since the BRS sum rules can test the form factors except for the contribution 
of overall factors such as the wave-function renormalization contribution,  
we divide  $F_{i,\tau}^{(1)}$ 
into the one which is the contributions from the $W$-boson wave-function 
renormalization ($F_{i,\tau}^{\rm (1) ext}$), 
and the other is the rest ($F_{i,\tau}^{\rm (1) int}$). 
Eq.~(\ref{ff01}) is then rewritten as   
\begin{eqnarray}
  F_{i,\tau} = F_{i,\tau}^{\rm (0)} + F_{i,\tau}^{\rm (1) int} 
             + F_{i,\tau}^{\rm (1) ext} 
             \equiv \widetilde{F}_{i,\tau} + F_{i,\tau}^{\rm (1) ext},   
    \label{ftilde}
\end{eqnarray}
where $\widetilde{F}_{i,\tau}$ include all the one-loop as well as tree level 
contributions except for the corrections of external $W$-boson lines. 
This part will be tested by the BRS sum rules in Sec.~5.1, 
while the overall normalization will be verified by using the decoupling 
property of the sfermion contributions in the large mass limit 
in Sec.~5.2.  For the BRS test we have to calculate all 16 of 
the $\widetilde{F}_{i,\tau}$ ($i = 1$ - $16$) for each $\tau$,  
while we have only to calculate the $F_{i,\tau}^{\rm (1) ext}$ for 
physical external $W$ lines ($i = 1$ - $9$). 
The form factors for the physical process, 
$F_{i,\tau}$ ($i = 1$ - $9$), are obtained by adding the 
$F_{i,\tau}^{\rm (1)ext}$ ($i = 1$ - $9$) to 
$\widetilde{F}_{i,\tau}$ ($i = 1$ - $9$) by Eq.~(\ref{ftilde}). 
Let us consider each part of the form factors in order. 

First, $\widetilde{F}_{i,\tau}$ are expressed by  
\begin{eqnarray}
\nonumber
&&\makebox[-1cm]{} 
\widetilde{F}_{i,\tau}(s,t)  =  \frac{\hatesq}{s}\Bigg\{\bigg[Q_e \Big(
        1 - \pitggg(s) + \Gamma_1^{\,e}(s)\Big) + T^3_e\, 
        \overline{\Gamma}\mbox{}^{\,e}_2(s) 
        \bigg]f_i^{\gamma\,(0)} 
        +  Q_e f_i^{\gamma\,(1)}(s) \Bigg\}
\\ \nonumber && 
\makebox[-0.5cm]{} 
        + \frac{\hatgsq}{s - (m_W^2/\hat{c}^2)}
        \Bigg\{\bigg[\big(T^3_e-\hatssq Q_e\big)
        \Big( 1 + 
             \frac{\Delta}{s - m_W^2/\hat{c}^2}
 - \pitzzz(s) + \Gamma_1^{\,e}(s) \Big) 
        + T^3_e \Big( \hat{c}^2 \overline{\Gamma}_2^{\,e}(s)  
+   \Gamma_3^{\,e}(s) \Big) \\ \nonumber&& 
\makebox[1.6cm]{}  
+ \Gamma_4^{\,e}(s) \bigg]f_i^{Z\,(0)} 
        + \big(T^3_e-\hatssq Q_e\big) f_i^{Z\,(1)}(s)
        - \frac{\hats}{\hatc} \bigg[ Q_e \hatcsq f_i^{Z\,(0)} 
        + \big(T^3_e-\hatssq Q_e\big) f_i^{\gamma\,(0)} \bigg] \pitgzg(s)
           \Bigg\}
\\ && 
\makebox[-0.5cm]{} 
        + \frac{T^3_e\hatgsq}{2t} 
\Big[ f_i^{t\,(0)} + \Gamma^{e\nu W}(t) + \overline{\Gamma}^{e\nu W}(t) \Big] 
        + F_{i,\tau}^{[\rm Box]}(s,t)\;,
\label{big-F}
\end{eqnarray}
where $i = 1$ - $16$.
We here have already expanded the $Z$-boson propagator according to 
Eq.~(\ref{expmz}).

The quantities $\Pi^{V_1V_2}_{T,V_3}(\qsq)$ 
where $V_i$ is $\gamma$, $Z$ or $W$ are defined by\cite{hhkm94}  
\begin{equation}
\label{defn-pitv}
\Pi^{V_1V_2}_{T,V_3}(\qsq) = 
\frac{\Pi^{V_1V_2}_T(\qsq) - \Pi^{V_1V_2}_T(m^2_{V_3})}  {\qsq - m^2_{V_3}}\;,
\end{equation}
where $\Pi_T^{V_1 V_2}$ are the propagator correction functions 
for the vector bosons.   
We present the sfermion one-loop contribution to $\Pi_{T}^{V_1 V_2}(q^2)$ 
in Appendix B. 

The vertex coefficients $f_i^{V}$ are  
divided into the tree contribution and the one-loop vertex contribution 
according to Eq.~(\ref{ff01}),   
\begin{equation}
 f_i^V(s) = f_i^{V\,(0)} + f_i^{V\,(1)}(s)\;,
\end{equation}
where $V = \gamma$ and $Z$.  
The nonzero tree-level values, $f_i^{V\,(0)}$, have been given in 
Table~\ref{table-smf0}.  
In the one-loop sfermion effects, the triangle-type and the sea-gull-type
diagrams for the $VWW$ trilinear gauge vertices contribute to
$f_i^{V\,(1)}(s)$, which are calculated in Appendix B.2. 
The triangle-type diagrams contribute only to 
$f_1^{V\,(0)}$, $f_2^{V\,(0)}$ and $f_3^{V\,(0)}$ among  
the {\it physical} from factor coefficients $f_1^{V\,(0)}$ - $f_9^{V\,(0)}$ 
and the sea-gull-type diagrams only contribute to 
the {\it unphysical} from factor coefficients.     

The vertex functions for the $Vee$ vertex, denoted by 
$\Gamma_1^{\,e}$, $\overline{\Gamma}\mbox{}^{\,e}_2$, $\Gamma_3^{\,e}$ and 
$\Gamma_4^{\,e}$ also appear in 
$e^-e^+\rightarrow f\overline{f}$ amplitudes\cite{hhkm94}. 
The vertex functions 
$\Gamma^{\,e\nu W}$ and $\overline{\Gamma}\mbox{}^{\,e\nu W}$ 
appear in charged current processes; they contain $\nu e W$ vertex corrections 
as well as two-point function corrections 
for the external electrons and $W$ bosons and the internal neutrino 
propagator. 
Finally, the 
$F_{i,\tau}^{[\rm Box]}$ terms account for contributions of box diagrams.
In the limit of heavy SUSY particles except for squarks and sleptons, 
all these vertex and box corrections are small and we can set them to zero.

Next, as for the part of the corrections to external $W$-boson lines,  
$F_{i,\tau}^{\rm (1)ext}$, we have only to discuss the cases in which   
all the external $W$ boson are physical ($\lambda$ or $\overline{\lambda} = 
0, \pm 1$);  
\begin{eqnarray}
&&\makebox[-1cm]{}  
F_{i,\tau}^{\rm (1)ext}(s,t)  = \Bigg[  \frac{\hatesq}{s} 
              Q_e f_i^{\gamma\,(0)} 
        + \frac{\hatgsq}{s - (m_W^2/\hat{c}^2)}
        \big(T^3_e-\hatssq Q_e\big)
                       f_i^{Z\,(0)} 
        + \frac{T^3_e\hatgsq}{2t} 
 f_i^{t\,(0)}  \Bigg] 
  \delta Z_W , 
\label{big-Fext}
\end{eqnarray}
where $i = 1$ - $9$ and 
$\delta Z_W$ is the wavefunction-renormalization-correction factor of physical 
$W$-bosons with the helicities $\lambda$ or $\overline{\lambda} = 0, \pm$, 
and its sfermion one-loop contribution is given in Appendix B. 

We note that the sfermion one-loop contributions 
do not contribute to $F_{4,\tau}$, $F_{6,\tau}$, $F_{7,\tau}$ 
and $F_{9,\tau}$.  This fact may be explained from 
the viewpoint of the $C$, $P$ and $CP$ property of the form factors 
in Table~\ref{table-c-p-cp}. 
Because all the contributions of the gauge-boson two-point functions 
are accompanied by the tree-level form-factor coefficients 
$f_i^{V(0)}$ and $f_i^{t(0)}$ of Table~\ref{table-smf0},  
nonzero contributions to $F_{4,\tau}$, $F_{6,\tau}$, $F_{7,\tau}$ and 
$F_{9,\tau}$ can arise only from the triangle-type diagrams\footnote{
There is no contribution of the sea-gull type diagrams to $F_1$-$F_9$. 
(See Appendix B.2.)}. 
Since each $\tilde{f}_i^\ast$-$\tilde{f}_j$-$V$ or 
$\tilde{u}_i^\ast$-$\tilde{d}_j$-$W$ 
coupling is clearly $P$ even, 
these triangle diagrams have $P=+$.   
For the $CP$-violating phases that appear in the 
$\tilde{f}_i^\ast$-$\tilde{f}_j$-$V$ and  
$\tilde{u}_i^\ast$-$\tilde{d}_j$-$W$ couplings, we can easily see 
that these are cancelled out in each sfermion triangle diagram.  
Thus, these diagrams can contribute only to the form 
factors with ($P=+$, $C=+$, $CP=+$), hence  
$F_{4,\tau} = F_{6,\tau} =F_{7,\tau} =F_{9,\tau}=0$ 
hold for the sfermion one-loop contributions.

\subsection{$e^-e^+ \rightarrow W^\mp\chi^\pm$ and 
            $e^-e^+ \rightarrow \chi^-\chi^+$}
\label{subsec-ff-eewx}

\begin{figure}[t]
\begin{center}
\leavevmode\psfig{file=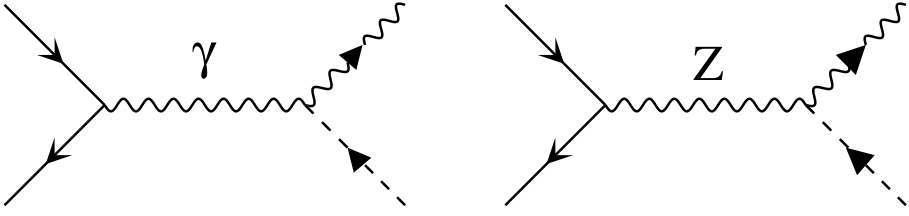,angle=0,height=3cm,silent=0}
\end{center}
\caption{Tree-level Feynman diagrams contributing to $e^-e^+ \rightarrow 
W^-\chi^+$.  The arrows in the charged boson lines 
indicate the flow of negative 
electric charge.  By reversing the direction of these arrows we obtain the 
graphs for $e^-e^+ \rightarrow \chi^-W^+$.}
\label{fig-eewx-tree}
\end{figure}

\hspace*{12pt}
First, we calculate one-loop sfermion contributions to 
$\HHbar_i$ in Sec.~2.2. The results will be used for the test 
of the one-loop form factors of $\eeww$. 
We have only to calculate $\HHbar_i$ except for 
the corrections to the external $W$- and $\chi$-boson lines.  
The form factors $\HHbar_i$ are then expressed by 
\begin{eqnarray}
\nonumber
  \lefteqn{\HHbar\mbox{}_{i,\tau}(s,t) = 
\frac{\hatesq}{s}\Bigg\{\bigg[ Q_e \Big( 1
- \pitggg(s) + \Gamma_1^{\,e}(s)
\Big) + T^3_e \overline{\Gamma}\mbox{}^{\,e}_2(s) \bigg]
\hhbar\mbox{}_i^{\gamma\,(0)} 
+ Q_e \hhbar\mbox{}_i^{\gamma\,(1)}(s) \Bigg\}}
&&\\
\nonumber && \makebox[-1.5cm]{} 
+ \frac{\hatgsq}{s  - \mwsq/\hat{c}^2}
\Bigg\{ \bigg[ (T^3_e-\hatssq Q_e)\Big( 
 1  + 
\frac{\Delta}{s-m_W^2/\hat{c}^2}
  - \pitzzz(s) + \Gamma_1^{\,e}(s)
\Big) 
+ T^3_e \Big( \hat{c}^2 \overline{\Gamma}_2^{\,e}(s) + \Gamma_3^{\,e}(s) \Big)
\nonumber \\&&
\makebox[-1.5cm]{}
+ \Gamma_4^{\,e}(s)\bigg] \hhbar\mbox{}_i^{Z\,(0)}  + 
(T^3_e-\hatssq Q_e) \hhbar\mbox{}_i^{Z\,(1)}(s)  
- \frac{\hats}{\hatc} 
\bigg[ (T^3_e-\hatssq Q_e)\hhbar\mbox{}_i^{\gamma\,(0)} 
+ Q_e \hat{c}^2 \hhbar\mbox{}_i^{Z\,(0)} \bigg]
\pitgzg(s) \Bigg\} \nonumber \\
&&+ \frac{\hatgsq T_e^3}{2 t} \GammaGammabar^{e\nu\chi}
+ \HHbar\mbox{}_{i,\tau}^{\,\boxes}(s,t) \;,
\label{big-H}
\end{eqnarray}
where $\HHbar\mbox{}_{i,\tau}$, $\hhbar\mbox{}_i$ and   
$\GammaGammabar^{e\nu\chi}$ should be read as 
$H_{i,\tau}$, $h_i$ and $\Gamma^{e\nu\chi}$ for 
$e^-e^+ \rightarrow W^-\chi^+$, and 
$\ov{H}_{i,\tau}$, $\ov{h}_i$ and $\ov{\Gamma}^{e\nu\chi}$ for 
$e^-e^+ \rightarrow \chi^-W^+$, respectively. 
At the tree-level, Feynman graphs in Fig.~\ref{fig-eewx-tree} contribute. 
The 
expansion of $H_{i,\tau}(s,t)$ introduces the vertex form factors $h_i^\gamma$ 
and $h_i^Z$,  while the expansion of $\overline{H}_{i,\tau}(s,t)$ introduces 
$\overline{h}_i^\gamma$ and $\overline{h}_i^Z$. They are written as 
the sum of the tree and one-loop contributions by
\begin{equation}
\hhbar\mbox{}_i^V(s) = \hhbar\mbox{}_i^{V\,(0)} 
+ \hhbar\mbox{}_i^{V\,(1)}(s)\;,
\end{equation}
for $V = \gamma$, $Z$. 
The tree-level coefficients ${h}_i^{\,V\,(0)}$ are given by 
$h_i^{\,\gamma\,(0)} = - \overline{h}_i^{\,\gamma\,(0)} 
= \delta_{i\; 1}$ and 
$h_i^{\, Z (0)} = 
-\overline{h}_i^{\, Z (0)} = -({\hatssq}/{\hatcsq}) \delta_{i\; 1}$. 
The $\hhbar\mbox{}_i^{V\,(1)}(s)$ come from the one-loop 
1PI $VW\chi$ vertex corrections, and their sfermion contributions 
are shown in Appendix C.1. 
The other form factor coefficients (the vertex- and the 
box-diagrams which connect with initial $e^\pm$)  
turn out to be zero for the sfermion contribution.

Second, the sfermion one-loop contributions to ${R}_{\tau}$ 
defined in Sec.~2.2 are expressed by 
\begin{eqnarray}
  {R}_{\tau}(s,t) &=&  - \frac{\hatesq}{s}  \bigg[ Q_e \Big( 1
- \pitggg(s) + \Gamma_1^{\,e}(s) \Big) 
+ T^3_e \overline{\Gamma}\mbox{}^{\,e}_2(s)  
+ Q_e r^{\gamma\,(1)}(s)  \bigg] \nonumber \\
&&- \frac{\hatgzsq }
          {s  - \mwsq/\hat{c}^2}
\Bigg\{ \bigg[ (T^3_e-\hatssq Q_e) \left(\frac{1}{2} - \hatssq \right) \Big( 
 1  + 
\frac{\Delta}{s-m_W^2/\hat{c}^2}
  - \pitzzz(s) + \Gamma_1^{\,e}(s)
\Big) \nonumber \\
&& \makebox[-1.5cm]{} 
+ T^3_e \Big( \hat{c}^2 \overline{\Gamma}_2^{\,e}(s) + \Gamma_3^{\,e}(s) \Big)
+ \Gamma_4^{\,e}(s)\bigg]  \nonumber \\
&& \makebox[-1.5cm]{} 
  + (T^3_e-\hatssq Q_e)  r^{Z\,(1)}(s)  
- \frac{\hats}{\hatc} 
\bigg[ (T^3_e-\hatssq Q_e) + Q_e \left(\frac{1}{2} - \hatssq \right)  \bigg]
\pitgzg(s) \Bigg\} + R_{\tau}^{\,\boxes}(s,t) \;,
\end{eqnarray}
where we do not include the corrections to 
the external $\chi$ lines by the same reason as for the $\HHbar_{i,\tau}$.
For the sfermion effects, we have to calculate  
the $V\chi\chi$-vertex corrections parametrised by $r^{V(1)}$. 
The results are given in Appendix~C.   
All the other form-factor coefficients (for the one-loop vertices and 
the boxes that contain initial $e^\pm$ lines)  
turn out to be zero for the sfermion contribution.

\section{Tests of the one-loop calculation}
\cleqn

\hspace*{12pt}
The purpose of this paper is to evaluate the sfermion one-loop effects 
on $\eeww$. To obtain trustworthy results, we test 
our one-loop calculation by using the BRS invariance, the decoupling 
theorem, and the high energy behaviors of the theory.  
We show these procedure in order.

\subsection{The BRS test for the $e^-e^+ \rightarrow W^-W^+$ form factors}

\subsubsection{The BRS sum rules}

\hspace*{12pt}
The global BRS symmetry in the quantized electroweak gauge theories 
gives identities between a process with scalar-polarized $W$ bosons 
and one where the $W$ bosons are replaced by the Goldstone bosons 
$\chi$\cite{gkn}. 
Regarding to our process $\eeww$, we have two kind of the 
identities\cite{brs}; 
we call the first one as the single BRS identity where only one external 
$W$ boson is replaced by the $\chi$ boson, and the second one as the double 
BRS identity where both the $W$ bosons are replaced. 
From these identities for amplitudes, we obtain useful sum rules among 
the form factors which we introduced in Sec.~2;  
$\widetilde{F}_{i,\tau}$ ($i = 1$ - $16$), $\HHbar_{i,\tau}$ 
($i = 1$ - $4$) and $R_{\tau}$.  
Note that $\widetilde{F}_{1,\tau}$ - $\widetilde{F}_{9,\tau}$ 
are common with the physical $\eeww$ amplitude (See Eq.~(\ref{ftilde})). 

A set of the sum rules is obtained from the BRS identity between 
$\eeww$ and $e^-e^+ \rightarrow W^-\chi^+$ amplitudes\cite{brs}.
\begin{subequations}
\label{eq-expbrs1}
\begin{eqnarray}
  -2 \gamma^2 \,
  \Big\{ \widetilde{F}_{3,\tau}(s,t) - i\, \widetilde{F}_{4,\tau}(s,t) \Big\}
  + 4 \delta^2\, \widetilde{F}_{8,\tau}(s,t) +  \widetilde{F}_{13,\tau}(s,t) 
   &=& C_{\rm mod}^{BRS}  {H}_{1,\tau}(s,t) \;, \;\;
\label{eq-expbrs1a}
\\  
  - \widetilde{F}_{1,\tau}(s,t) 
  + 2 \gamma^2\, \widetilde{F}_{2,\tau}(s,t) 
  + \frac{1}{2} \widetilde{F}_{3,\tau}(s,t) 
  + \frac{i}{2} \widetilde{F}_{4,\tau}(s,t) 
  + \widetilde{F}_{14,\tau}(s,t) 
   &=& C_{\rm mod}^{BRS} {H}_{2,\tau}(s,t)\;, \;\;
\label{eq-expbrs1b}
\\ 
  - \frac{1}{2} \widetilde{F}_{5,\tau}(s,t) 
  - \frac{i}{2} \widetilde{F}_{6,\tau}(s,t)
  - \frac{\tau}{2} \widetilde{F}_{8,\tau}(s,t)
  + 2 \delta^2\, \widetilde{F}_{9,\tau}(s,t)
  + \widetilde{F}_{15,\tau}(s,t) 
   &=& C_{\rm mod}^{BRS} {H}_{3,\tau}(s,t)\;, \;\;
\label{eq-expbrs1c}
\end{eqnarray}
\end{subequations}
where 
\begin{eqnarray}
  \gamma^2 = \frac{s}{4 m_W^2} , \;\;\;\; {\rm and} \;\;\;\;\;
  \delta^2 = \frac{s + 2 t - 2 m_W^2}{4 m_W^2} , 
\end{eqnarray}
and
\begin{eqnarray}
C_{\rm mod}^{BRS} = \frac{\hat{m}_W^{}}{m_W^{}}.   \label{cmod-brs}
\end{eqnarray}
These sum rules can be used for a non-trivial test of the one-loop 
form factors for the physical amplitudes except for the part of 
the wavefunction renormalization factors. 
Another set of sum rules can be obtained from the `single' BRS identity 
between the $W^-W^+$ and $\chi^-W^+$ production. In practice, once one of 
the two sets of the `single' BRS sum rules is used to verify the accuracy 
of a calculation, the other set is redundant.
In addition, the sum rules among the $\eeww$, $e^-e^+\to W^\mp\chi^\pm$, 
and $e^-e^+\to \chi^-\chi^+$ form factors are also obtained from 
the `double' BRS identity, which may be useful for an independent BRS test 
of $\widetilde{F}_1$, $\widetilde{F}_2$,  $\widetilde{F}_3$ and 
$\widetilde{F}_8$ .

\subsubsection{Numerical tests of the one-loop results for $F_{i,\tau}$ 
by using the BRS sum rules}

\hspace*{12pt}
The most practical application of the BRS sum rules is the numerical test 
of the program. In our formalization, they should hold exactly for the form 
factors calculated at the one-loop level.
Our computational program has been tested to satisfy them with an 
excellent agreement.

\begin{table}[t]
\label{tab:brs1}
\begin{center}
\begin{tabular}{l|l} \hline\hline
\multicolumn{2}{l}{First BRS sum rule ($\tau=-1$)} \\ \hline
$\sqrt{s}$ &  Left-hand-side of Eq.~(\ref{eq-expbrs1a})\\
           & Right-hand-side of Eq.~(\ref{eq-expbrs1a})\\ \hline
%
%
 200GeV    &        $-$0.1385496590672218 $\times 10^{-5}$ \\
           &        $-$0.1385496590672223 $\times 10^{-5}$ \\ \hline
%
%
  500GeV   & $-$0.2654648169991279 $\times 10^{-6}$ 
             $-$0.3685263974480902 $\times 10^{-8}$ $i$ \\
           & $-$0.2654648169991285 $\times 10^{-6}$ 
             $-$0.3685263974480899 $\times 10^{-8}$ $i$ \\ \hline
%
%
  1000GeV  & $-$0.6682526871892199 $\times 10^{-7}$ 
             $-$0.6849932023212376 $\times 10^{-8}$ $i$ \\
           & $-$0.6682526871892053 $\times 10^{-7}$ 
             $-$0.6849932023212378 $\times 10^{-8}$ $i$ \\ \hline
%
%
2000GeV& $-$0.1434490310523539 $\times 10^{-7}$ 
         $-$0.1201404359981963 $\times 10^{-8}$ $i$ \\
       & $-$0.1434490310523237 $\times 10^{-7}$ 
         $-$0.1201404359981958 $\times 10^{-8}$ $i$ \\ \hline\hline 
\end{tabular}
\caption{The BRS test of the sfermion one-loop effects on the 
         $\eeww$ form factors 
         by using the first BRS sum rule~(\ref{eq-expbrs1a}).  
         As for the MSSM parameters, Case 28 of Table~10 in Sec.~6 is used.}
\end{center}
\end{table}
\begin{table}[t]  
\label{tab:brs2}
\begin{center}
\begin{tabular}{l|l} \hline \hline
\multicolumn{2}{l}{Second BRS sum rule ($\tau=-1$)} \\ \hline
$\sqrt{s}$ &  Left-hand-side of Eq.~(\ref{eq-expbrs1b})\\
           & Right-hand-side of Eq.~(\ref{eq-expbrs1b})\\ \hline
%
%
200GeV & $-$0.2438990547345640 $\times 10^{-9}$ \\
       & $-$0.2438990547345377 $\times 10^{-9}$ \\ \hline
%
%
500GeV & $-$0.7616498712364096 $\times 10^{-10}$  
         $-$0.2662690301056969 $\times 10^{-10}$ $i$ \\
       & $-$0.7616498712364688 $\times 10^{-10}$ 
         $-$0.2662690301056939 $\times 10^{-10}$ $i$ \\ \hline
%
%
1000GeV & \hspace{2.mm} 0.1916470996088027 $\times 10^{-12}$ 
          $-$0.2445806925474609 $\times 10^{-10}$ $i$ \\
        & \hspace{2.mm} 0.1916470996084505 $\times 10^{-12}$ 
          $-$0.2445806925474613 $\times 10^{-10}$ $i$ \\ \hline
%
%
2000GeV & \hspace{2.mm} 0.3363019521319420 $\times 10^{-11}$ 
            +0.1791648684551098 $\times 10^{-11}$ $i$ \\
        & \hspace{2.mm} 0.3363019521319535 $\times 10^{-11}$ 
            +0.1791648684551017 $\times 10^{-11}$ $i$ \\ \hline \hline
\end{tabular}
\caption{The BRS test of the sfermion one-loop effects on the 
         $\eeww$ form factors 
         by using the second BRS sum rule~(\ref{eq-expbrs1b}).  
         As for the MSSM parameters, Case 28 of Table~10 in Sec.~6 is used.}
\end{center}
\end{table}

We here test our one-loop calculation of the form factors 
numerically by using the BRS sum rules~(\ref{eq-expbrs1}). 
In our evaluation of the scalar one-loop integral functions, 
we partly use the Fortran FF-package\cite{ff}.  
As a sample MSSM parameter choice, we take Case 28 that is defined later
in Table~10 in Sec.~6\footnote{
As for the renormalization scheme, in addition to the method defined in
Sec.~4, the expansion by the coupling constants in the SM is employed 
which will be introduced a little later in Sec.~5.2. }. 
We see in Tables~3 and 4 that the first sum rule~(\ref{eq-expbrs1a})  
and the second one~(\ref{eq-expbrs1b}) hold to better than 11 digits 
accuracy at $\sqrt{s}$ between 200GeV and 2000GeV, respectively.  
As for the third one~(\ref{eq-expbrs1c}), it turns out that the both sides 
are zero, so it is rather trivial.  

\subsection{The test by using the decoupling theorem}

\hspace*{12pt}
The second useful instrument for the test is the decoupling property 
of the sfermion one-loop contributions at the large mass limit, 
where the sfermion effects should decouple from the observable 
and the model should be regarded as the SM effectively by the argument of 
the decoupling theorem\cite{decoupling,dobado}.  
To see this property at each order of perturbation, a consistent 
renormalization scheme must be taken, by which the one-loop result 
in the MSSM is coincident with that in the SM at the large mass limit.  
In the $\overline{\rm MS}$ scheme, the perturbation is performed
by the $\msbar$ couplings of the MSSM. 
In order to obtain the one-loop expression which reduces to the SM 
results exactly in the large SUSY-mass limit, we expand the one-loop    
amplitudes by the SM $\msbar$ couplings by using Eqs.~(\ref{eq-mssm}).  
By dropping the higher-order (${\cal O}(\hat{g}^6_{\rm SM})$) terms, 
the decoupling of the one-loop effects can be made exact.

\subsubsection{The expansion by the coupling constants in the SM}

\hspace*{12pt}
In the one-loop calculation of the sfermion effects in Sec.~4,  
we took $\hatesq(\mu)$, $\hatssq(\mu)$ and $m^{}_W$ as the input 
parameters. The $W$-boson mass, $m^{}_W$, is determined by the precision 
data as in Eq.~(\ref{mwvalue}). From the equations \eq{eq-emssm} and 
\eq{eq-gmssm}, $\hatesq (\mu)$ and $\hatgsq (\mu)$ include higher order 
sfermion effects beyond the one-loop level. 
In order to test the decoupling property analytically, 
we expand the original one-loop amplitudes by the SM coupling constants 
$\hatesq_{\rm SM} (\mu)$ and $\hatgsq_{\rm SM} (\mu)$:
\begin{subequations}
\label{eq-bysm}
\begin{eqnarray}
\!\!\!\!\!\!\!\!
\hat{e}^2 (\mu) &=& \hat{e}^2_{\rm SM} (\mu)
\left\{ 1 + 
\frac{\hat{e}^2_{\rm SM}(\mu)}{16\pi^2}
\left[
\frac{16}{3}
\log{\mu^2}-  \sum_{{f}} N_c^f\frac{Q^2_{{f}}}{3}
\left( \log{m_{\tilde{f}_1}^2} + \log{m_{\tilde{f}_2}^2} \right)
\right]\right\}, \label{eq-bysm1}\\
\!\!\!\!\!\!\!\!
\hat{g}^2 (\mu) &=& \hat{g}^2_{\rm SM} (\mu)
\left\{ 1 + 
\frac{\hat{g}^2_{\rm SM} (\mu)}{16\pi^2}
\left[2\log{\mu^2}-\sum_{{f}} N_c^f\frac{T^3_{{f_L}} Q_{{f}}}{3} 
\left(\cos^{2} \theta_{\tilde{f}} \log{m^2_{\tilde{f}_1}}
    + \sin^{2} \theta_{\tilde{f}} \log{m^2_{\tilde{f}_2}}
\right)\right]\right\}. \label{eq-bysm2}
\end{eqnarray}
\end{subequations}
%

Hereafter, we perform this procedure in all our calculation. 
All the form factors we have presented previously are now expanded  
by the SM couplings by using Eqs.~\eq{eq-bysm}, and we retain 
only terms of ${\cal O} (\hatgsq_{\rm SM})$ and ${\cal O} (\hatg^4_{\rm SM})$. 
We then will find below that the one-loop sfermion contributions vanish 
exactly in the limit of infinitely heavy sfermion masses.     
In Sec.~7, we discuss the difference of the magnitude between 
the amplitude expanded by the SM couplings and that in terms of the 
MSSM couplings without such additional expansion.  
Since we do not include the SM loop contributions\cite{eeww-smrc} 
in our analysis, the ${\cal O} (\hatg^4_{\rm SM})$ terms are solely 
coming from the sfermion one-loop contributions, which decouple in the 
heavy sfermion mass limit ($s/m^2_{\tilde{f}} \ll  1$) and 
grow logarithmically at high energies ($s/m^2_{\tilde{f}}\gg 1$).  
As for the scale dependence of the SM $\msbar$ couplings, we set 
$\mu=\sqrt{s}$ for brevity.

\subsubsection{The decoupling limit}

\hspace*{12pt}
The consistent calculation according to the above procedure  
allows us to observe the exact decoupling in the large sfermion-mass 
limit.
In the original expression of the amplitudes which are expressed in terms 
of the MSSM $\msbar$ couplings, the amplitudes behave as 
\begin{eqnarray}
   \delta {\cal M}^{\rm sfermion-loop} 
  \sim  A + B \frac{s}{m_{\tilde{f}}^2} 
+ {\cal O} \left( \frac{s^2}{m_{\tilde{f}}^4} \right) , 
\label{dec}
\end{eqnarray}
in the large sfermion mass limit,   
where the constant term, $A$, remains nonzero as terms of ${\cal O}(\hatg^6)$ 
do not cancel exactly.  
Contrary by taking the SM coupling constants as the expansion 
parameter and by truncating the expansion at the 
${\cal O} (\hat{g}_{\rm SM}^4)$ 
terms, the term $A$ in \eq{dec} becomes zero.
This property of the exact decoupling in our scheme can be used for the 
excellent test of the form factors including the overall normalization 
factors that have not yet been tested in the BRS sum rules.

\begin{figure}[t]
\begin{center}
\leavevmode\psfig{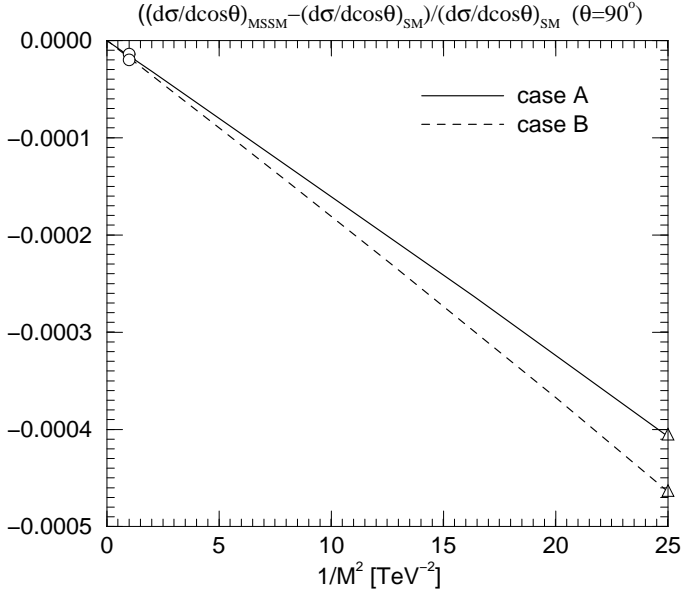} 
\caption{The test of the decoupling of the sfermion contribution.
         The deviation of the helicity-summed differential 
         cross section from the SM value versus $1/M^2$ is shown at 
         $\sqrt{s} = 200$GeV and the scattering angle 
         $\theta = 90^{\circ}$, where 
         $M$ is the scale of the sfermion masses.  
         The solid line is for the case where 
         $M$ is the mass of all the sfermions (Case A). 
         The dashed line is for the case where 
         $m_{\tilde{u}_L^{}} = m_{\tilde{u}_L^{}} = M$, 
         $m_{\tilde{u}_R^{}} = 1.1 M$, 
         $m_{\tilde{d}_R^{}} = 1.2 M$, 
         $m_{\tilde{e}_L^{}} = m_{\tilde{\nu}_L^{}} = 1.3 M$ and  
         $m_{\tilde{e}_R^{}} = 1.4 M$ (Case B).  
         The symbol $\bigtriangleup$ ($\bigcirc$) 
         on the lines is  the point of 
         $M=200$ ($1000$) GeV.}
\label{fig:decoup}
\end{center}
\end{figure}

Fig.~7 shows the sfermion contribution in the helicity-summed 
differential cross section as a function of $1/{m_{\tilde{f}}^2}$ 
at $\sqrt{s} = 200$GeV and at the large scattering angle 
$\theta = 90^{\circ}$.        
We test the decoupling property in Case A and Case B; 
all the sfermion masses are set to $M$ in Case A, and the sfermion 
masses are taken as  
         $m_{\tilde{u}_L^{}} = m_{\tilde{d}_L^{}} = M$, 
         $m_{\tilde{u}_R^{}} = 1.1 M$, 
         $m_{\tilde{d}_R^{}} = 1.2 M$, 
         $m_{\tilde{e}_L^{}} = m_{\tilde{\nu}_L^{}} = 1.3 M$ and  
         $m_{\tilde{e}_R^{}} = 1.4 M$ in Case B.   
In Fig.~7, we see that the both lines of Case A and Case B include 
the origin; the term $A$ in Eq.~\eq{dec} is certainly
zero in our calculation of the sfermion one-loop contributions. 

\subsection{The high energy limit}

\hspace*{12pt}
From the full expression of the one-loop helicity amplitudes  
of $\eeww$, compact analytic formulas in the high energy limit 
can be calculated, which are very useful for the test 
of the computational program of the full one-loop calculation. 
In this limit, there is the equivalence between $e^-e^+ \to W^-_LW^+_L$ 
and $e^-e^+ \rightarrow \chi^- \chi^+$, where $W^\pm_L$ denote the 
longitudinally polarized $W$ bosons.  
The equivalence holds even at the loop levels with some modification, by 
which we can confirm the analytic formulas mentioned above.
At high energy, the $00$, $+-$ and $-+$ helicity sets of the $W$-boson 
pair are important as discussed in Sec.~3.  
Since the $+-$ and $-+$ helicity amplitudes do not suffer from subtle 
gauge-theory cancellation, we discuss the 00 helicity amplitude here. 

\subsubsection{Analytic high-energy expressions of 
               $e^-e^+ \to W^-_LW^+_L$}

\hspace*{12pt}
By using high-energy formulas of integral functions (Appendix D),  
we obtain the high energy expression of the 00 helicity amplitude of 
$\eeww$:
\begin{eqnarray}
    {\cal M}(e^-e^+\rightarrow W^-_LW^+_L) = M_\tau^{00}
&\equiv& Q_e M_X^{00} + T_e^3 M_Y^{00}  ,   
\end{eqnarray}
where we denote $W_L$ as a longitudinally polarized 
($\lambdalambdabar = 0$) $W$-boson, and 
$M^{00}_X$ and $M^{00}_Y$ are expressed in terms of the $\msbar$
coupling constants of the MSSM by 
\begin{subequations}
\begin{eqnarray}
  M_X^{00} &=& \frac{\hatesq(\mu)}
                    {2 \hatcsq(\mu)} \nonumber 
+ \sum_{\rm generation} \frac{\hatg^4}{16 \pi^2} 
     \frac{{\hats}^4}{12 \hatc^4} 
 \left[ 3  \left\{  \frac{11}{9} 
        \ln \frac{s}{\mu^2}   
 - \frac{88}{27} \right\} 
 +\left\{    \ln \frac{s}{\mu^2}
        - 8    \right\}
 \right] \nonumber\\
&& + \hatgsq \frac{\hatssq}{2 \hatcsq} 
   \left( \delta Z_W + \frac{\Pi^{WW}_T(m_W^2)}{m_W^2} \right) 
+ {\cal O}\left( \frac{m_W^2}{s}\right) , \label{mx001}  \\
  M_Y^{00} &=& \frac{\hatgsq(\mu)}{\hatcsq(\mu)} 
       \left( \frac{1}{2} - \hatssq(\mu) \right)
 + \sum_{\rm generation}  \frac{\hatg^4}{16 \pi^2} 
     \frac{1}{12 \hatc^4} 
  \left[ 3 \left\{ \left( 1 - 2 \hatssq - \frac{2}{9} \hats^4 \right) 
             \ln \frac{s}{\mu^2} \right.\right. \nonumber  \\
&& \left.\left. +   
   \left( - 1 + 2 \hatssq + \frac{2}{9} \hats^4 \right) \frac{8}{3}
   \right\} + \left\{ \left( 1 - 2 \hatssq - 2 \hats^4 \right) 
             \ln \frac{s}{\mu^2} 
+ \left( - 1 +  \hatssq + 2 \hats^4 \right) \frac{8}{3}  \right\}
                    \right] 
\nonumber \\
&& + \hatgsq \frac{1}{2 \hatc^4}
\left( \delta Z_W + \frac{\Pi^{WW}_T(m_W^2)}{m_W^2} \right) 
+ {\cal O}\left( \frac{m_W^2}{s}\right) , \label{my001}
\end{eqnarray}
\end{subequations}
where the first curly bracket $\{ \;\;\; \}$ in RHS of each equation 
comes from the squark effects and the second one from the slepton
effects, respectively.
By using the Eqs.~\eq{eq-bysm1} and \eq{eq-bysm2} we obtain the expression
in terms of the SM coupling constants up to ${\cal O}(\hatg_{\rm SM}^4)$:  
\begin{subequations}
\label{he-exsm}
\begin{eqnarray}
  M_X^{00} &=& \frac{\hatesq_{\rm SM}} {2 \hatcsq_{\rm SM}} 
 + \sum_{\rm generation}  \frac{\hatg^4_{\rm SM}}{16 \pi^2} 
     \frac{\hats^4_{\rm SM}}{12 \hatc^4_{\rm SM}} 
 \left[ 3 \left\{  
        \left( \frac{8}{9} - \frac{2}{3} \cos^2 \theta_{\tilde{u}} \right) 
             \ln \frac{s}{m_{\tilde{u}_1}^2} 
   +   
        \left( \frac{8}{9} - \frac{2}{3} \sin^2 \theta_{\tilde{u}} \right) 
             \ln \frac{s}{m_{\tilde{u}_2}^2} \right. \right. \nonumber \\
&& \;\;\;\;\;\;\;\;\;\;\;\;\;\;\;\;\;\;\;\;\;\;\;    \left.
+
        \left( \frac{2}{9} - \frac{1}{3} \cos^2 \theta_{\tilde{d}} \right) 
             \ln \frac{s}{m_{\tilde{d}_1}^2}  
    +   
        \left( \frac{2}{9} - \frac{1}{3} \sin^2 \theta_{\tilde{d}} \right) 
             \ln \frac{s}{m_{\tilde{d}_2}^2}  
- \frac{88}{27}   \right\}   \nonumber \\
&& \left. \;\;\;\;\;\;\;\;\;\;\;\;\;\;\;\;\;\;\;\;\;\;\; +  
   \left\{ 
        \left( 1 -  \cos^2 \theta_{\tilde{e}} \right) 
           \ln \frac{s}{m_{\tilde{e}_1}^2}
        + 
        \left( 1 -  \sin^2 \theta_{\tilde{e}} \right) 
          \ln \frac{s}{m_{\tilde{e}_2}^2} 
      - 8    \right\} \right] \nonumber\\
&&   + \hatgsq_{\rm SM} \frac{\hatssq_{\rm SM}}{2 \hatcsq_{\rm SM}} 
\left( \delta Z_W + \frac{\Pi^{WW}_T(m_W^2)}{m_W^2} \right) 
+ {\cal O}\left( \frac{m_W^2}{s}\right), 
\label{mx002}\\
M_Y^{00} &=& \frac{\hatgsq_{\rm SM}}{\hatcsq_{\rm SM}} 
       \left( \frac{1}{2} - \hatssq_{\rm SM} \right) 
 + \sum_{\rm generation}  \frac{\hatg^4_{\rm SM}}{16 \pi^2} 
     \frac{1}{12 \hatc^4_{\rm SM}} \left[ 
 3 \left\{ \left( 
        \left( 1 - 2 \hatssq_{\rm SM} + 2 \hats^4_{\rm SM} \right) 
                    \frac{2}{3} \cos^2 \theta_{\tilde{u}} 
             - \frac{8}{9} \hats^4_{\rm SM}     \right) 
             \ln \frac{s}{m_{\tilde{u}_1}^2} \right. \right. \nonumber \\ 
&& \;\;\;\;\;\;\;\;\;\;\;\;\;\;\;
     +  \left( 
        \left( 1 - 2 \hatssq_{\rm SM} + 2 \hats^4_{\rm SM} \right) 
                     \frac{2}{3} \sin^2 \theta_{\tilde{u}} 
             - \frac{8}{9} \hats^4_{\rm SM}     \right) 
             \ln \frac{s}{m_{\tilde{u}_2}^2}   \nonumber \\
&& \;\;\;\;\;\;\;\;\;\;\;\;\;\;\;
+ \left( 
        \left( 1 - 2 \hatssq_{\rm SM} + 2 \hats^4_{\rm SM} \right) 
                     \frac{1}{3} \cos^2 \theta_{\tilde{d}} 
             - \frac{2}{9} \hats^4_{\rm SM}     \right) 
             \ln \frac{s}{m_{\tilde{d}_1}^2} \nonumber \\ 
&& \;\;\;\;\;\;\;\;\;\;\;\;\;\;\;
  \left. +  \left( 
        \left( 1 - 2 \hatssq_{\rm SM} + 2 \hats^4_{\rm SM} \right) 
                     \frac{1}{3} \sin^2 \theta_{\tilde{d}} 
             - \frac{2}{9} \hats^4_{\rm SM}     \right) 
             \ln \frac{s}{m_{\tilde{d}_2}^2}     
    +    
   \left( - 1 + 2 \hatssq_{\rm SM} + \frac{2}{9} \hats^4_{\rm SM}
       \right) \frac{8}{3} 
   \right\}   \nonumber \\
&& \;\;\;\;\;\;\;\;\;\;\;\;\;\;\;
 + \left\{ \left( 
        \left( 1 - 2 \hatssq_{\rm SM} + 2 \hats^4_{\rm SM} \right) 
                   \cos^2 \theta_{\tilde{e}} 
             - 2  \hats^4_{\rm SM}     \right) 
             \ln \frac{s}{m_{\tilde{e}_1}^2} \right. \nonumber \\ 
&& \;\;\;\;\;\;\;\;\;\;\;\;\;\;\;
    \left. \left. +  \left( 
        \left( 1 - 2 \hatssq_{\rm SM} + 2 \hats^4_{\rm SM} \right) 
                    \sin^2 \theta_{\tilde{e}} 
             - 2 \hats^4_{\rm SM}     \right) 
             \ln \frac{s}{m_{\tilde{e}_2}^2}  
     + \left( - 1 +  \hatssq_{\rm SM} + 2 \hats^4_{\rm SM} \right)\frac{8}{3}  
       \right\} \right] 
\nonumber \\
&& + \hatgsq_{\rm SM} \frac{1}{2 \hatc^4_{\rm SM}}
\left( \delta Z_W + \frac{\Pi^{WW}_T(m_W^2)}{m_W^2} \right) 
+ {\cal O}\left( \frac{m_W^2}{s}\right) . 
  \label{my002}
\end{eqnarray}
\end{subequations}

\subsubsection{The equivalence theorem}

\hspace*{12pt}
By the similarity of the polarization vectors between 
longitudinal and scalar  $W$ bosons, 
the relation  
\begin{eqnarray}
   {\cal M}(e^-e^+ \rightarrow W^-_LW^+_L) = 
\left\{ i C_{\rm mod}^{ET} \right\}^2 
 {\cal M}(e^-e^+ \rightarrow \chi^-\chi^+) 
\left( 1 + {\cal O}(m_W/\sqrt{s}) \right) , \label{et} 
\end{eqnarray}
is induced from the single and double BRS identities.
The modification factor $C_{\rm mod}^{ET}$, 
which is not unity beyond the tree level, is expressed at one-loop by  
\begin{eqnarray}
  C_{\rm mod}^{ET} = \left( \frac{Z_W}{Z_\chi} \right)^{\frac{1}{2}}
                 C_{\rm mod}^{BRS}  ,
\end{eqnarray}
in our scheme\footnote{
In the generic renormalization scheme, $C_{\rm mod}^{ET}$ 
  should be expressed by 
    $C_{\rm mod}^{ET} = (Z_W/Z_\chi)^{1/2}
                 Z_{m_W}  \hat{C}_0(m_W^2, \xi)$,  where 
    $\hat{C}_0(m_W^2, \xi)$ depends on the gauge parameter $\xi$\cite{et2} 
    and it is 1 in our scheme.}.
The relation \eq{et} is what we call the equivalence theorem\cite{et,et2}.  

We here show that the high energy expressions of the one-loop amplitudes 
\eq{mx001} and \eq{my001} are tested by using the equivalence theorem. 
To this aim, we also calculate the high energy expression of the amplitude 
of $e^-e^+ \rightarrow \chi^-\chi^+$.   
The amplitude for $e^-e^+ \rightarrow \chi^-\chi^+$ is expressed 
at high energy by    
\begin{eqnarray}
    {\cal M}(e^-e^+\rightarrow \chi^-\chi^+)
    &\equiv& Q_e M_X^{\chi\chi} + T_e^3 M_Y^{\chi\chi} 
   + {\cal O}\left( \frac{m_W^2}{s}\right) , 
\end{eqnarray}
where $M_X^{\chi\chi}$ and $M_Y^{\chi\chi}$ are expressed in terms of 
the MSSM couplings by 
\begin{subequations}
\begin{eqnarray}
  M_X^{\chi\chi} &=& - \frac{\hatesq(\mu)}
                    {2 \hatcsq(\mu)} 
- \sum_{\rm generation} \frac{\hatg^4}{16 \pi^2} 
     \frac{\hats^4}{12 \hatc^4} 
 \left[ 3  \left\{  \frac{11}{9} 
                      \ln \frac{s}{\mu^2}   
                    - \frac{88}{27} \right\} 
 +\left\{   \ln \frac{s}{\mu^2}
           - 8    \right\} 
 \right] + {\cal O} \left( \frac{\mwsq}{s} \right), \label{mxxx} \\
  M_Y^{\chi\chi} &=& - \frac{\hatgsq(\mu)}{\hatcsq(\mu)} 
       \left( \frac{1}{2} - \hatssq(\mu) \right)
 - \sum_{\rm generation}  \frac{\hatg^4}{16 \pi^2} 
     \frac{1}{12 \hatc^4} 
  \left[ 3 \left\{ \left( 1 - 2 \hatssq - \frac{2}{9} \hats^4 \right) 
             \ln \frac{s}{\mu^2} +    
   \left( - 1 + 2 \hatssq + \frac{2}{9} \hats^4 \right) \frac{8}{3} 
   \right\} \right.   \nonumber \\
&& \left.
\;\;\;\;\;\;\;\;\;\;\;\;\;\;\;\;\;\;\;\;\;\;\;\;\;\;\;\;\;\;
 + \left\{ \left( 1 - 2 \hatssq - 2 \hats^4 \right) 
             \ln \frac{s}{\mu^2} 
+ \left( -1 + 2 \hatssq + 2 \hats^4 \right) \frac{8}{3} \right\}
 \right] + {\cal O} \left( \frac{\mwsq}{s} \right). \label{myxx}
\end{eqnarray}
\end{subequations}
On the other hand, $C_{\rm mod}^{BRS}$ of \eq{cmod-brs} 
is expressed by     
\begin{eqnarray}
  C_{\rm mod}^{BRS} &=& 1 + \frac{1}{2 m_W^2} \Pi_T^{WW}(m_W^2). \label{c2}
\end{eqnarray} 
For the Goldstone wavefunction factors, we set $Z_\chi  = 1$  
when the sfermion masses are neglected. 
Thus, the analytic expression of the modification factor 
$C_{\rm mod}^{ET}$ is calculated as 
\begin{eqnarray}
  C_{\rm mod}^{ET} = 1 + \frac{1}{2} 
 \left( \delta Z_W + \frac{\Pi_T^{WW}(m_W^2)}{m_W^2} \right). \label{cet}
\end{eqnarray}
By using \eq{cet} and the high energy expression of the amplitudes  
of $\eeww$ (Eqs.~\eq{mx001} and \eq{my001}), 
and of $e^-e^+ \rightarrow \chi^-\chi^+$ (Eqs.~\eq{mxxx} and \eq{myxx}), 
one can see that the leading equation of 
the equivalence theorem \eq{et} 
holds among the high energy expressions of the one-loop amplitudes.  

\subsubsection{Numerical test for the high energy behavior}

\begin{figure}[th]
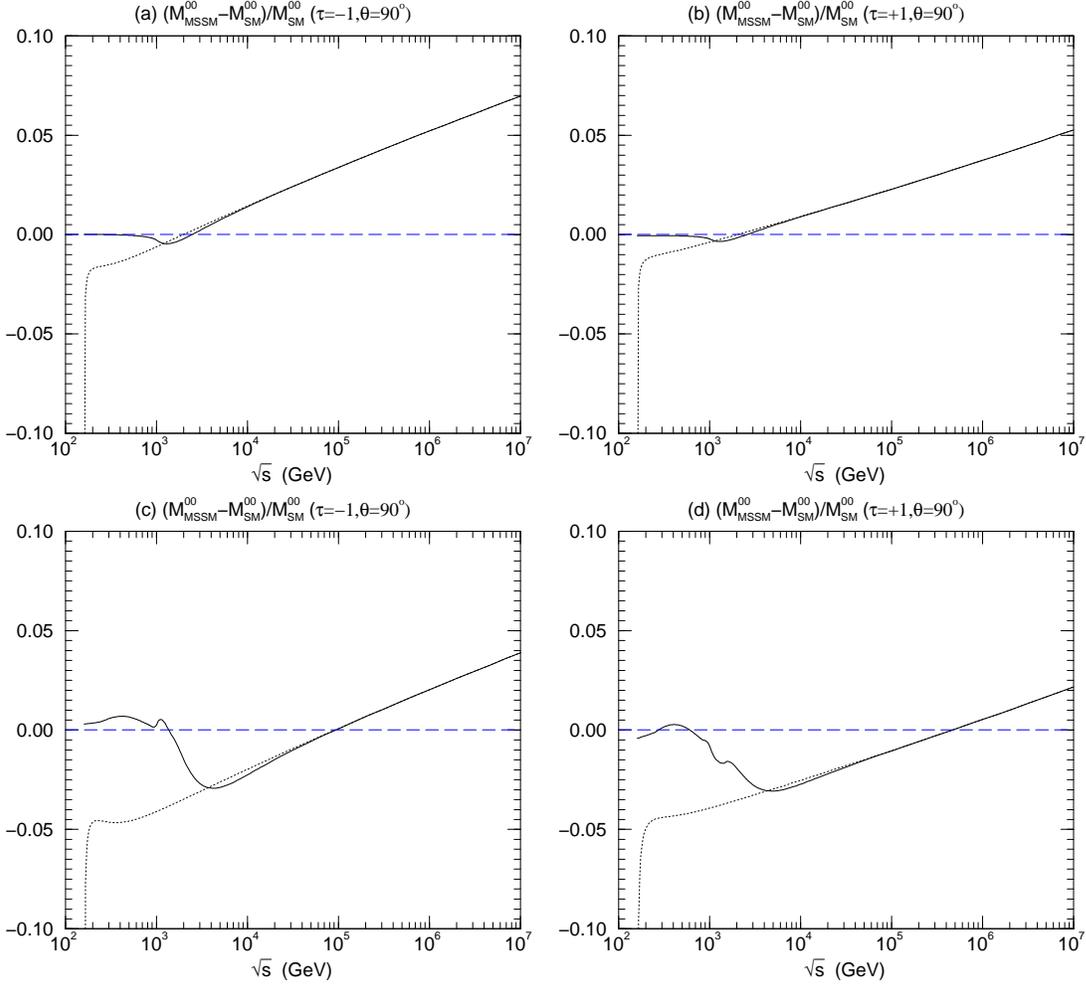

\begin{center}
$\begin{array}{rl}
\leavevmode\psfig{figure=m00_HAdy.epsi,width=7cm} &
\leavevmode\psfig{figure=m00_HAdx.epsi,width=7cm} \\
\leavevmode\psfig{figure=m00_HBdy.epsi,width=7cm} &
\leavevmode\psfig{figure=m00_HBdx.epsi,width=7cm}
\end{array}$
\caption{The high-energy behaviors of the sfermion one-loop contribution  
         to $M_\tau^{00}$ at $\theta = 90^{\circ}$.
         The solid lines show  the full calculation 
         of the sfermion one-loop contribution, while  
         the dotted lines are of the high-energy 
         analytic formulas~(\ref{he-exsm}).
The parameters are set to  
$m_{\tilde{Q}}=m_{\tilde{L}}=m_{\tilde{U}}
 =m_{\tilde{D}}=m_{\tilde{E}}=500$GeV for $A_f^{eff}=0$
in (a) and (b), and for $A_f^{eff}=1500$GeV in (c) and (d).  
}
\label{fig:high-en}
\end{center}
\end{figure}

\hspace*{12pt}
In Fig.~\ref{fig:high-en}, we compare the full expressions of 
the one-loop 00 helicity ($W_L^-W_L^+$ production) amplitude  and 
the high-energy expressions given by Eqs.~(\ref{mx002}) 
and (\ref{my002}). 
Figs.~8(a) and 8(c) show the amplitudes of $W_LW_L$ production 
from left-handed electrons $\tau=-1$,  
whereas Figs.~8(b) and 8(d) 
show those from right-handed electrons ($\tau=+1$).  
%
In each figure, the full one-loop results are shown by solid lines, 
and dotted lines represent the high-energy analytic results of 
Eqs.~(\ref{he-exsm}).
The cases with $m_{\tilde{Q}}=m_{\tilde{L}}=m_{\tilde{U}}
 =m_{\tilde{D}}=m_{\tilde{E}}=500$GeV 
are shown in Figs.~8(a) and 8(b) ($A_f^{eff}=0$GeV), and 
in Figs.~8(c) and 8(d) ($A_f^{eff}=1500$GeV).  
Here the parameter $A_f^{eff}$ give the left-right mixing matrix elements  
of the sfermion mass matrices; see Eq.~(\ref{eq:afeff}) of Appendix A.  
With the large value of $A_f^{eff}$ shown in Figs.~8(c) and 8(d), 
there appears significant mixing in the stop sector, 
and the mass eigenvalues become 
$m_{\tilde{t}_1} = 130$GeV and 
$m_{\tilde{t}_2} = 736$GeV for $\tan\beta=2$\footnote{
Since we use the effective parameter $A_f^{eff}$ of Eq.~(\ref{eq:afeff}) 
to represent the left-right mixing term, the sfermion parameters 
depend on $\tan\beta$ only through the $\cos2\beta$ terms in the 
diagonal mass-squark matrix elements.  
The $\tan\beta$ dependence from these terms are small, and we set 
$\tan\beta=2$ in all the numerical results presented in this paper.}.   
The high-energy prediction of Eqs.~(\ref{he-exsm}) is useful in understanding 
the normalization of the corrections at high energies for the case with 
the large mass mixing.  
We find large negative correction above 
the stop-pair ($\tilde{t}$-$\bar{\tilde{t}}$) production 
thresholds except before the asymptotic regime sets in. 
In the formulas~(\ref{he-exsm}), 
this negative correction comes from the constant part,  
\begin{eqnarray}
        \delta Z_W + \frac{\Pi^{WW}_T(m^2_W)}{m^2_W},  \label{const-term}
\end{eqnarray}
whose magnitude grows when there is a significant mass mixing in 
the sfermion sector.  
In the full amplitudes the stop contributions around the threshold 
compensate for this large negative constant, which makes
the correction small below the $\tilde{t}_2$-$\bar{\tilde{t}}_2$ threshold.  
%

%
Therefore, the equivalence theorem is useful to confirm the high-energy 
formulas of the (00) amplitude, which are then used to test the stability 
of our numerical program at high energies and the overall normalization 
corrections to the amplitudes.  
  
\subsection{Summary of the tests}

\hspace*{12pt}
We have tested our one-loop calculation of the form factors 
by using the three methods; 
1) the BRS sum rules for the test of our one-loop form factors 
   except for the overall normalization corrections,  
2) the decoupling theorem for the test of such overall normalization of our 
   amplitudes,    
3) the high-energy analytic expressions of the amplitudes for the test of 
   the stability of the numerical program 
   at high energies where subtle gauge-theory cancellation occurs.  
In the following section, we study the magnitude of the one-loop sfermion 
contribution to the $\eeww$ helicity amplitudes in various 
the MSSM parameter sets.

\section{The numerical evaluation of $\eeww$}
\cleqn

\subsection{The helicity amplitudes}

\hspace*{12pt}
As discussed in Sec.~3, among all the tree-level helicity amplitudes,  
$M^{+-}_{\tau=-1}$,  $M^{-+}_{\tau=-1}$ and $M^{00}_{\tau=\mp 1}$
are significant for all energies: see 
Fig.~\ref{fig:amp2vstot}. 
The one-loop sfermion contributions to these helicity 
amplitudes should be examined in detail. 
On the other hand, $M^{++}_{\tau=\mp 1}$ and $M^{--}_{\tau=\mp 1}$ reduce 
by ${\cal{O}}(1/s)$, while $M^{0+}_{\tau=\mp 1}$, 
$M^{-0}_{\tau=\mp 1}$, $M^{0-}_{\tau=\mp 1}$ and $M^{+0}_{\tau=\mp 1}$
decrease as ${\cal{O}}(1/\sqrt{s})$. Therefore, the one-loop contributions 
to these helicity amplitudes can be almost neglected at high energies. 
Note that 
only $M^{0+}_{\tau=-1}$ and $M^{-0}_{\tau=-1}$ are, however, larger than 
$M^{00}_{\tau=-1}$ for $\sqrt{s} < 274$GeV, so that the one-loop
contributions to these helicity amplitudes may also be valuable to 
discuss at low energies.
In the following, we show the sfermion one-loop contributions 
to $M^{+-}_{\tau=-1}$ and  $M^{-+}_{\tau=-1}$, 
   $M^{00}_{\tau=\mp 1}$, 
   $M^{0+}_{\tau=-1}$ and $M^{-0}_{\tau=-1}$,  
at the large scattering angle ($\theta=90^\circ$).   
The net sfermion contributions in each helicity amplitude are given by    
\begin{eqnarray}
 \frac{{M_{\tau}^{\lambda \overline{\lambda}}}_{\rm MSSM}
       - {M_{\tau}^{\lambda \overline{\lambda}}}_{\rm SM}}
  {{M_{\tau}^{\lambda \overline{\lambda}}}_{\rm SM}}, 
\end{eqnarray}
where ${M_{\tau}^{\lambda \overline{\lambda}}}_{\rm MSSM}$ are the helicity 
amplitudes of the MSSM in which only sfermion contributions are considered, 
and 
${M_{\tau}^{\lambda \overline{\lambda}}}_{\rm SM}$ are those of the SM. 

We consider the twenty cases in Tables~\ref{tab:set1}, \ref{tab:set2}, 
\ref{tab:set3} and \ref{tab:set4} as parameter choice, 
which are categorized in the 4 groups. The first two are the following.    
\begin{description}
\item[Set 1:] 
Case~1 - Case~5 in Table~\ref{tab:set1},  
in which only the sleptons are light, and all the squarks are heavy 
enough. 
The mass eigenstates $\tilde{l}_1$ and $\tilde{l}_2$ coincident with 
$\tilde{l}_L^{}$ and $\tilde{l}_R^{}$, respectively. 
\item[Set 2:]
Case~6 - Case~10 in Table~\ref{tab:set2}, in which only the squarks 
of the first and the second generations are light and all the other 
sfermions are heavy enough to decouple, where 
the mass eigenstates $\tilde{q}_1$ and $\tilde{q}_2$ coincident with 
$\tilde{q}_L^{}$ and $\tilde{q}_R^{}$, respectively.   
\end{description}
The rest two groups are for the cases where only squarks in the third 
generation, which we refer to the $(\tilde{t}, \tilde{b}$) sector,  
are light. In this sector, the large mass mixing between 
$\tilde{t}_L^{}$ and $\tilde{t}_R^{}$ is 
possible by the large top quark mass. 
\begin{description}
\item[Set 3:]
Case~11 - Case~15 in Table~\ref{tab:set3}, where 
the $(\tilde{t}, \tilde{b}$) sector is considered 
without the $\tilde{t}_L^{}$-$\tilde{t}_R^{}$ mixing.
\item[Set 4:]
Case~16 - Case~20 in Table~\ref{tab:set4}, where 
the $(\tilde{t}, \tilde{b}$) sector is considered 
with the large mass mixing between $\tilde{t}_L^{}$ and $\tilde{t}_R^{}$ 
with the angle $\tilde{\theta}_t \sim \pi/4$. 
\end{description}
In each parameter set, the sfermion masses that we directly do not consider  
should be regarded to be sufficiently large.  
Since we adopt the SM couplings as our expansion parameters so that 
the decoupling of such heavy particles is exact at the one-loop level, 
the contribution of these heavy sfermions can be removed simply by 
dropping their explicit contribution.  

\begin{table}[b]
\begin{center}
\begin{tabular}{l|rrrrr}
 {\bf Set 1:} & Case 1 & Case 2 & Case 3 & Case 4 
              & Case 5   \\ \hline
\multicolumn{6}{l}{Input parameters }           \\ \hline
$m_{\tilde{L}}$ & 100 & 200 & 300 & 500 &1000 \\
$m_{\tilde{E}}$ & 100 & 200 & 300 & 500 &1000 \\
$A_{f}^{eff}$&   0 &   0 &   0 &   0 &   0 \\ 
\hline  
\multicolumn{6}{l}{Output parameters} \\ \hline
$m_{\tilde{\nu}_e}\!=\!m_{\tilde{\nu}_{\mu}}\!=\!m_{\tilde{\nu}_{\tau}}$
  &  85 & 193 & 295 & 497 &  999 \\
$m_{\tilde{e}_1}\!=\!m_{\tilde{\mu}_1}\!\sim\!m_{\tilde{\tau}_1}$ 
  & 105 & 203 & 302 & 501 & 1001 \\
$m_{\tilde{e}_2}\!=\!m_{\tilde{\mu}_2}\!\sim\!m_{\tilde{\tau}_2}$
  & 109 & 205 & 303 & 502 & 1001 \\
\hline\hline
\end{tabular}
\end{center}
\caption{The cases where only sleptons are light.
The squarks are taken to be sufficiently heavy. }
\label{tab:set1}
\end{table}

\begin{table}[p]
\begin{center}
\begin{tabular}{l|rrrrr}
{\bf Set 2}: & Case 6 & Case 7 & Case 8 & Case 9 
            & Case 10   \\ \hline
\multicolumn{6}{l}{Input parameters }           \\ \hline
$m_{\tilde{Q}}$ & 100 & 200 & 300 & 500 &1000 \\
$m_{\tilde{U}}$=  
$m_{\tilde{D}}$ & 100 & 200 & 300 & 500 &1000 \\
$A_{\tilde{f}}^{eff}$&   0 &   0 &   0 &   0 &   0  \\ 
\hline  
\multicolumn{6}{l}{Output parameters} \\ \hline
$m_{\tilde{u}_1}\!=\!m_{\tilde{c}_1}$ &  91 & 196 & 297 & 498 &  999 \\
$m_{\tilde{u}_2}\!=\!m_{\tilde{c}_2}$ &  93 & 197 & 298 & 499 &  999 \\
$m_{\tilde{d}_1}\!=\!m_{\tilde{s}_1}$ & 111 & 206 & 304 & 502 & 1001 \\
$m_{\tilde{d}_2}\!=\!m_{\tilde{s}_2}$ & 103 & 202 & 301 & 501 & 1000 \\
\hline\hline 
\end{tabular}
\end{center}
\vspace*{-0.4cm}
\caption{Cases where only squarks from the first-two generations are light.
         The other sfermions are taken to be sufficiently heavy. }
\label{tab:set2}
\begin{center}
\begin{tabular}{l|rrrrr}
{\bf Set 3:} & Case 11 & Case 12 & Case 13 & Case 14 
             & Case 15   \\ \hline
\multicolumn{6}{l}{Input parameters }           \\ \hline
$m_{\tilde{Q}}$ & 100 & 200 & 300 & 500 &1000 \\
$m_{\tilde{U}}$=  
$m_{\tilde{D}}$ & 100 & 200 & 300 & 500 &1000 \\
$A_{\tilde{f}}^{eff}$&   0 &   0 &   0 &   0 &   0  \\
\hline  
\multicolumn{6}{l}{Output parameters} \\ \hline
$m_{\tilde{t}_1}$ & 197 & 263 & 345 & 528 & 1014 \\
$m_{\tilde{t}_2}$ & 199 & 263 & 346 & 529 & 1015 \\
$m_{\tilde{b}_1}$ & 111 & 206 & 304 & 502 & 1001 \\
$m_{\tilde{b}_2}$ & 103 & 202 & 301 & 501 & 1000 \\
\hline\hline 
\end{tabular}
\end{center}
\vspace*{-0.4cm}
\caption{5 sets of mass parameters of $\tilde{t}$ and 
$\tilde{b}$ are listed in which no mass mixing between 
$\tilde{t}_L^{}$-$\tilde{t}_R^{}$ is considered. 
The sfermions which we do not consider are sufficiently large. 
}
\label{tab:set3}
\begin{center}
\begin{tabular}{l|rrrrr}
{\bf Set 4:}  & Case 16 & Case 17 & Case 18 & Case 19 
              & Case 20   \\ \hline
\multicolumn{6}{l}{Input parameters }           \\ \hline
$m_{\tilde{Q}}$ & 100 & 200 & 300 & 400 &500 \\
$m_{\tilde{U}}$=  
$m_{\tilde{D}}$ & 100 & 200 & 300 & 400 &500 \\
$A_{\tilde{f}}^{eff}$& 168 & 339 & 625 & 1025 & 1539 \\ \hline
\hline  
\multicolumn{6}{l}{Output parameters} \\ \hline
$m_{\tilde{t}_1}$ & 100 & 100 & 100 & 100 &  100 \\
$m_{\tilde{t}_2}$ & 262 & 358 & 478 & 607 & 741 \\
$m_{\tilde{b}_1}$ & 111 & 206 & 304 & 403 & 502 \\
$m_{\tilde{b}_2}$ & 103 & 202 & 301 & 401 & 501 \\
$\cos\theta_{\tilde{t}}$ &0.710&0.708&0.708&0.708&0.707 \\
\hline\hline 
\end{tabular}
\end{center}
\vspace*{-0.4cm}
\caption{Cases where the squarks of the $(\tilde{t},\tilde{b})$ sector 
are light with the large $\tilde{t}_L^{}$-$\tilde{t}_R^{}$ mass mixing.
The mass of $\tilde{t}_1^{}$ is fixed to be 100GeV in these cases.
The sfermions which we do not consider are sufficiently large. 
}
\label{tab:set4}
\end{table}

\subsubsection{The sfermion one-loop contributions to 
$M_\tau^{+-}$ and $M_\tau^{-+}$}

\hspace*{12pt}
The tree-level helicity amplitudes $M^{+-}_{\tau=-1}$ and $M^{-+}_{\tau=-1}$  
are the largest of all the helicity amplitudes at the scattering angle 
$\theta=90^\circ$. 
Since the $(+-)$ and $(-+)$ helicity-set processes contain only the 
$t$-channel diagrams, these amplitudes contain only wavefunction 
renormalization factors as the sfermion one-loop contribution. 
Thus, they are almost independent of $\sqrt{s}$ 
and determined by the logarithmic function of the sfermion
masses and the $W$-boson mass.  
In Table~\ref{tab:m+-}, we list the sfermion corrections to the 
$M^{+-}_{\tau=-1}$ at $\theta = 90^\circ$ for 
$\sqrt{s} = 200$GeV and $1000$GeV.  
As we expected, we see that the corrections are insensitive to $\sqrt{s}$.
The magnitude of the sfermion contributions to this helicity amplitude is 
 rather small for all cases we consider. 
The contributions of the stop-sbottom sector with a large mass mixing 
give the biggest contribution to $M^{+-}_{\tau=-1}$, 
where the deviation from the SM is less than $+0.1$\% (Case 20). 
The corrections to  $M^{-+}_{\tau=-1}$ are 
the same as those to $M^{+-}_{\tau=-1}$ at $\theta = 90^\circ$.

\begin{table}[t]
\begin{center}
\begin{tabular}{|r|r|r||r|r|r|}
\hline
Case & \multicolumn{2}{l|}{$\frac{M^{+-}_{\rm MSSM}-M^{+-}_{\rm
SM}}{M^{+-}_{\rm SM}}$ } & 
Case & \multicolumn{2}{l|}{$\frac{M^{+-}_{\rm MSSM}-M^{+-}_{\rm
SM}}{M^{+-}_{\rm SM}}$} \\ \hline
& $\sqrt{s}=200$GeV & 1000GeV & & $\sqrt{s}=200$GeV & 1000GeV \\ \hline
1  & $-4.5\!\times\!10^{-4}$ & $-4.4\!\times\!10^{-4}$ & 
11 & $-1.6\!\times\!10^{-4}$ & $-1.5\!\times\!10^{-4}$\\
2  & $-1.1\!\times\!10^{-4}$ & $-1.1\!\times\!10^{-4}$ & 
12 & $-1.1\!\times\!10^{-4}$ & $-1.1\!\times\!10^{-4}$\\
3  & $-4.8\!\times\!10^{-5}$ & $-4.7\!\times\!10^{-5}$ & 
13 & $-6.4\!\times\!10^{-5}$ & $-6.2\!\times\!10^{-5}$\\
4  & $-1.7\!\times\!10^{-5}$ & $-1.7\!\times\!10^{-5}$ & 
14 & $-2.7\!\times\!10^{-5}$ & $-2.7\!\times\!10^{-5}$\\
5  & $-4.3\!\times\!10^{-6}$ & $-4.2\!\times\!10^{-6}$ & 
15 & $-7.4\!\times\!10^{-6}$ & $-7.2\!\times\!10^{-6}$\\
6  & $-1.5\!\times\!10^{-4}$ & $-1.5\!\times\!10^{-4}$ & 
16 & $-8.8\!\times\!10^{-5}$ & $-8.5\!\times\!10^{-5}$\\
7  & $-4.2\!\times\!10^{-5}$ & $-4.1\!\times\!10^{-5}$ & 
17 & $ 2.0\!\times\!10^{-4}$ & $ 2.0\!\times\!10^{-4}$\\
8  & $-1.8\!\times\!10^{-5}$ & $-1.8\!\times\!10^{-5}$ & 
18 & $ 4.7\!\times\!10^{-4}$ & $ 4.6\!\times\!10^{-4}$\\
9  & $-6.8\!\times\!10^{-6}$ & $-6.7\!\times\!10^{-6}$ & 
19 & $ 6.9\!\times\!10^{-4}$ & $ 6.7\!\times\!10^{-4}$\\
10 & $-1.7\!\times\!10^{-6}$ & $-1.7\!\times\!10^{-6}$ & 
20 & $ 8.6\!\times\!10^{-4}$ & $ 8.4\!\times\!10^{-4}$\\
\hline
\end{tabular}
\end{center}
\vspace*{-0.4cm}
\caption{The sfermion one-loop contributions to the $M^{+-}_{\tau=-1}$ 
amplitude are listed in Case~1 - Case~20.}
\label{tab:m+-}
\end{table}

\subsubsection{The sfermion one-loop contributions to $M_{\tau}^{00}$}

\begin{figure}[t]
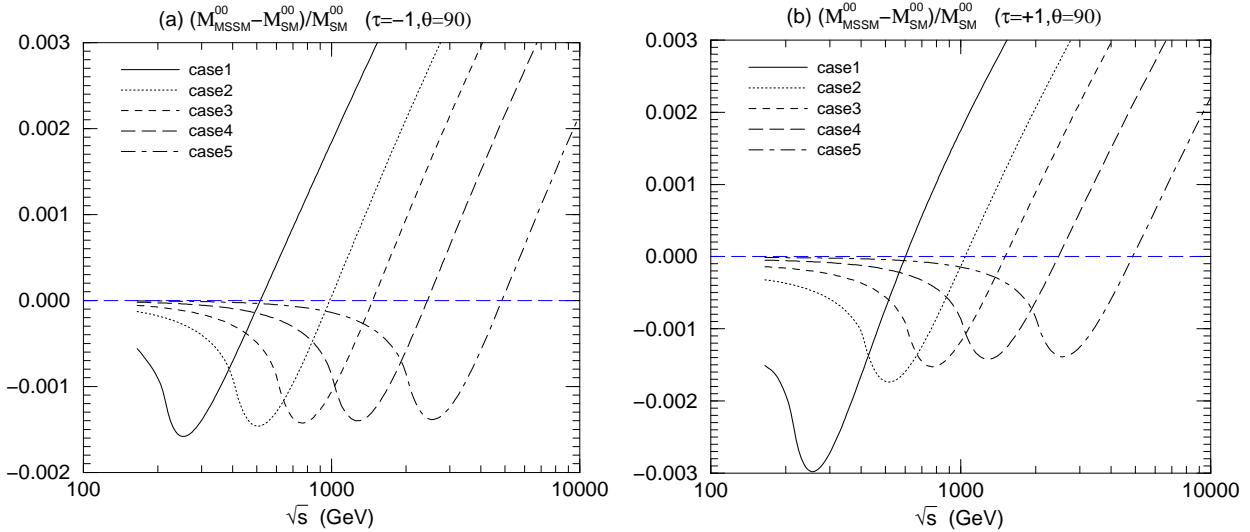

\begin{center}
$\begin{array}{rl}
\leavevmode\psfig{figure=amp00_lep_minus.epsi,width=8cm}& 
\leavevmode\psfig{figure=amp00_lep_plus.epsi,width=8cm} 
\end{array}$
\end{center}
\caption{The sfermion one-loop corrections to $M^{00}_\tau$ 
for Set~1 (Case~1 - Case~5), in which only the sleptons are light. 
}
\label{fig:amp00-1}
\end{figure}

\begin{figure}[t]
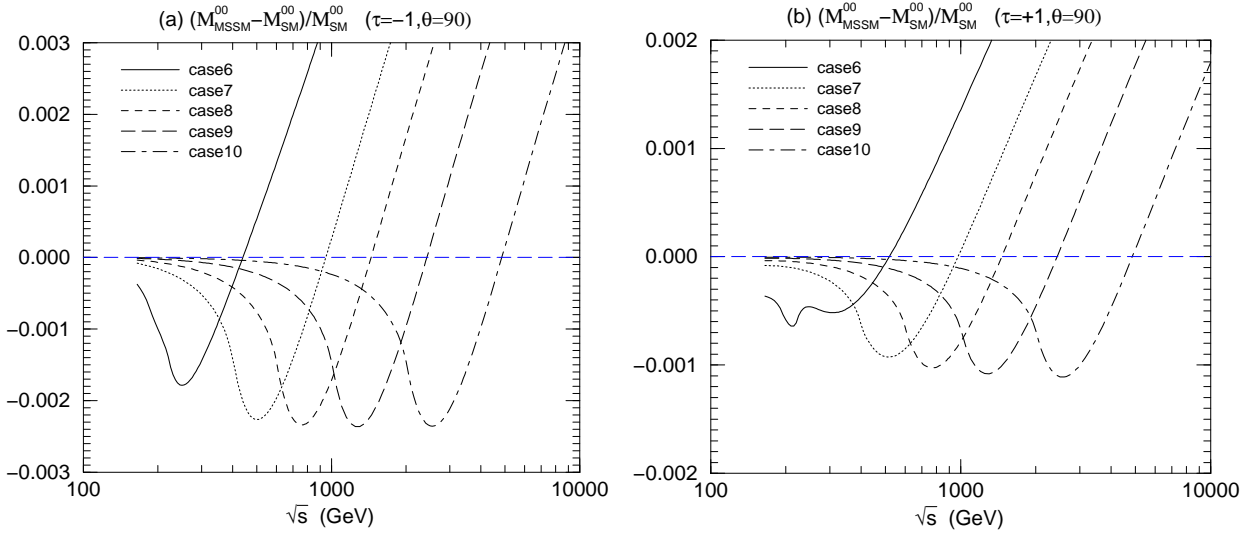

\begin{center}
$\begin{array}{rl}
\leavevmode\psfig{figure=amp00_qu_2gene_minus.epsi,width=8cm}&
\leavevmode\psfig{figure=amp00_qu_2gene_plus.epsi,width=8cm} 
\end{array}$
\end{center}
\caption{The sfermion one-loop corrections to $M^{00}_{\tau}$ 
are shown for Set~2 (Case~6 - Case~10), in which  
only the  squarks of the first-two generations are light 
}
\label{fig:amp00-2}
\end{figure}

\begin{figure}[tp]
\begin{center}
$\begin{array}{cc}
\leavevmode\psfig{figure=amp00_tb_nomix_minus.epsi,width=8cm}& 
\leavevmode\psfig{figure=amp00_tb_nomix_plus.epsi,width=8cm} 
\end{array}$
\end{center}
\caption{The sfermion one-loop contributions to $M^{00}_\tau$  
are shown for Set~3 (Case~11 - Case~15), in which  only the squarks 
of the third generation, $\tilde{t}$ and $\tilde{b}$, are light. 
The mixing between $\tilde{t}_L^{}$ and $\tilde{t}_R^{}$ 
is zero in these cases.
}
\label{fig:amp00-3}
%
\begin{center}
$\begin{array}{cc}
\leavevmode\psfig{figure=amp00_tbsame_minus.epsi,width=8cm}&
\leavevmode\psfig{figure=amp00_tbsame_plus.epsi,width=8cm} 
\end{array}$
\end{center}
\caption{The sfermion one-loop contributions to $M^{00}_\tau$ 
for Set~4 (Case~16 - Case~20), in which  only the squarks 
of the third generation, $\tilde{t}$ and $\tilde{b}$, are light with 
the nonzero mass mixing between $\tilde{t}_L^{}$ and $\tilde{t}_R^{}$. 
$m^{}_{\tilde{t}_1^{}}$=100GeV is fixed in these cases.
}
\label{fig:amp00-4}
\end{figure}

\hspace*{12pt}
We next consider the sfermion 
one-loop contributions to  $M^{00}_{\tau}$ at $\theta = 90^\circ$, 
which are expected to have rich structures because 
$M^{00}_\tau$ receive contributions from all the sfermion one-loop diagrams.
The slepton contributions to $M^{00}_{\tau=\mp 1}$ are described by 
using Set~1 in Table~\ref{tab:set1} and the results are shown in 
Figs.~\ref{fig:amp00-1}(a) and \ref{fig:amp00-1}(b).   
See the $\tau=-1$ amplitude in Fig.~\ref{fig:amp00-1}(a) first.  
The slepton corrections contribute destructively to the SM amplitudes 
below the threshold of the sneutrino-pair production.  
The magnitude of the negative deviation from the SM reaches to its 
maximum slightly above the threshold, and the maximal deviation is 
about $-0.15\%$ in amplitudes, which is almost independent of the 
slepton masses.  
Second, for the $\tau=+1$ amplitudes shown in Fig.~\ref{fig:amp00-1}(b), 
the larger deviation is observed in Case~1, but it may be difficult 
to observe this effect because of the smallness of the $\tau=+1$ amplitudes: 
see Fig.~5. 

The contributions of the squarks of the first-two generations, 
Set~2 of Table~\ref{tab:set2}, are shown in Figs.~\ref{fig:amp00-2}(a) 
and \ref{fig:amp00-2}(b). The qualitative behavior is quite similar to 
that of the slepton contributions of Set~1.
The magnitude of the squark contributions per a generation is larger 
than that of the slepton contributions for the similar mass sets.
In Fig.~\ref{fig:amp00-2}(a), 
the corrections to the $\tau=-1$ amplitude amount to $-0.24$\% 
at the peak slightly above the thresholds of the $\tilde{u}$-pair 
productions and the $\tilde{c}$-pair productions. 
In Fig.~\ref{fig:amp00-2}(b), the contributions to the $\tau=+1$ amplitude 
in these cases are similar to those in Set~1, but the magnitude is smaller.

The contributions of the ($\tilde{t},\tilde{b}$) sector are rather 
interesting. The cases without the $\tilde{t}_L^{}$-$\tilde{t}_R^{}$ 
mass mixing are given in Set~3 of Table~\ref{tab:set3}. 
See the curve of Case 11 in Fig.~\ref{fig:amp00-3}(a), where the 
thresholds of the $\tilde{t}$-pair production are 394-398GeV 
and those of the sbottom-pair production are 206-222GeV. 
The corrections to the $\tau=-1$ amplitude are positive around the first 
threshold of the sbottom-pair productions and the deviation 
from the SM amounts to $+0.25$\% in amplitudes 
at the first peak above the $\tilde{b}$-pair thresholds.
Around the thresholds of the $\tilde{t}$-pair productions, the correction 
rapidly reduces and the deviation changes its sign 
from positive to negative due to the constant part \eq{const-term} 
in the amplitude. 
Beyond the negative peak around 1000GeV, where the deviation 
amounts to $-0.17$\% in amplitudes, the correction behaves asymptotically 
according to the analytic high-energy formulas discussed in Sec.~5.3.
The qualitative characteristics in Case~11 are common with the 
other 4 cases of Set~3 (Case~12 - Case~15), but the corrections 
at low energy becomes smaller as the masses of stops and sbottoms 
are set to be larger.
For the $\tau=+1$ amplitudes in Fig.~\ref{fig:amp00-3}(b), 
corrections around the first threshold of the sbottom-pair production 
are all negative. The corrections are larger 
than in the $\tau=-1$ amplitudes because the constant term 
of \eq{const-term} enlarges negative squark contributions.

Finally, we show the contributions of the $(\tilde{t},\tilde{b})$ sector 
with the $\tilde{t}_L^{}$-$\tilde{t}_R^{}$  mixing described 
in Set~4 (Case~16 - Case~20) of Table~\ref{tab:set4}, where {\it maximal} 
mixing ($\theta_{\tilde{t}} \sim \pi/4$) takes place.    
The mass of $\tilde{t}_1^{}$ is fixed to be 100GeV, 
and the other squarks of this sector 
$\tilde{t}_2$, $\tilde{b}_1$ and  $\tilde{b}_2$ are varied widely. 
First, see the $\tau=-1$ amplitude in Fig.~\ref{fig:amp00-4}(a). 
The largest sfermion contributions are observed in Case 20, where the 
smallness of $m_{\tilde{t}_1} = 100$GeV comes from the mass mixing.  
For this case, the corrections are positive around the 
$\tilde{t}_1$-pair threshold. At the first peak above this  
threshold, the deviation from the SM prediction 
can be about $+0.7$\% in amplitudes, and it amounts to $+0.9$\% 
at the second peak just above 
the threshold of the $\tilde{b}$-pair production. 
The deviation then goes to be negative drastically 
due to the negative constant term \eq{const-term}. 
The term \eq{const-term} is enlarged by the mass difference 
between $\tilde{t}_1$ and the others, 
so that the correction reaches $-3.9$\% in amplitudes (Case~20) 
before the asymptotic behavior is observed. 
In Case~16 - Case~19, the corrections behave in the same way as in Case 20,  
but the smaller corrections are observed because 
the mass difference between $\tilde{t}_1$ and the others is smaller.    
For the $\tau=+1$ amplitude, as similar to Set~3, the corrections 
below the threshold of $\tilde{t}_1$-pair production are negative, 
but the large mass difference between $\tilde{t}_1$ and 
the others makes the correction positive around the first 
peak above the threshold of the $\tilde{t}_1$-pair production. 

Therefore the sfermion one-loop corrections to $M^{00}_{\tau}$ are  
sensitive to the sfermion parameter choice.   
The typical magnitude of the sfermion one-loop contribution, however, 
is a few times $0.1$\% in amplitudes in each part.  
The deviations from the SM prediction at low energies tend to be negative, 
but the larger mass splitting between $\tilde{t}_1$ and the other
squarks of the $(\tilde{t},\tilde{b})$ sector can induce the larger 
positive corrections.

\subsubsection{The sfermion one-loop corrections to 
$M_{\tau=-1}^{0+}$ and $M_{\tau=-1}^{-0}$}

\begin{figure}[t]
\begin{center}
\leavevmode\psfig{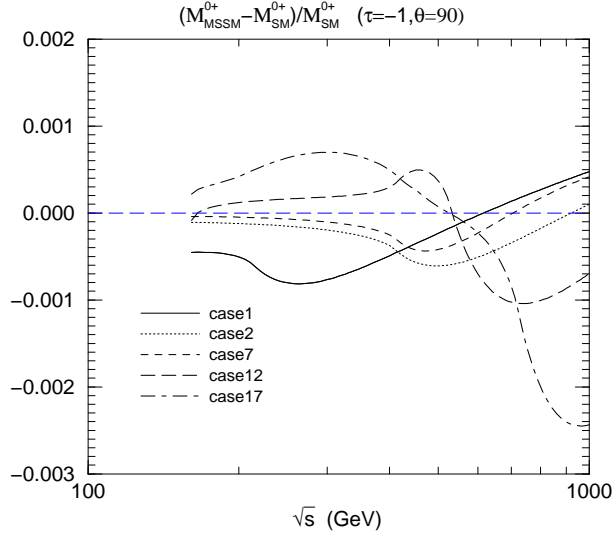}
\end{center}
\caption{The sfermion one-loop contributions to 
$M^{0+}_{\tau=-1}$ are shown in various MSSM parameter sets; 
Case~1, Case~2 in Table~5, Case~7 in Table~6, 
Case~12 in Table~7 and Case~17 in Table~8.
}
\label{fig:amp0+}
\end{figure}

\hspace*{12pt}
The one-loop corrections to $M_{\tau=-1}^{0+}$ and $M_{\tau}^{-0 }$ 
may be valuable to study only for low energies where the tree-level 
amplitudes are large; see Fig.~\ref{fig:amp2vstot}.  
We here present a figure (Fig.~\ref{fig:amp0+}) for the sfermion one-loop 
contributions to $M_{\tau=-1}^{0+}$, where the results for the following
5~cases are shown;
Case 1 and Case 2 in Table~\ref{tab:set1}, 
Case 7            in Table~\ref{tab:set2}, 
Case 12           in Table~\ref{tab:set3} and 
Case 17           in Table~\ref{tab:set4}. 

In Fig.~\ref{fig:amp0+}, we find the similar characteristics of the 
corrections to $M_{\tau=-1}^{00}$ which we have already discussed in detail. 
The deviation from the SM prediction by the slepton contributions 
(Case~1 and Case~2) and by the squark contributions from the 
first-two generations (Case~7) are negative at low energies.  
On the other hand, the deviation by the ($\tilde{t}, \tilde{b}$) sector 
(Case~12 and Case~17) is positive at low energies. 
In the curve of Case 17 in Fig.~\ref{fig:amp0+}, the typical effects of the 
stop mass mixing seen in the study of $M_{\tau=-1}^{00}$ are 
also observed in $M_{\tau=-1}^{0+}$.
The magnitude of the deviation from the SM prediction is smaller than that of 
$M_{\tau=-1}^{00}$ for each cases.  

\subsection{The one-loop corrections to cross sections}

\hspace*{12pt}
In this subsection, we study the corrections to the helicity-summed 
differential cross section at the large scattering angle
($\theta=90^\circ$).  
The sfermion one-loop contributions to the helicity-summed 
differential cross section may be  
measured by 
\begin{eqnarray}
   \frac{\left(\frac{d \sigma}{d \cos \theta}\right)_{\rm MSSM} - 
         \left(\frac{d \sigma}{d \cos \theta} \right)_{\rm SM}}
        {\left(\frac{d \sigma}{d \cos \theta} \right)_{\rm SM}     } ,
\end{eqnarray}
where $(d \sigma/d \cos \theta)_{\rm MSSM}$ and 
$(d \sigma/d \cos \theta)_{\rm SM}$ are the helicity-summed 
differential cross sections in the MSSM and the SM, respectively. 
The scattering angle is fixed at $\theta = 90^{\circ}$ in the following. 

\begin{table}[t]
\begin{center}
\begin{tabular}{|l|rrrrr|rrrr|} \hline
      & $\!\!\!${\small Case\,21}$\!$ & $\!${\small Case\,22} $\!$
      &       $\!${\small Case\,23}$\!$ & $\!${\small Case\,24} $\!$ 
      &       $\!${\small Case\,25}$\!$ & $\!${\small Case\,26} $\!$
      &       $\!${\small Case\,27}$\!$ & $\!${\small Case\,28} $\!$ 
      &       $\!${\small Case\,29}$\!$  \\ \hline
\multicolumn{10}{|l|}{Input parameters }          \\ \hline
$m_{\tilde{Q}}$       & 1000 &  250 &  250 &  250 & 1000 
                      &  600 &  600 &  600 &  600 \\
$m_{\tilde{U}}$=$m_{\tilde{D}}$
                      & 1000 &  250 &  250 &  250 & 1000 
                      &  540 &  540 &  540 &  540 \\
$m_{\tilde{L}}$       &  100 & 1000 &  100 &  250 &  500 
                      &  540 &  540 &  540 &  540 \\
$m_{\tilde{E}}$       &  100 & 1000 &  100 &  250 &  100
                      &  540 &  540 &  540 &  540 \\
$A^{e\!f\!f}_t\!\!=\!A^{e\!f\!f}_b\!\!=\!A^{e\!f\!f}_\tau\!\!$  
                      &    0 &    0 &    0 &    0 &    0
                      &    0 & 1000 & 1800 & 1950 \\ \hline
\multicolumn{10}{|l|}{Output parameters }          \\ \hline
$m_{\tilde{u}_1}\!=\!m_{\tilde{c}_1}$      
                      &  999 &  246 &  246 &  246 &  999 
                      &  599 &  599 &  599 &  599 \\
$m_{\tilde{u}_2}\!=\!m_{\tilde{c}_2}$        
                      & 1000 &  248 &  248 &  248 & 1000 
                      &  539 &  539 &  539 &  539 \\
$m_{\tilde{d}_1}\!=\!m_{\tilde{s}_1}\!\sim\!m_{\tilde{b}_1}$       
                      & 1001 &  254 &  254 &  254 & 1001 
                      &  602 &  602 &  602 &  602 \\
$m_{\tilde{d}_2}\!=\!m_{\tilde{s}_2}\!\sim\!m_{\tilde{b}_2}$
                      & 1000 &  251 &  251 &  251 & 1000 
                      &  540 &  540 &  540 &  540 \\
$m_{\tilde{e}_1}\!=\!m_{\tilde{\mu}_1}\!\sim\!m_{\tilde{\tau}_1}$
                      &  107 & 1001 &  107 &  253 &  501 
                      &  541 &  541 &  541 &  541 \\
$m_{\tilde{e}_2}\!=\!m_{\tilde{\mu}_2}\!\sim\!m_{\tilde{\tau}_2}$
                      &  106 & 1001 &  106 &  252 &  106 
                      &  541 &  541 &  541 &  541 \\
$m_{\tilde{\nu}_e}\!=\!m_{\tilde{\nu}_{\mu}}\!=\!m_{\tilde{\nu}_{\tau}}$
                      &   86 &  999 &   86 &  245 &  497 
                      &  538 &  538 &  538 &  538 \\
$m_{\tilde{t}_1}$     & 1014 &  302 &  302 &  302 & 1014 
                      &  624 &  421 &  196 &  111 \\
$m_{\tilde{t}_2}$     & 1015 &  304 &  304 &  304 & 1015 
                      &  567 &  730 &  820 &  835 \\
$\cos\theta_{\tilde{t}}$     
                      &    1 &    1 &    1 &     1 &    1 
                      &    1 & 0.637& 0.668& 0.671 \\
\hline \hline  
\end{tabular}
\end{center}
\vspace*{-0.4cm}
\caption{The parameter sets for the study of combined effects of 
sfermion contributions, which satisfies the constraint from the 
direct search experiments. 
In Case~21 - Case~25, cases without the mass mixing are assumed,  
while the effects of the $\tilde{t}_1$-$\tilde{t}_2$ mixing  
are studied by using Case~26 - Case~29. 
}
\label{tab:cross}
\end{table}

Here, in order to examine combined effects including all the sfermion 
contributions, we dare to assume another sets of the sfermion parameters, 
where all the sfermion masses are not larger than about 1000GeV and 
the results from the direct search experiments are taken into account.  
The results from the direct search experiments\cite{pdg99} give lower 
bounds of the sfermion masses; 
all the slepton masses should not be smaller than about 100GeV, while 
the squarks except for the stops should be heavier than about 200GeV.
As for the stop mass, it can still be about 100GeV.  
The sfermion mass parameter sets that we examine are defined in 
Table~\ref{tab:cross}.  

\subsubsection{The cases without the mass mixing}

\hspace*{12pt}
In the first 5 cases (Case~21 - Case~25) in Table~\ref{tab:cross}, 
we include all the sfermions but we do not consider the mass mixing 
by setting all the $A^{eff}_{f}$ to be zero. 
In Case~21, the light sleptons with rather heavy squarks are assumed,  
while the light squarks with heavy sleptons are assumed in Case~22. 
In Case~23, we can study the case where all the sfermions are light   
but their masses are consistent with the data from the direct search 
experiments. 
The case of the complete degeneracy of the input SUSY  
mass parameters $m_{\tilde{Q}}$, $m_{\tilde{U}}$, $m_{\tilde{D}}$, 
$m_{\tilde{L}}$ and $m_{\tilde{E}}$ is described as Case~24. 
Finally the case where only right-handed slepton is light 
and the others have heavier masses is represented by Case~25. 

\begin{figure}[t]
\begin{center}
\leavevmode\psfig{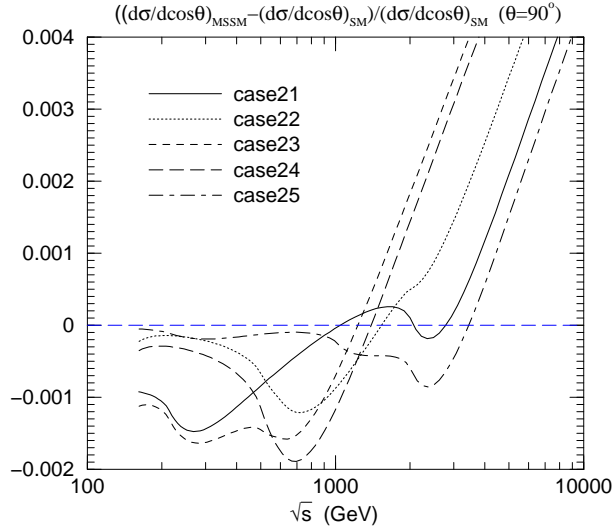}
\end{center}
\caption{The sfermion one-loop corrections to the  
helicity-summed differential cross section are shown 
in Case~21 - Case~25 in Table~10.
The mixing between $\tilde{t}_L^{}$ and $\tilde{t}_R^{}$ 
is zero in these cases.
}
\label{fig:cro-nomix1}
\end{figure}

The corrections are shown in Fig.~\ref{fig:cro-nomix1}.   
First, see the curve of Case~21.
At low energies, the slepton contributions are dominant and thus 
the corrections are negative around the thresholds of the slepton-pair 
productions. The deviation amounts at most to $-0.15$\% 
at the first peak.   
The corrections become slightly positive below 
the thresholds of the squark-pair productions. 
Since the peak above these thresholds is negative, the combined squark 
effects are negative in this case. 
Second, see the curve of Case~22 where the combined squark effects 
can be seen. The corrections below and around the threshold of the 
squark-pair productions are also destructive and the deviation is at 
most about $-0.1$\%.
Third, in the Case~23, both sleptons and squarks have small masses and 
they are set slightly above the lower bounds from the direct
search experiment.  The corrections are approximately 
the sum of those of Case~21 and Case~22 at low energies. 
Forth, the large combined corrections are found in the Case~24, 
where the sfermions are almost mass-degenerate and all the thresholds 
of the sfermion-pair productions are between 490 - 610 GeV.  
The deviation reaches to $-0.2$\% at the negative peak 
slightly above the thresholds. 
Finally in Case~25, the effect of the right-handed squarks is very small 
around the first threshold of the slepton-pair production (212~GeV); 
{\rm i.e.} the most part of the slepton contributions in Case~21 
comes from the left-handed sleptons. 

Therefore, in these cases without the mass mixing, 
the combined contributions to the cross section are negative 
below and around the lowest threshold of the sfermion-pair production. 
Even in the light squark case (Case~22), the positive contributions 
which we have observed in $M^{00}_{\tau=-1}$ and $M^{0+}_{\tau=-1}$ 
are smaller than the summed negative contributions from  
the squarks of the first-two generations and the negative corrections 
in $M^{+-}_{\tau=-1}$. 

\subsubsection{The effects of the mass mixing}

\begin{figure}[t]
\begin{center}
  \leavevmode\psfig{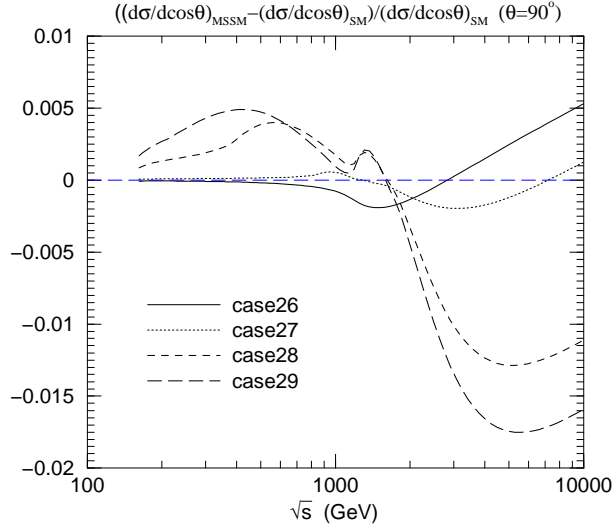} 
\end{center}
\caption{The sfermion one-loop corrections to the  
helicity-summed differential cross section are shown 
in Case~26 - Case~29 of Table~10. 
The mixing between $\tilde{t}_L^{}$ and $\tilde{t}_R^{}$  
appears in Case~27 - Case~29.
}
\label{fig:cro-mix}
\end{figure}

\hspace*{12pt}
The latter 4 cases (Case~26 - Case~29) in Table~\ref{tab:cross} 
are introduced in order to observe the sfermion mass-mixing effects. 
$A^{eff}_{t}(=A^{eff}_b=A^{eff}_\tau)$ is varied 
as $0$, $1000$GeV, $1800$GeV and $1950$GeV    
in Case~26, Case~27, Case~28 and Case~29, respectively. 
The larger mass splitting between $\tilde{t}_1$ and $\tilde{t}_2$ 
takes place for larger $A^{eff}_{t}$. 
In Case~25, since we put $A^{eff}_{t}=0$ the mass difference between 
$\tilde{t}_1$ and $\tilde{t}_2$ is the smallest, 
while in the Case~28 with $A^{eff}_{t}=1950$GeV the large stop 
mass-splitting takes place  
where $m_{\tilde{t}_1} =111$GeV and $m_{\tilde{t}_2}=836$GeV.   
The results in these cases are shown in Fig.~\ref{fig:cro-mix}.

In Case~26, there is nothing new because of $A^{eff}_t = 0$ and 
the curve behaves qualitatively in the same way as Cases~24 
in Fig.~\ref{fig:cro-nomix1}.  
In Case~27 - Case~29, the combined sfermion contributions 
become positive because of the effect of nonzero $A^{eff}_t$ values. 
The behaviors of the corrections  are very similar 
to those of $M^{00}_{\tau=-1}$ and $M^{0+}_{\tau=-1}$ with the mass mixing.  
For large $A^{eff}_t$, the positive corrections maximally 
reach to near $+0.5$\% (Case~28 and Case~29).  
The large negative correction due to the negative constant 
term \eq{const-term} before the asymptotic behavior sets in 
is also observed, which is one of the interesting characteristics 
with large mass-mixing cases seen in $M^{00}_{\tau=-1}$ and
$M^{0+}_{\tau=-1}$.  

Therefore, in the cases with the large mass mixing, 
the magnitude of the positive contributions of 
the $(\tilde{t},\tilde{b})$ sector is much larger than 
that of the combined negative corrections from all the other sfermions. 
In the next section, we will see that such large corrections 
due to the large mass mixing are almost excluded by the 
constraints from the electroweak precision data. 

\subsection{Constraints on the sfermion sector 
            by the electroweak precision data }

\hspace*{12pt}
We have examined in previous subsections the sfermion effects only    
taking into account the direct search results as experimental constraints. 
We here consider the constraints from the precision measurement 
and give a constraint on sfermion effects on $\eeww$. 
The stringent experimental constraints on the MSSM parameters 
are obtained from the electroweak experiments, especially on $Z$-pole 
experiments, the $m^{}_W$ measurements and the low-energy neutral 
current experiments. 
The latest data on the $Z$ parameters\cite{cern-ep99} and the 
$W$-boson mass\cite{mwvalue} are studied in the framework of the MSSM 
in Ref.~\cite{ch99} and we use them here. 
We also consider the external constraints   
\bsub
\bea
\alphas(\mz)         &=& 0.119 \;\;   \pm 0.002 , \\ 
1/\alpha(\mz)      &=& 128.90       \pm 0.09 , \\
 m_t                 &=& 174.3 \;\;   \pm 5.1  \;{\rm GeV},  
\eea
\esub
referring to Ref.~\cite{pdg99} for $\alpha_s(\mz)$ and $m_t$, and 
Ref.~\cite{alpha_ej} for $1/\alpha(\mz)$.
The new physics contributions to the three oblique parameters 
$\Delta S_Z$, $\Delta T_Z$ and $\Delta \mw$ of Ref.~\cite{ch99}, which
we here express by $(S_Z)_{\rm new}$, $(T_Z)_{\rm new}$ and  
$(m_W)_{\rm new}^{}$, are then constrained as 
\bsub
\bea
(S_Z)_{\new} &=& -0.082 \pm 0.114 ,\\
(T_Z)_{\new} &=& -0.179 \pm 0.146 ,\\
(\mw)_{\new} &=& \;\;\; 0.118 \pm 0.057 , 
\eea
\esub
where the correlation between the first-two errors is 
$\rho_{\rm corr} = 0.80$.    
Here we choose the reference value of the SM-Higgs-boson mass as 
$m^{}_H = 117$GeV, the best fit value in the SM. 

\begin{figure}[t]
\begin{center}
\leavevmode\psfig{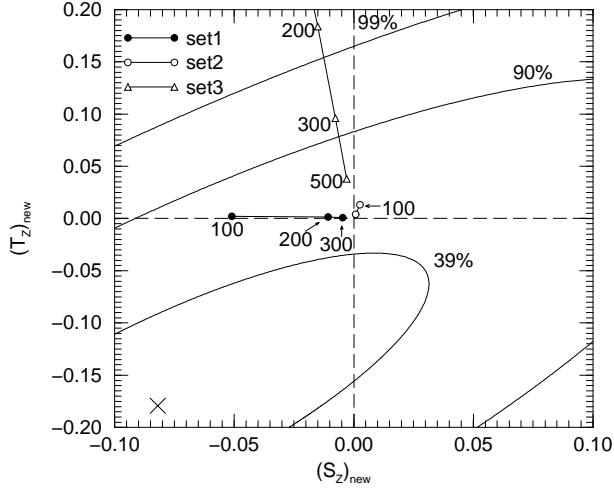}
\end{center}
\caption{The constraints of $(S_Z)_{\rm new}$ and 
                            $(T_Z)_{\rm new}$ for 
Set~1 (Table~\ref{tab:set1}), Set~2 (Table~\ref{tab:set2}) 
and Set~3 (Table~\ref{tab:set3}) from the precision data. 
The contour show the $(S_Z)_{\rm new}$ 
and $(T_Z)_{\rm new}$ fit to the all electroweak data. 
The point $(S_Z)_{\rm new}=(T_Z)_{\rm new} =0$ 
corresponds to the SM prediction. 
The numbers 100, 200, ... etc in the figure are values of 
$m_{\tilde{L}}$ in the unit of GeV for Set~1, 
and those of $m_{\tilde{Q}}$ for Set~2 and Set~3.}
\label{eq:ST1}
\end{figure}
\begin{figure}[t]
\begin{center}
\leavevmode\psfig{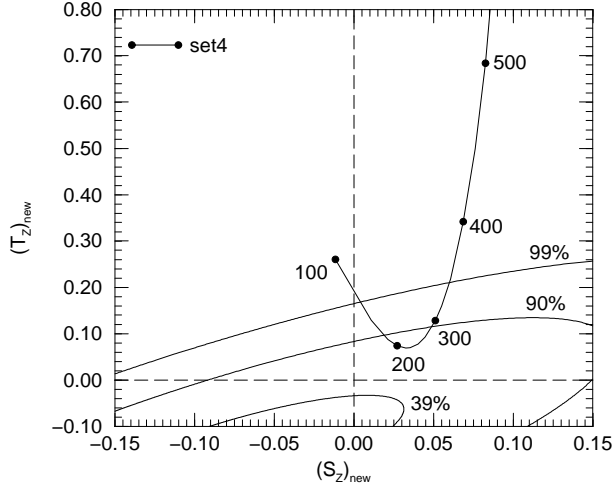}
\end{center}
\caption{The constraints of $(S_Z)_{\rm new}$ and 
$(T_{Z})_{\rm new}$ for Set~4 (Table~\ref{tab:set4})
from the precision data. 
The point of $(S_Z)_{\rm new}=(T_Z)_{\rm new}= 0$ 
corresponds to the SM prediction.
The numbers 100, 200, ... etc in the figure are values of 
$m_{\tilde{Q}}$ in the unit of GeV in each case of Set~4.}
\label{eq:ST2}
\end{figure}

On the other hand, the new physics contributions      
$(S_Z)_{\rm new}$, $(T_Z)_{\rm new}$ and  
$(m_W)_{\rm new}^{}$ are parametrized in terms of  
the new physics contributions to the $S$, $T$, $U$ and $R$ 
parameters by\cite{HHM98}
\bsub
\begin{eqnarray}
  (S_Z)_{\rm new} &=& S_{\rm new} + R_{\rm new} ,  \\
  (T_Z)_{\rm new} &=& T_{\rm new} + 1.49 R_{\rm new} - 
         \frac{({\delta_G})_{\rm new}}{\alpha} ,      \\ 
  (m_W)_{\rm new} &=& 
  - 0.288 S_{\rm new} + 0.418 T_{\rm new} + 0.337 U_{\rm new} 
  - 0.126 \frac{ ({\delta_G})_{\rm new}}{\alpha},
\end{eqnarray}
\esub
where $(\delta_G)_{\rm \new}$ are 
the new physics contributions to the muon decay parameter, whose 
sfermion one-loop corrections are calculated in Ref.~\cite{hmy95}, and 
the $S_{\rm new}$, $T_{\rm new}$, $U_{\rm new}$ and $R_{\rm new}$ are 
calculated by\cite{hhkm94}
\bsub
\begin{eqnarray}
  S_{\rm new} &=& 16 \pi {\rm Re} 
    \left[ {\Pi}^{3Q}_{T,\gamma}(m_Z^2) 
                   - {\Pi}^{33}_{T, Z}(0)  \right] , \\
  T_{\rm new} &=&
        \frac{4 \sqrt{2} G_F}{\alpha}  
    \left[ {\Pi}^{33}_{T}(0) - {\Pi}^{11}_{T}(0)  \right] , \\
  U_{\rm new} &=&  16 \pi {\rm Re} 
    \left[ {\Pi}^{33}_{T, Z}(0) - {\Pi}^{11}_{T, W}(0)  \right] , \\
  R_{\rm new} &=&  16 \pi 
     \left[ {\Pi}^{33}_{T, Z}(0) - {\Pi}^{33}_{T, Z}(q^2)  
        - 2 \hatssq 
      \left\{  {\Pi}^{3Q}_{T, Z}(0) - {\Pi}^{3Q}_{T, Z}(q^2)   
      \right\} 
\right.  \nonumber \\
&& \left. 
+ 4 \hats^4 
      \left\{  {\Pi}^{QQ}_{T, Z}(0) - {\Pi}^{QQ}_{T, Z}(q^2)   
      \right\} \right] , 
\end{eqnarray}
\esub
where the sfermion one-loop corrections to the propagator functions 
are given in Appendix B. 

\begin{table}[t]
\begin{center}
\begin{tabular}{||r|llr||r|llr||}
\hline
Case & $(S_Z)_{\rm new}$ & 
       $(T_Z)_{\rm new}$ & $\Delta\chi^2$ &
Case & $(S_Z)_{\rm new}$ & 
       $(T_Z)_{\rm new}$ & $\Delta\chi^2$ \\ \hline
21 &         $-$0.052  & 0.012    & 3.4  &
26 &         $-$0.0035 & 0.028    & 2.5  \\
22 &         $-$0.010  & 0.13     & 7.2  &
27 &\hspace{3.3mm}0.0046 & 0.014  & 2.0  \\
23 &         $-$0.061  & 0.13     & 11  &
28 &\hspace{3.3mm}0.043  & 0.42   & 30  \\
24 &         $-$0.017  & 0.13     & 8.1  &
29 &\hspace{3.3mm}0.080  & 0.81   & 90  \\
25 &         $-$0.0038 & 0.0082   & 2.0  &
   & & & \\
\hline
\end{tabular}
\end{center}
\caption{The values of $(S_Z)_{\rm new}$, $(T_Z)_{\rm new}$ 
and $\Delta\chi^2$ for Case~21 - Case~25 of Table~10. 
}
\label{tab:case-chi}
\end{table}

Let us examine how the sfermion parameter sets in Sec.~6.1 
are constrained by the precision data. 
We here note that Case~6 and Case~7 of Set~2 in Table~\ref{tab:set2},  
Case~11 and Case~12 of Set~3 in Table~\ref{tab:set3}, and 
Case~16 and Case~17 of Set~4 in Table~\ref{tab:set4} are 
excluded by the results from direct search experiments. 
In Figs.~\ref{eq:ST1} and \ref{eq:ST2}, the contributions to 
$(S_Z)_{\rm new}$ and $(T_Z)_{\rm new}$ are shown for Set~1 - Set~4 of 
Tables~\ref{tab:set1} - \ref{tab:set4} in Sec.~6.1, respectively. 
The origin of the plot shows the SM prediction at $m^{}_H = 117$GeV.     
First, in Fig.~\ref{eq:ST1}, the three series of the points for the cases 
without the mass mixing (Case~1 - Case~15) are shown according to the 
three corresponding categories 
Set~1 (Case~1 - Case~5), Set~2 (Case~6 - Case~10), 
and Set~3 (Case~11 - Case~15) of 
Tables~\ref{tab:set1}, \ref{tab:set2} and \ref{tab:set3}. 
We see that all the cases of 
Set~1 (cases where only sleptons are light) and 
Set~2 (cases where only the squarks of the first-two generations are light) 
are in the 90\%~CL region.
As for the cases of Set~3 (cases where only the squarks of 
$(\tilde{t},\tilde{b})$ sector are light without the mass mixing), 
the point of cases moves outside of the 99$\%$~CL region 
as the mass splitting between $\tilde{t}$ and $\tilde{b}$ grows\footnote{
The mass splitting between $\tilde{t}$ and $\tilde{b}$ 
comes from the fermion mass difference between ${t}$ and ${b}$ 
in these cases. This effect relatively becomes larger for the 
case with smaller SUSY masses $m_{\tilde{Q}}$, $m_{\tilde{U}}$ 
and $m_{\tilde{D}}$}.
Second, in Fig.~\ref{eq:ST2}, the constraint on the $(\tilde{t},\tilde{b})$ 
sector with the mass mixing is shown 
for Set~4 (Case~16 - Case~20) of Table~\ref{tab:set4}. The points of 
Case~16, Case~19 and Case~20 are  outside of the 99\%~CL region 
and thus they are almost excluded by the precision data. 
In Case~19 and Case~20,  the large mass splitting occurs 
between $\tilde{t}$ and $\tilde{b}$ because of the large off-diagonal 
coefficient $m_t A^{eff}_t$. 

Next, we consider the constraints 
on the MSSM parameter sets (Case~21-Case~29) of Table~\ref{tab:cross} 
in Sec.~6.2 that are consistent with the results from the direct 
search experiments. 
These cases have been introduced in order to study the combined 
contributions of all the sfermions to the helicity-summed 
differential cross section.  
The values of 
$(S_Z)_{\rm new}$, 
$(T_Z)_{\rm new}$ and $\Delta \chi^2$, 
where $\Delta \chi^2 \equiv \chi^2 - \chi^2_{\rm min}$,  
for these cases are given in Table~\ref{tab:case-chi}. 
Case~21, Case~25, Case~26 and Case~27 have their $\Delta \chi^2$ values 
less than 4.6 (90\%~CL). 
On the other hand, 
the $\Delta \chi^2$ values of Case~23, Case~28 and Case~29 are 
much larger than 9.2 (99\%~CL). 
Therefore, Case~23, Case~28 and Case~29 are almost excluded. 

The cases allowed by the direct search experiments but strongly 
constrained by the precision data are Case~19 and Case~20 of 
Table~\ref{tab:set4}, and Case~23, Case~28 and Case~29 of 
Table~\ref{tab:case-chi}. Case~23 is the case with the light squarks.
The others are those with the large 
$\tilde{t}_L$-$\tilde{t}_R$ mixing.
They all have the large mass splitting between $\tilde{t}$ and $\tilde{b}$, 
which indicates the breakdown of the $SU(2)_V$ custodial symmetry at 
the sfermion sector, so that they are strongly constrained 
by the precision data.

\subsection{Summary of the numerical results}

\hspace*{12pt}
In this section, the results of the sfermion one-loop  
contributions to the $\eeww$ helicity amplitudes and also those 
to the helicity-summed differential cross sections at the large 
scattering angle have been presented under various sfermion 
parameter choices. 

The contributions from each sfermion sector have been 
examined to the amplitudes  $M^{+-}_{\tau=-1}$, 
$M^{00}_{\tau=\mp 1}$ and $M^{0+}_{\tau=-1}$ in the various parameter 
sets of Tables~\ref{tab:set1} - \ref{tab:set4}.    
First, for the $M^{+-}_{\tau=-1}$, which is the biggest at tree level 
for large scattering angles,  
the sfermion one-loop contributions come only from the wave-function 
renormalization, and the corrections are found to be rather simple as 
seen in Table~\ref{tab:m+-}.  The magnitude of the deviation from the SM 
value is very small; it is at most $\pm 0.05$\% in amplitudes 
under the constraints from the direct search experiments 
and the precision data.   
Second, the rich structure of the sfermion contributions is found in 
$M^{00}_\tau$.  The deviation from the SM value tend to be negative below 
and around the lowest threshold of 
sfermion-pair productions when its corresponding fermion is light; see
Figs.~\ref{fig:amp00-1} and  \ref{fig:amp00-2}. 
The corrections due to the sleptons and the first-two-generation  
squarks do not receive severe bounds from the precision test. By including 
the results of the direct search experiments, the corrections amount at 
most to $-0.15$\% in amplitudes by the slepton contributions and 
$-0.25$\% by the squark contributions from the first-two generations. 
The corrections to $M_{\tau=+1}^{00}$ are the same order as 
those to $M_{\tau=-1}^{00}$, but they may be more difficult to be measured 
because of the smallness of the tree-level contribution.
The large positive contributions at low energies are also observed  
in the $M^{00}_{\tau=-1}$ helicity amplitude by the one-loop effects from 
the $(\tilde{t},\tilde{b})$ sector, where the large mass-splitting between 
$\tilde{t}$ and $\tilde{b}$ can occur due to the large top-quark mass; 
see the cases in Figs.~\ref{fig:amp00-3}(a) and \ref{fig:amp00-4}(a). 
The mass mixing between $\tilde{t}_L^{}$ and $\tilde{t}_R^{}$ enhances such 
positive corrections in the $\tau=-1$ helicity amplitude as seen in 
Figs.~\ref{fig:amp00-4}. 
In Sec.~6.3, it has, however, been found that the MSSM parameter sets which 
give the large mass splitting between $\tilde{t}$ and $\tilde{b}$ are
inconsistent with the precision data. 
Therefore, the large positive corrections by the $(\tilde{t},\tilde{b})$ 
sector to $M^{00}_{\tau}$ in Fig.~\ref{fig:amp00-4}(a) are strongly bounded.  
Still, $+0.3$\% of the deviation may be possible in amplitudes 
between the first and the 
second thresholds: see Case~18 in Fig.~\ref{fig:amp00-4}(a).  
Third, for $M^{0+}_{\tau=-1}$, which is substantial for only low energies,  
the structure of the sfermion contributions is similar 
to that of $M^{00}_{\tau=-1}$, but the magnitude is rather smaller  
(Fig.~\ref{fig:amp0+}). 

Next, we have examined the combined sfermion one-loop contributions to the 
helicity-summed differential cross section at the large scattering angle 
by assuming the various MSSM parameter sets in which all the masses of 
sfermions are not larger than ${\cal O}(1)$ TeV. 
In the case where all the sfermion masses are degenerate at slightly above 
the lower bound of the squarks, the summed negative contributions to the 
cross section amount to $-0.2$\% at the negative peak above 
the thresholds of the sfermion-pair productions. 
The positive corrections due to the quarks of the $(\tilde{t},\tilde{b})$ 
sector become at most $+0.1$\% at the first peak. 

In summary, the rich structure of the sfermion contributions has been 
observed in $M^{00}_{\tau=-1}$. 
At low energies, negative corrections to $M^{00}_{\tau=-1}$  
indicate the effects of the sleptons and the squarks of the first-two 
generations, while the squarks from the third generation give the 
positive corrections. 
These positive corrections are enlarged by the nonzero mixing between 
$\tilde{t}_L$ and $\tilde{t}_R$. 
Since such mixing is constrained by the precision data,  
the magnitude of the corrections to $M^{00}_{\tau=-1}$ 
is at most $-0.4$\% and $+0.3$\% in amplitudes at low energies. 
In terms of the ${00}$-helicity differential cross section, 
these values are counted by multiplying the factor 2. 
On the other hand, in the corrections to the helicity-summed differential 
cross section, the magnitude of the corrections becomes smaller.
The magnitude of the corrections to the helicity-summed cross section 
is at most a few times $0.1$\%. 
These results recall to us the importance of measuring the decaying 
$W$-boson polarizations\cite{hpzh87}.

\section{Discussion and Conclusion}
\label{sec-conclusions}
\cleqn

\hspace*{12pt}
In this paper, we have investigated the sfermion one-loop contributions to 
the $\eeww$ helicity amplitudes in the MSSM. 
The calculation has been thoroughly tested by using various methods; 
especially by   
(i) the exact satisfaction of the BRS sum rules;   
(ii) the clear observation of the decoupling property of the sfermion 
   effects in the low energy limit (the heavy sfermion mass limit);   
(iii) the coincidence in the high energy limit between the 
   results from the full calculation program and 
   the analytic expression of the high energy amplitude 
   that has been verified by the one-loop version of the equivalence theorem.
%
The BRS sum rules among the form factors have been constructed so as to 
hold exactly in our calculational scheme.  
The demonstration of the BRS test when there are left-right 
mass mixings in the third generation sfermions is new in this paper, 
while Ref.~\cite{brs} showed the BRS sum rules for the sfermion one-loop 
contribution only for non-mixing cases.   
The agreement in the BRS tests has given us confidence on our one-loop 
calculation of the form factors except for the overall normalization 
contribution. 
%
The $\overline{\rm MS}$ scheme has been employed in our calculation. 
In addition, all the results have been expanded by the SM 
$\overline{\rm MS}$ couplings, $\hatesq_{\rm SM}(\mu)$ and 
$\hatgsq_{\rm SM}(\mu)$ so that we could see the exact decoupling of 
the sfermion effects in the low-energy limit after all the higher order 
terms of ${\cal O}(\hatg_{\rm SM}^6)$ in amplitude are eliminated. 
The exact decoupling property of the sfermions effects can be 
used as an excellent test of the amplitudes including the overall 
normalization factors. 

We note here that the use of the SM couplings as the expansion parameters 
of the MSSM amplitudes is fully justified at around and below the SUSY 
particle production threshold.  
In this paper, we adopted the SM couplings as the expansion parameters 
even at higher energies above the thresholds, where the use of the 
MSSM $\ov{\rm MS}$ coupling could re-sum the logarithms of the type 
$\log {s/m_{\rm SUSY}^2}$.  
We compared the results of the amplitudes expanded in terms of the 
MSSM couplings and those expanded in terms of the SM couplings, and 
found that their numerical difference is at most $0.013$\% or less  
in $M_{\tau=-1}^{00}$ for the energies below a few TeV. 
This means that the error in the deviation from the SM
prediction can be as large as 15\%.    

In this paper, we have not calculated the full one-loop effects of the SM 
particles. Instead, we estimate the SM amplitudes by setting $\mu = \sqrt{s}$ 
in the SM $\ov{\rm MS}$ couplings. This may or may not be a valid 
approximation to the full SM amplitudes\cite{lemoine80,philippe,eeww-smrc} 
at high energies.  
We therefore presented all our results for the SUSY corrections 
in the form of the relative correction to the SM predictions.  

With the numerical program established by passing through 
all the tests above, we have analyzed  the sfermion one-loop  
contributions to each helicity amplitude and also those to the 
helicity-summed differential cross section of $\eeww$.  The summary of the 
numerical study has been given in Sec.~6.4. 
The $00$ helicity amplitude, $M_{\tau=-1}^{00}$, is one of 
the most appropriate amplitudes for the study of the sfermion one-loop 
contributions.  The magnitude of the correction in 
$M_{\tau=-1}^{00}$ at low energies becomes large in the following cases;  
(1) the light sfermion with no-mixing cases 
   (destructive interference with the SM amplitude),  
(2) the large mass mixing cases in the $(\tilde{t},\tilde{b})$ sector 
   (constructive interference with the SM amplitude). 
The experimental results of the sfermion direct search give lower bounds 
on the sfermion masses. 
By including the electroweak precision data, the  
$(\tilde{t},\tilde{b})$ sector especially with the large mass mixing    
is strongly constrained.  
Under all these experimental constraints, 
the deviation from the SM amplitude 
at the peak slightly after the first thresholds of 
sfermion-pair productions can be at most $-0.8$\% (in cases (1)) 
and $+0.6$\% (in cases (2)) in the differential cross section of 
the helicity amplitude $M^{00}_{\tau=-1}$ at the large scattering angle 
($\theta=90^\circ$).
Although these characteristics of the $M^{00}_{\tau=-1}$ have been observed 
in the helicity-summed differential cross section, the magnitude of the 
corrections is smaller; typically a few times 0.1\%. 
Therefore, it is important to measure the decaying $W$-boson
polarizations in exploring the sfermion sector through their indirect 
effects on $\eeww$. 
 
In conclusion, the sfermion one-loop contributions are small 
(about a few times $\pm 0.1$\% level) in the helicity-summed  
cross section under the constraint from the direct search results 
and the electroweak precision tests. In some of the helicity amplitudes 
such as that for the longitudinally-polarized $W$-boson pair, 
the corrections of near $-0.8$\% and $+0.6$\% 
in observables may be possible. 
One-loop effects from the other sector of the MSSM will be reported  
elsewhere\cite{hku00}.  

\vspace*{2mm}
\noindent
{\bf Note added}\\
After the completion of this work, we received a preprint\cite{kniehl}.   
We confirmed the agreement with their work  
in the analytic results of the sfermion one-loop contributions 
in the gauge-boson vacuum polarizations and the trilinear vertices.\\

\noindent
{\Large \it Acknowledgements}\\
\cleqn
The authors would like to thank Gi-Chol Cho for his valuable contribution 
in the early stage of the collaboration.    
S.~K.\ was supported by the Alexander von Humboldt Foundation, and 
Y.~U.\ was supported by Deutsche Forschungsgemeinschaft under 
Contract KL 1266/1-1.


\appendix
\section{The Lagrangian}
\label{app-lagrangian}
\cleqn

\subsection{Physical masses and mixing angles}
\label{subsec-masses}

\hspace*{12pt}
We begin by discussing the sfermion mass-matrices.  
We will ignore mixing between generations, hence we need to discuss 
only one generation which contains the up-type squark, $\scup$, 
the down-type squark, $\scdown$, a charged slepton, $\sclepton$, 
and its associated sneutrino, $\scnu$.  The left-handed squarks, 
$\scupl$ and $\scdownl$, form an SU(2) doublet which we denote by $\scq$.  
Similarly, $\scnul$ and $\scleptonl$ form the doublet $\scl$.  
As for the right-handed sfermion $\scupr$, $\scdownr$ and $\scleptonr$ 
are SU(2) singles which we denote by $\scu$, $\scd$ and $\sce$, respectively.

The mass matrix for the up-type squarks and the down-type squarks 
can be written as
\begin{equation}\label{up-mixing-matrix}
M^2_\scup = \left( \begin{array}{cc}
m_\scq^2 + \mzsq\cos 2\beta (T^3_{u_L} - \hatssq Q_u) + m_u^2 
   & -m_u|A_u^\ast + \mu\cot\beta|\phasepossu \\
-m_u|A_u + \mu^\ast\cot\beta|\phasenegsu & m^2_\scu + 
\mzsq\cos 2\beta \hatssq Q_u + m_u^2 
\end{array} \right)\;,
\end{equation}
\begin{equation}\label{down-mixing-matrix}
M^2_\scdown = \left( \begin{array}{cc}
m_\scq^2 + \mzsq\cos 2\beta (T^3_{d_L} - \hatssq Q_d) + m_d^2 
   & -m_d|A_d^\ast + \mu\tan\beta|\phasepossd \\
-m_d|A_d + \mu^\ast\tan\beta|\phasenegsd & 
m^2_\scd + \mzsq\cos 2\beta \hatssq Q_d + m_d^2 
\end{array} \right)\;.
\end{equation}
where $m_\scq$, $m_\scu$ and $m_\scd$ are explicit SUSY-breaking masses 
for the doublet $\scq$ and the singlets $\scupr$ and $\scdownr$, 
respectively.  The off-diagonal 
elements depend upon $A_u$, the coefficient of the trilinear SUSY-breaking 
term. The parameter $\mu$ is the coefficient of the quadratic Higgs term, 
$H_1\cdot H_2$, in the superpotential, and $\tan\beta$ is the ratio of the 
vacuum expectation values for the two Higgs doublets.  
%
The the sneutrino $\scnu$ has only a left-handed state, whose mass is given by
\begin{equation}\label{scnu-mass}
m^2_\scnul = m^2_\scl + \mzsq \cos 2\beta (T^3_{\nu_L} - \hatssq Q_\nu)
\;.
\end{equation}
The mass-matrix for the down type sleptons is
\begin{equation}\label{e-mixing-matrix}
M^2_\sclepton = \left( \begin{array}{cc}
m_\scl^2 + \mzsq\cos 2\beta(T^3_{e_L} - \hatssq Q_e) + m_e^2 
   & -m_e|A_e^\ast+\mu\tan\beta|\phasepossel \\
-m_e|A_e+\mu^\ast\tan\beta|\phasenegsel & 
m^2_\sce + \mzsq\cos 2\beta \hatssq Q_e + m_e^2 
\end{array} \right)\;.
\end{equation}
In this paper, we refer the off diagonal elements to 
\begin{subequations}
\label{eq:afeff}
\begin{eqnarray}
 A^{eff}_{d,e} = A^\ast_{d,e} + \mu \cot \beta , \;\;\;\; 
 {\rm and}\;\;\;\;
 A^{eff}_{u}   = A^\ast_{u} + \mu \tan \beta .
\end{eqnarray}
\end{subequations}

After diagonalizing these matrices and finding their eigenvalues, 
the lighter of the two is denoted as $m^2_\scferone$, while 
the heavier is then $m^2_\scfertwo$.  
The mass matrix is diagonalized according to
\begin{equation}
 S^\dagger_\scfer M^2_\scfer S_\scfer = 
{\rm diag}(m^2_\scferone,m^2_\scfertwo)\;,
\end{equation}
and the physical eigenstates are given by
\begin{eqnarray}
\label{mixing-complex}
\left( \begin{array}{c} \scferl \\ \scferr \end{array} \right)
& = & S_\scfer 
\left( \begin{array}{c} \scferone \\ \scfertwo \end{array} \right)\;. 
\end{eqnarray}
The mixing matrix  $S_\scfer$ may be parametrized as
\begin{equation}
S_\scfer = 
\left( \begin{array}{cc} \cosf & \sinf\phasepossf
\\ - \sinf\phasenegsf & \cosf \end{array} \right)
\;,\makebox[1cm]{}
S_\scfer^\dagger = 
\left( \begin{array}{cc} \cosf & - \sinf\phasepossf
\\ \sinf\phasenegsf & \cosf \end{array} \right)
\;,
\end{equation}
where $0 \leq \cos \theta_{\tilde{f}} \leq 1$ and 
      $0 \leq \sin \theta_{\tilde{f}} \leq 1$.  
Because the mass-matrices are Hermitian, the eigenvalues are real.
To prevent the breaking of SU(3) color or electric charge, none of the 
squared masses can be negative.  If the explicit SUSY-breaking mass terms 
are sufficiently large, then the diagonal elements are positive for all
values of $\tan\beta$.  The most stringent condition, $m_\scl > m_Z/\sqrt{2}$,
comes from the requirement $m^2_{\scnul} > 0$ in the large $\tan\beta$
limit. If $\tan\beta = 1$, then $\cos 2\beta = 0$ and the diagonal terms are 
positive even in the limit where these mass terms vanish.  
As for the third family the off-diagonal entries can also be large.  
Assuming that the diagonal elements are positive, the
condition
\begin{equation}
(M^2_\scfer)_{11} (M^2_\scfer)_{22} 
> (M^2_\scfer)_{12} (M^2_\scfer)_{21} \;
\end{equation}
must be imposed 
to guarantee that $m^2_\scferone>0$ and $m^2_\scfertwo>0$.  

\subsection{Sfermion--gauge-boson interactions}
\label{subsec-ffv}

\hspace*{12pt}
The interactions of one gauge boson with two sfermions are given by
\begin{eqnarray}
\nonumber
{\cal L}_{V\scfer\scfer} & = &
   i \bigg\{ g^\gamma_{\scfer_i\scfer_j} A^\mu + g^Z_{\scfer_i\scfer_j}
Z^\mu 
   \bigg\}\scfer_i^\ast \bdd_\mu \scfer_j 
 + ig^Z_{\scnul\scnul} \scnul^\ast \bdd_\mu \scnul Z^\mu 
\\ \label{lagr-vff}&& \mbox{} 
 + \bigg\{ i g^{W}_{\scup_i\scdown_j} \scup^\ast_i \bdd_\mu \scdown_j 
 W^{+\,\mu} + {\rm h.c.} \bigg\}
 + \bigg\{i g^{W}_{\scnul\sclepton_i} \scnul^\ast \bdd_\mu \sclepton_i
 W^{+\,\mu} + {\rm h.c.}\bigg\} \;,
\end{eqnarray}
where summation over $\scfer = \scup$, $\scdown$ and $\sclepton$ and 
$i,j = 1,2$ is implied.   The couplings are
then 
given by
\begin{subequations}
\label{coupling-vff}
\begin{eqnarray}
\label{coupling-aff}
g^{\gamma}_{\scfer_i\scfer_j} & = & \left[ S_\scfer^{\dagger} 
\left( \begin{array}{cc} -\hate Q_f & 0 \\ 0 & -\hate Q_f
\end{array} \right) S_\scfer \right]_{ij} 
= \left( \begin{array}{cc} -\hate Q_f & 0 \\ 
                        0 & -\hate Q_f \end{array} \right)_{ij} \;,\\
\nonumber
g^{Z}_{\scfer_i\scfer_j} & = & \left[ S_\scfer^{\dagger}  
\left( \begin{array}{cc} 
-\hatgz( T^3_{f_L}-\hatssq Q_f) & 0 \\ 
0 & -\hatgz ( -\hatssq Q_f ) 
\end{array} \right) S_\scfer \right]_{ij} \\
\label{coupling-zff}
&& \makebox[2cm]{} = \mbox{} -\hatgz
\left( \begin{array}{cc}
(T^3_{f_L}\cosfsq -\hatssq Q_f) &  T^3_{f_L}\sinf\cosf\phasepossf
\\ 
 T^3_{f_L}\sinf\cosf\phasenegsf & (T^3_{f_L}\sinfsq -\hatssq Q_f)
\end{array} \right)_{ij} \;,\\
\label{coupling-znn}
g^Z_{\scnul\scnul} & = & -\hatgz (T^3_{\nu_L} - \hatssq Q_\nu) \;,\\
\label{coupling-wff}
g^{W}_{\scup_i\scdown_j} 
& = & \left[ S^{\dagger}_\scup
\left( \begin{array}{cc} -\hatg/\sqrt{2} & 0 \\ 0 & 0 
\end{array} \right) S_\scdown \right]_{ij} \!\!=  
-\frac{\hatg}{\sqrt{2}}
\left( \begin{array}{cc} \cosu\cosd & \cosu\sind\phasepossd 
\\ \sinu\cosd\phasenegsu & \sinu\sind \phasenegdiff
\end{array} \right)_{ij}\!\!,\makebox[5mm]{}\\
\label{coupling-wnn}
g^W_{\scnul\sclepton_i} 
& = & \left[ \left( \begin{array}{cc} -\frac{\hatg}{\sqrt{2}} & 0
\end{array}
\right) S_\sclepton 
\right]_i = -\frac{\hatg}{\sqrt{2}} \left( \begin{array}{cc}
\cosl & \sinl\phasepossel  \end{array} \right)_i \;.
\end{eqnarray}
\end{subequations}
For the $\scnul$ there is no mixing.  The other couplings 
are expressed as $2\times 2$ matrices.

The $VV\scfer\scfer$ seagull-type terms are given by
\begin{eqnarray}
\nonumber
\lefteqn{{\cal L}_{VV\scfer\scfer} = \scfer_i^\ast \scfer_j 
   \bigg\{ g^{\gamma\gamma}_{\scfer_i\scfer_j} A_\mu A^\mu 
 + g^{\gamma Z}_{\scfer_i\scfer_j} A_\mu Z^\mu + g^{ZZ}_{\scfer_i\scfer_j} 
   Z_\mu Z^\mu + g^{WW}_{\scfer_i\scfer_j}  W^-_\mu W^{+\,\mu} \bigg\}}&&
\\\nonumber && \mbox{}
 + g^{ZZ}_{\scnul\scnul} \scnul^\ast\scnul Z_\mu Z^\mu
 + g^{WW}_{\scnul\scnul} \scnul^\ast\scnul W^+_\mu W^{-\,\mu}
 + \bigg\{ \Big[ g^{W\gamma}_{\scup_i\scdown_j}A^\mu 
 + g^{WZ}_{\scup_i\scdown_j}Z^\mu\Big] \scup_i^\ast \scdown_j W^+_\mu 
  + {\rm h.c.}\bigg\}
\\ \label{lagr-vvff} && \makebox[0cm]{} 
 + \bigg\{ 
   \Big[ g^{W\gamma}_{\scnul\sclepton_i}A^\mu 
 + g^{WZ}_{\scnul\sclepton_i}Z^\mu\Big]\scnul^\ast \sclepton_i W^+_\mu 
  +{\rm h.c.} \bigg\}\;,
\end{eqnarray}
with the following couplings:
\begin{subequations}
\label{coupling-vvff}
\begin{eqnarray}
\label{coupling-ggff}
   g^{\gamma\gamma}_{\scfer_i\scfer_j} & = & \left[ S_\scfer^{\dagger} 
\left( \begin{array}{cc} \hatesq Q_f^2 & 0 \\ 0 & \hatesq Q_f^2 
\end{array} \right) S_\scfer \right]_{ij}
= \left( \begin{array}{cc} \hatesq Q_f^2 & 0 \\ 
   0 & \hatesq Q_f^2 \end{array} \right)_{ij} \;,\\
\nonumber
g^{ZZ}_{\scfer_i\scfer_j} & = & \left[ S_\scfer^{\dagger}  
\left( \begin{array}{cc} 
\hatgzsq (T^3_{f_L} - \hatssq Q_f)^2 & 0 \\ 
0 & \hatgzsq (- \hatssq Q_f)^2
\end{array} \right) S_\scfer\right]_{ij} \\
\label{coupling-zzff}
&= & \mbox{}\hatgzsq
\left( \begin{array}{cc} 
   \Big[T^3_{f_L} (T^3_{f_L}-2\hatssq Q_f)\cosfsq + \hatssq
Q_f^2
   \Big] & T^3_{f_L}(T^3_{f_L}-2\hatssq Q_f)\sinf\cosf\phasepossf
\\
   T^3_{f_L}(T^3_{f_L}-2\hatssq Q_f)\sinf\cosf\phasenegsf & 
   \Big[T^3_{f_L}(T^3_{f_L}-2\hatssq Q_f)\sinfsq + \hatssq
Q_f^2
   \Big] 
\end{array} \right)_{ij}\!\!,\makebox[7mm]{}\\
\label{coupling-zznn}
g^{ZZ}_{\scnul\scnul} & = & \hatgzsq(T^3_{\nu_L}-\hatssq Q_\nu)^2\;,\\
\nonumber
g^{\gamma Z}_{\scfer_i\scfer_j} & = & \left[S_\scfer^{\dagger} 
\left( \begin{array}{cc} 
  2\hate\hatgz Q_f(T^3_{f_L}-\hatssq Q_f) & 0 \\ 
  0 & 2\hate\hatgz Q_f(-\hatssq Q_f)
\end{array} \right) S_\scfer \right]_{ij}\\ 
\label{coupling-gzff}
& = & 2\hate\hatgz\left( \begin{array}{cc}
 Q_f(T^3_{f_L}\cosfsq-\hatssq Q_f) &
 Q_f T^3_{f_L}\sinf\cosf\phasepossf \\   
 Q_f T^3_{f_L}\sinf\cosf \phasenegsf
         & Q_f(T^3_{f_L}\sinfsq-\hatssq Q_f) 
\end{array} \right)_{ij}\;,\\
\label{coupling-wwff}
   g^{WW}_{\scfer_i\scfer_j} & = & \left[S_\scfer^{\dagger} 
\left( \begin{array}{cc} 
\hatgsq/2 & 0 \\ 0 & 0 \end{array} \right) S_\scfer \right]_{ij}
= \frac{\hatgsq}{2} \left( \begin{array}{cc} 
\cosfsq & \sinf\cosf\phasepossf \\ 
\sinf\cosf\phasenegsf & \sinfsq \end{array} \right)_{ij} \;,\\
\label{coupling-wwnn}
g^{WW}_{\scnul\scnul} & = & \frac{1}{2}\hatgsq\;,\\
\nonumber
g^{\gamma W}_{\scup_i\scdown_j} & = & 
\left[S_\scup^{\dagger}  \left( \begin{array}{cc} 
  \hatg\hate (Q_u+Q_d) /\sqrt{2} & 0 \\ 0 & 0
\end{array} \right) S_\scdown \right]_{ij}\\
& = &
\label{coupling-gwff}
\frac{\hate\hatg}{\sqrt{2}}(Q_u+Q_d)
\left( \begin{array}{cc}
 \cosu\cosd & \cosu\sind\phasepossd \\ 
 \sinu\cosd\phasenegsu & \sinu\sind\phasenegdiff
\end{array} \right)_{ij}\;,\\
\label{coupling-gwnn}
g^{\gamma W}_{\scnul\sclepton_i} & = & \left[\left( 
  \frac{\hatg\hate}{\sqrt{2}}(Q_\nu+Q_e) \;\; 0 \right)
  S_\sclepton \right]_i = 
  \frac{\hatg\hate}{\sqrt{2}}(Q_\nu+Q_e)\left(\cosl \;\; \sinl 
  \phasepossel \right)_i \;,\\
\nonumber
g^{ZW}_{\scup_i\scdown_j} & = & \left[S_\scup^{\dagger} 
\left( \begin{array}{cc} 
  -\hatg\hatgz\hatssq (Q_u+Q_d) /\sqrt{2} & 0 \\ 0 & 0
\end{array} \right) S_\scdown \right]_{ij}\\
\label{coupling-zwff}
 & = & -\frac{\hatg\hatgz\hatssq}{\sqrt{2}}(Q_u+Q_d)
\left( \begin{array}{cc}
 \cosu\cosd & \cosu\sind\phasepossd \\ 
 \sinu\cosd\phasenegsu & \sinu\sind\phasenegdiff
\end{array} \right)_{ij}\;,\\
\label{coupling-zwnn}
g^{Z W}_{\scnul\sclepton_i} & = & \left[\left( 
  -\frac{\hatg\hatgz\hatssq}{\sqrt{2}}(Q_\nu+Q_e) \;\; 0 \right)
  S_\sclepton \right]_i = 
  -\frac{\hatg\hatgz\hatssq}{\sqrt{2}}(Q_\nu+Q_e)
  \left(\cosl \;\; \sinl\phasepossel \right)_i \;.
\end{eqnarray}
\end{subequations}

\subsection{Sfermion--Goldstone-boson interactions}
\label{subsec-ffx}

\hspace*{12pt}
We present
only the portion of the Lagrangian which contains the interactions 
between one charged Goldstone boson and two scalar fermions:
\begin{equation}
\label{lagr-xff}
{\cal L}_{\chi\scfer\scfer} = 
\bigg\{ i g^\chi_{\scup_i\scdown_j} \scup_i^\ast \scdown_j \chi^+
+ {\rm h.c.}\bigg\} + \bigg\{i g^\chi_{\scnul\sclepton_i} \scnul^\ast 
\sclepton_i \chi^+
+ {\rm h.c.} \bigg\}\;,
\end{equation}
where summation over $i,j = 1, 2$ is implied, and the couplings are given
by
\begin{subequations}
\label{coupling-xff}
\begin{eqnarray}
\label{coupling-xud}
g^\chi_{\scup_i\scdown_j} & = & \left[ 
\frac{\hatg}{\sqrt{2}\hat{m}^2_W}S_\scup^{\dagger} 
\left( \begin{array}{cc} 
\mwsq\cos 2\beta + m_u^2 - m_d^2 & 
m_d\Big|A_d^\ast + \mu\tan\beta\Big|\phasepossd \\
-m_u\Big|A_u + \mu^\ast\cot\beta\Big|\phasenegsu & 0 \end{array} \right)
S_\scdown \right]_{ij} \;,\\
\label{coupling-xne}
g^\chi_{\scnul\sclepton_i} & = & \left[ 
\frac{\hatg}{\sqrt{2}\hat{m}^2_W} 
\left( \begin{array}{cc} 
\mwsq\cos 2\beta - m_e^2 & m_e\Big|A_e^\ast +
\mu\tan\beta\Big|\phasepossel
\end{array} \right)
S_\sclepton \right]_{i} \;.
\end{eqnarray}
\end{subequations}
The overall phase factors for $g^\chi_{\scup_i\scdown_j}$ exactly parallel
the phase factors for $g^W_{\scup_i\scdown_j}$ reflected in 
Eq.~(\ref{coupling-wff}), while the overall phase factors for 
$g^\chi_{\scnul\sclepton_i}$ mimic those of $g^W_{\scnul\sclepton_i}$ in 
Eq.~(\ref{coupling-wnn}).  


\section{Sfermion effects on the form factors}
\label{app-eeww-ff}
\cleqn

\subsection{Two-point functions}

\hspace*{12pt}
For the photon propagator,
\begin{equation}
\label{pitgg}
\pitgg (\qsq) = \frac{\hatesq}{16\pi^2} \scfersumi N_c^f Q^2_f
                     B_5(\qsq;m_\scferi,m_\scferi) \;.
\end{equation}
where $N_c^f = 3$ for squarks and $N_c^f = 1$ for sleptons.
The function 
$B_5(\qsq; m_1,m_2)$\cite{hhkm94} on RHS is related to the familiar
notation of Ref.~\cite{pv79} by 
\begin{equation}
B_5(\qsq; m_1,m_2) = A(m_1) + A(m_2) - 4
B_{22}(\qsq; m_1,m_2)\;. 
\end{equation}
For the $\gamma Z$, $ZZ$ and $WW$ two-point functions, we obtain 
\begin{eqnarray}
\pitgz (\qsq) & = & \frac{\hate\hatgz}{16\pi^2} \scfersumude 
                    N_c^f Q_f \bigg\{ 
                    (T^3_{f_L} \cosfsq - \hatssq Q_f )
                     B_5(\qsq; m_\scferone,m_\scferone) \nonumber \\
&&                     + (T^3_{f_L} \sinfsq - \hatssq Q_f )
                     B_5(\qsq;m_\scfertwo,m_\scfertwo) \bigg\}
\label{pitgz}\;, \\
\pitzz (\qsq) & = & \frac{\hatgzsq}{16\pi^2} 
\sum_{\scfer = \scup,\scdown,\sclepton}N_c^f \bigg\{
  (T^3_{f_L} \cosfsq - \hatssq Q_f )^2 
  B_5(\qsq; m_\scferone,m_\scferone) 
+(T^3_{f_L} \sinfsq - \hatssq Q_f )^2 
  B_5(\qsq; m_\scfertwo,m_\scfertwo)\nonumber\\
&&  + 2 (T^3_{f_L}\sinf\cosf)^2 
  B_5(\qsq; m_\scferone,m_\scfertwo) \bigg\}
  + \frac{\hatgzsq}{16\pi^2}(T^3_{\nu_L} - \hatssq Q_\nu )^2 
  B_5(\qsq; m_\scnul,m_\scnul)\;, 
\label{pitzz} \\
\pitww (\qsq) & = & \frac{\hatgsq/2}{16\pi^2}\bigg\{ 
   3\cosusq\,\cosdsq B_5(\qsq; m_\scupone,m_\scdownone)
  +3\cosusq\,\sindsq B_5(\qsq; m_\scupone,m_\scdowntwo)
\\ \nonumber && \makebox[1cm]{}
  +3\sinusq\,\cosdsq B_5(\qsq; m_\scuptwo,m_\scdownone)
  +3\sinusq\,\sindsq B_5(\qsq; m_\scuptwo,m_\scdowntwo)
\\ && \makebox[1cm]{}
  +\coslsq B_5(\qsq; m_\scnul,m_\scleptonone)
  +\sinlsq B_5(\qsq; m_\scnul,m_\scleptontwo) \bigg\}\;.
\label{pitww}
\end{eqnarray}
%
%
The one-loop sfermion contribution to 
the wavefunction renormalization factor of the physical $W$ boson 
is given by  
\begin{equation}
Z_{W}^\frac{1}{2} = 1 - \frac{1}{2} 
       \left.  \frac{d}{d q^2} \; \pitww(q^2) \right|_{q^2=\mwsq}\;,
\;\;{\rm and}\;\;
\delta Z_{W}^\frac{1}{2} = Z_{W}^\frac{1}{2}  - 1 \; .
\end{equation}

\subsection{Sfermion contributions to the $\eeww$}

\begin{figure}[t]
\begin{center}
\leavevmode\psfig{file=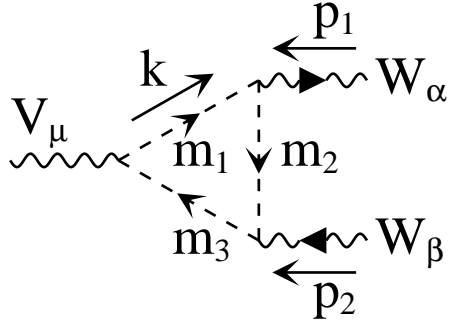,angle=0,width=6cm,silent=0}
\end{center}
\caption{Mass and momentum assignments for the calculation of the 
sfermion triangle graph are shown. The arrows in the $W$ lines indicate
the flow of a negative electric charge.}
\label{fig-scalar-triangle} 
\end{figure}

\begin{figure}[t]
\begin{center}
\leavevmode\psfig{file=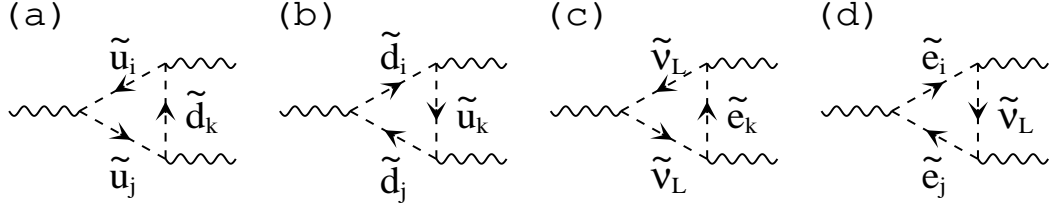,angle=0,width=14cm,silent=0}
\end{center}
\caption{Feynman graphs contributing to the $VWW$ vertex are shown.
The mass and momentum assignments are shown in 
Fig.~\ref{fig-scalar-triangle}. When $V=Z$, all graphs contribute.
In case of $V=\gamma$, graphs (c) do not contribute, and only
$i=j$ is allowed.
}
\label{fig-wwv-triangle}
\end{figure}

\hspace*{12pt}
Here we show the calcuration for the form factor coefficients 
$f_i^{V(1)}$ $(V=\gamma, Z)$ in Sec.~4.2. 
First, the triangle graphs are depicted in 
Fig.~\ref{fig-wwv-triangle}.  Mass and momentum assignments are as in 
Fig.~\ref{fig-scalar-triangle}.
For the evaluation of the loop integrals it is convenient to have the 
momenta incoming, hence we use $p_1 = -p$ and $p_2 = -\overline{p}$ where
$p$ and $\overline{p}$ were defined in Fig.~\ref{fig-eeww-blob}.
We obtain for the $\gamma WW$ vertex,
\begin{eqnarray}
\nonumber
\lefteqn{f_i^{\gamma\,(1)\,\rm SFT} = -\frac{1}{16\pi^2\hate}\bigg\{
3g^\gamma_{\scup_i\scup_i}\Big|g^W_{{\scup_i}{\scdown_j}}\Big|^2
     C^{\rm
SF}_{i}(p_1,p_2,m^2_{\scup_i},m^2_{\scdown_j},m^2_{\scup_i})}&&\\
&&\mbox{}
-3g^\gamma_{\scdown_i\scdown_i}\Big| g^W_{\scup_j\scdown_i} \Big|^2
     C^{\rm SF}_{i}(p_1,p_2,m^2_{\scdown_i},m^2_{\scup_j},m^2_{\scdown_i})
-g^\gamma_{\sclepton_i\sclepton_i}\Big|g^W_{\scnul\sclepton_i}\Big|^2
  C^{\rm SF}_{i}(p_1,p_2,m^2_{\sclepton_i},m^2_\scnul,m^2_{\sclepton_i})
\bigg\}\;,
\end{eqnarray}
where summation over $i,j = 1,2$ is implied, and the loop-integral 
coefficients $C^{\rm SF}_i$ are defined in Ref.~\cite{brs}.  
The photon couplings are real and the complex phases cancel between 
the two $W$-boson vertices. For the $ZWW$ vertex,
\begin{eqnarray}\nonumber
\lefteqn{f_i^{Z\,(1)\,\rm SFT} =
-\frac{1}{16\pi^2\hatgz\hatcsq}\bigg\{}&&\\
&&\nonumber\mbox{}
3g^Z_{\scup_k\scup_i}g^W_{{\scup_i}{\scdown_j}}
\Big(g^{W}_{{\scup_k}{\scdown_j}}\Big)^\ast
     C^{\rm SF}_{i}(p_1,p_2,m^2_{\scup_i},m^2_{\scdown_j},m^2_{\scup_k})
-3g^Z_{\scdown_i\scdown_k}g^W_{\scup_j\scdown_i}
\Big(g^W_{\scup_j\scdown_k}\Big)^\ast
     C^{\rm SF}_{i}(p_1,p_2,m^2_{\scdown_i},m^2_{\scup_j},m^2_{\scdown_k})
\\&&\mbox{}
+g^Z_{\scnul\scnul}\Big|g^W_{\scnul{\sclepton_k}}\Big|^2 
  C^{\rm SF}_{i}(p_1,p_2,m^2_\scnul,m^2_{\sclepton_k},m^2_\scnul)
-g^Z_{\sclepton_i\sclepton_k}g^W_{\scnul\sclepton_i}
\Big(g^W_{\scnul\sclepton_k}\Big)^\ast
  C^{\rm SF}_{i}(p_1,p_2,m^2_{\sclepton_i},m^2_\scnul,m^2_{\sclepton_k})
\bigg\},\makebox[5mm]{}
\end{eqnarray}
where summation over $i,j,k = 1,2$ is implied.  The complex phases cancel
between the three coupling factors.  The superscript `SFT' is chosen to
denote 
`sfermion triangle' contributions.

\begin{figure}[t]
\begin{center}
\leavevmode\psfig{file=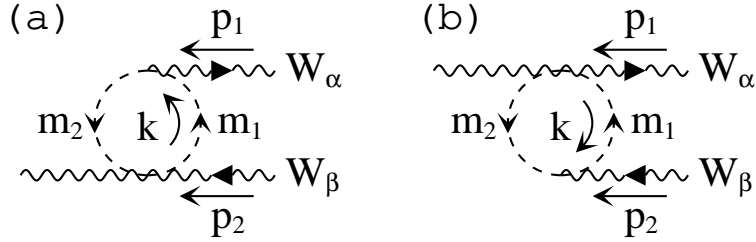,angle=0,width=10cm,silent=0}
\end{center}
\caption{Mass and momentum assignments for the calculation of the 
sfermion graphs containing seagull coupling are shown. 
The arrows in the $W$ lines indicate the flow of a negative 
electric charge.
}
\label{fig-scalar-wwv-sg} 
\end{figure}

\begin{figure}[t]
\begin{center}
\leavevmode\psfig{file=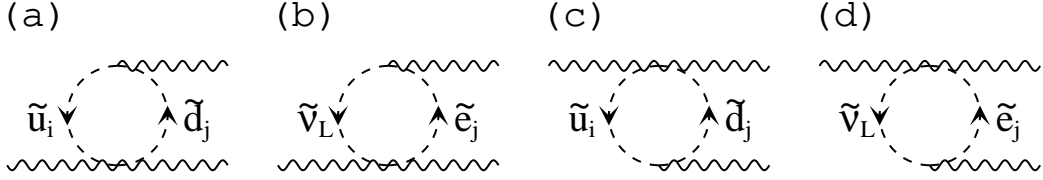,angle=0,width=14cm,silent=0}
\end{center}
\caption{Feynman graphs containing seagull coupling and
contributing to the $\gamma WW$ and $ZWW$ vertex are shown.
The mass and momentum assignments are shown in 
Fig.~\ref{fig-scalar-wwv-sg}. 
}
\label{fig-wwv-sg}
\end{figure}

The second category of vertex corrections are depicted in 
Fig.~\ref{fig-wwv-sg}.  We use the momentum assignments of
Fig.~\ref{fig-scalar-wwv-sg}.
The results for the $\gamma WW$ vertex and for $ZWW$ are summarized as 
\begin{subequations}
\begin{eqnarray}
\nonumber
f_{10}^{\gamma\,(1)\,\rm SFSG} & = & \frac{1}{16\pi^2\hate}\bigg\{ 
3\Big(g^{\gamma W}_{{\scup_i}{\scdown_j}}\Big)^\ast
g^W_{{\scup_i}{\scdown_j}}
(2B_1+B_0)(m_W^2; m_{\scdown_j},m_{\scup_i}) \\ &&\makebox[1.1cm]{}
+ \Big(g^{\gamma W}_{{\scnul}{\sclepton_j}}\Big)^\ast
g^W_{{\scnul}{\sclepton_j}}
(2B_1+B_0)(m_W^2; m_{\sclepton_j},m_\scnul)\bigg\}\;,
\\
f_{13}^{\gamma\,(1)\,\rm SFSG} & = &
- f_{10}^{\gamma\,(1)\,\rm SFSG}\;,\\
\nonumber
f_{10}^{Z\,(1)\,\rm SFSG} & = & \frac{1}{16\pi^2\hatgz\hatcsq}\bigg\{ 
3\Big(g^{ZW}_{{\scup_i}{\scdown_j}}\Big)^\ast
g^W_{{\scup_i}{\scdown_j}}
(2B_1+B_0)(m_W^2; m_{\scdown_j},m_{\scup_i}) \\ && \makebox[1.7cm]{}
+ \Big(g^{ZW}_{{\scnul}{\sclepton_j}}\Big)^\ast
g^W_{{\scnul}{\sclepton_j}}
(2B_1+B_0)(m_W^2; m_{\sclepton_j},m_\scnul)\bigg\}\;,
\\
f_{13}^{Z\,(1)\,\rm SFSG} & = &
- f_{10}^{Z\,(1)\,\rm SFSG}\;,
\end{eqnarray}
\end{subequations}
where superscript  `SFSG' represents `sfermion seagull-graph' contributions.
As is clear from the above 
expressions, the $W$ bosons are chosen to be on mass shell such that 
$p_1^2 = p_2^2 = m_W^2$.


\section{Sfermion effects on the $W^\mp\chi^\pm$ and $\chi^- \chi^+$ 
production}
\cleqn
 
\subsection{$e^-e^+ \rightarrow W^\mp \chi^\pm$}
\label{app-eewx-ff}

\hspace*{12pt}
The one-loop level vertex coefficients  
$\hhbar\mbox{}_i^{\gamma\,(1)}$ and $\hhbar\mbox{}_i^{Z\,(1)}$ 
receive contributions from the triangle graphs in 
Fig.~\ref{fig-vwx-vxw-triangle} and 
\begin{figure}[t]
\begin{center}
\leavevmode\psfig{file=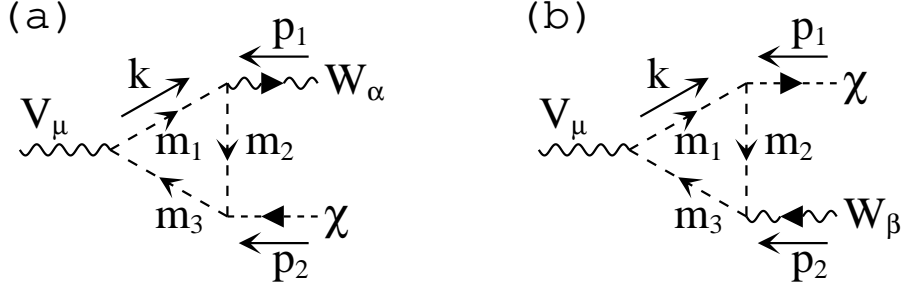,angle=0,width=12cm,silent=0}
\end{center}
\caption{Mass and momentum assignments for the calculation of the 
sfermion triangle graphs contributing to the $VW^\mp\chi^\pm$ vertices.
The arrows in the $W$ and $\chi$ indicate the flow of a negative
electric charge.
} 
\label{fig-scalar-vwx-vxw} 
\end{figure}
\begin{figure}[t]
\begin{center}
\leavevmode\psfig{file=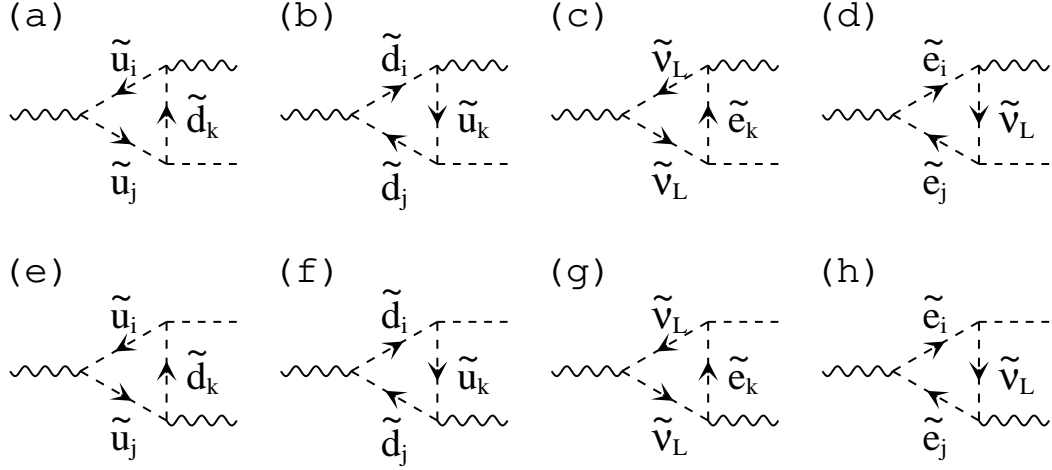,angle=0,width=14cm,silent=0}
\end{center}
\caption{
Feynman graphs contributing to the $VW^{\mp}\chi^{\pm}$ vertex are shown.
The mass and momentum assignments are shown in 
Fig.~\ref{fig-scalar-vwx-vxw}. 
Feynman graphs (a)-(d) contribute to
the $VW^+\chi^-$ vertex while graphs (e)-(h) contribute to the
$VW^-\chi^+$
vertex.
When $V=Z$, all graphs contribute.
In case of $V=\gamma$, graphs (c) and (g) do not contribute, and only
$i=j$ is allowed.
}
\label{fig-vwx-vxw-triangle} 
\end{figure}
the seagull-type vertices as depicted in Fig.~\ref{fig-vwx-vxw-sg}.
\begin{figure}[t]
\begin{center}
\leavevmode\psfig{file=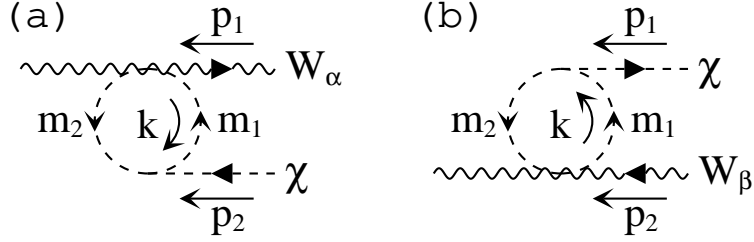,angle=0,width=10cm,silent=0}
\end{center}
\caption{
Mass and momentum assignments for the graphs containing seagull 
coupling and contributing to the  $\gamma W^\mp\chi^\pm$ and 
$ZW^\mp\chi^\pm$ vertex are shown. 
The arrows in the $W$ and $\chi$ indicate the flow of a negative
electric charge.
}
\label{fig-scalar-vwx-vxw-sg} 
\end{figure}
\begin{figure}[t]
\begin{center}
\leavevmode\psfig{file=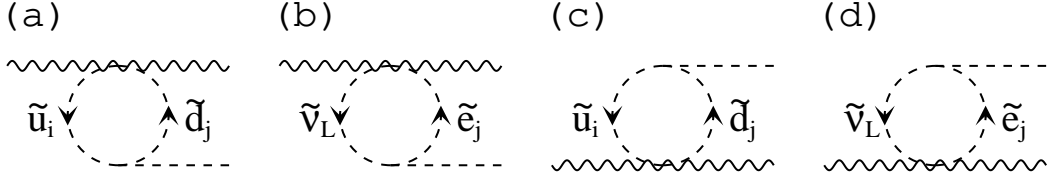,angle=0,width=14cm,silent=0}
\end{center}
\caption{
Feynman graphs containing seagull coupling and contributing 
to the $VW^{\mp}\chi^{\pm}$ vertex are shown.
The mass and momentum assignments are shown in 
Fig.~\ref{fig-scalar-vwx-vxw-sg}. 
Feynman graphs (a) and (b) contribute to
the $VW^+\chi^-$ vertex while graphs (c) and (d) contribute 
to the $VW^-\chi^+$
vertex.
}
\label{fig-vwx-vxw-sg} 
\end{figure}
We 
begin with the calculation of the triangle graphs with internal mass and 
momentum assignments as in Fig.~\ref{fig-scalar-vwx-vxw}.
We obtain 
\begin{subequations}
\begin{eqnarray}
\nonumber
\lefteqn{h_i^{\gamma\,(1)\,{\rm SFT}} = -\frac{1}{16\pi^2\hate}\bigg\{
g^\gamma_{\scup_j\scup_j} g^W_{\scup_j\scdown_k} 
\Big( g^\chi_{\scup_j\scdown_k} \Big)^\ast
c_i^{\rm SF}(p_1,p_2,m^2_{\scup_j},m^2_{\scdown_k},m^2_{\scup_j})}
&&\\ \label{hi-sft-gamma} &&\makebox[-0.8cm]{}
+ g^\gamma_{\scdown_j\scdown_j} g^W_{\scup_k\scdown_j} 
\Big( g^\chi_{\scup_k\scdown_j} \Big)^\ast
c_i^{\rm SF}(p_1,p_2,m^2_{\scdown_j},m^2_{\scup_k},m^2_{\scdown_j})
+ g^\gamma_{\sclepton_j\sclepton_j} g^W_{\scnul\sclepton_j} 
\Big(g^\chi_{\scnul\sclepton_j}\Big)^\ast
c_i^{\rm SF}(p_1,p_2,m^2_{\sclepton_j},m^2_{\scnul},
m^2_{\sclepton_j})
\bigg\}\;,\\\nonumber
\lefteqn{\overline{h}_i^{\gamma\,(1)\,{\rm SFT}} = 
\frac{1}{16\pi^2\hate}\bigg\{
g^\gamma_{\scup_j\scup_j} 
\Big( g^W_{\scup_j\scdown_k} \Big)^\ast 
g^\chi_{\scup_j\scdown_k}
\overline{c}_i^{\rm
SF}(p_2,p_1,m^2_{\scup_j},m^2_{\scdown_k},m^2_{\scup_j})}
&&\\ \label{hbari-sft-gamma} &&\makebox[-0.8cm]{}
+ g^\gamma_{\scdown_j\scdown_j} 
\Big( g^W_{\scup_k\scdown_j} \Big)^\ast
g^\chi_{\scup_k\scdown_j}
\overline{c}_i^{\rm
SF}(p_2,p_1,m^2_{\scdown_j},m^2_{\scup_k},m^2_{\scdown_j})
+ g^\gamma_{\sclepton_j\sclepton_j} 
\Big( g^W_{\scnul\sclepton_j} \Big)^\ast
g^\chi_{\scnul\sclepton_j}
\overline{c}_i^{\rm SF}(p_2,p_1,m^2_{\sclepton_j},m^2_{\scnul},
m^2_{\sclepton_j})
\bigg\}\;,
\end{eqnarray}
with summation over $j,k=1,2$,  and
\begin{eqnarray}
\nonumber
\lefteqn{h_i^{Z\,(1)\,{\rm SFT}} = -\frac{1}{16\pi^2\hatgz\hatssq}\bigg\{
}&&\\
\nonumber && \makebox[-0.2cm]{}
  g^Z_{\scup_j\scup_l} g^W_{\scup_j\scdown_k} 
\Big( g^\chi_{\scupl\scdown_k} \Big)^\ast
c_i^{\rm SF}(p_1,p_2,m^2_{\scup_j},m^2_{\scdown_k},
m^2_{\scup_l})
+ g^Z_{\scdown_j\scdown_l} g^W_{\scup_k\scdown_j} 
\Big( g^\chi_{\scup_k\scdown_l} \Big)^\ast
c_i^{\rm SF}(p_1,p_2,m^2_{\scdown_j},m^2_{\scup_k},
m^2_{\scdown_l})
\\
&&\makebox[-0.8cm]{}
+ g^Z_{\scnul\scnul}g^W_{\scnul\sclepton_k} 
\Big(g^\chi_{\scnul\sclepton_k}\Big)^\ast
c_i^{\rm SF}(p_1,p_2,m^2_{\scnul},m^2_{\sclepton_k},
m^2_{\scnul})
+ g^Z_{\sclepton_j\sclepton_l} g^W_{\scnul\sclepton_j} 
\Big( g^\chi_{\scnul\sclepton_l} \Big)^\ast
c_i^{\rm SF}(p_1,p_2,m^2_{\sclepton_j},m^2_{\scnul},
m^2_{\sclepton_l})\bigg\}\label{hi-sft-z}
\;,\\
\nonumber
\lefteqn{\overline{h}_i^{Z\,(1)\,{\rm SFT}} = 
\frac{1}{16\pi^2\hatgz\hatssq}\bigg\{ }&&\\
\nonumber && \makebox[-0.2cm]{}
  g^Z_{\scup_j\scup_l} 
\Big( g^W_{\scup_j\scdown_k} \Big)^\ast
g^\chi_{\scupl\scdown_k}
\overline{c}_i^{\rm SF}(p_2,p_1,m^2_{\scup_j},m^2_{\scdown_k},
m^2_{\scup_l})
+ g^Z_{\scdown_j\scdown_l} 
\Big( g^W_{\scup_k\scdown_j}\Big)^\ast 
g^\chi_{\scup_k\scdown_l}
\overline{c}_i^{\rm SF}(p_2,p_1,m^2_{\scdown_j},m^2_{\scup_k},
m^2_{\scdown_l})
\\
&&\makebox[-0.8cm]{}
+ g^Z_{\scnul\scnul}
\Big(g^W_{\scnul\sclepton_k}\Big)^\ast 
g^\chi_{\scnul\sclepton_k}
\overline{c}_i^{\rm SF}(p_2,p_1,m^2_{\scnul},m^2_{\sclepton_k},
m^2_{\scnul})
+ g^Z_{\sclepton_j\sclepton_l} 
\Big( g^W_{\scnul\sclepton_j} \Big)^\ast
  g^\chi_{\scnul\sclepton_l}
\overline{c}_i^{\rm SF}(p_2,p_1,m^2_{\sclepton_j},m^2_{\scnul},
m^2_{\sclepton_l})\bigg\}\;,\label{hbari-sft-z}
\end{eqnarray}
\end{subequations}
where $j,k,l = 1,2$, and the loop-integral factors 
$c_i^{\rm SF}$ and $\ov{c}_i^{\rm SF}$ are defined in Ref.~\cite{brs}.
Next, using the assignments of Fig.~\ref{fig-scalar-vwx-vxw-sg},
we calculate the contributions of the graphs in Fig.~\ref{fig-vwx-vxw-sg}
with the following results:
\begin{subequations}
\begin{eqnarray}
\makebox[-0.7cm]{}
h_1^{\gamma\,(1)\,{\rm SFSG}} & = & \frac{1}{16\pi^2m_W\hate} \bigg\{
\Big(g^{\gamma W}_{{\scup_j}{\scdown_i}}\Big)^\ast
g^{\chi}_{{\scup_j}{\scdown_i}}
B_0(m_W^2; m_{\scdown_i},m_{\scup_j})
+ \Big(g^{\gamma W}_{{\scnul}{\sclepton_i}}\Big)^\ast
g^{\chi}_{{\scnul}{\sclepton_i}}
B_0(m_W^2; m_{\sclepton_i},m_{\scnul}) \bigg\}\;,\\
\makebox[-0.7cm]{}
\overline{h}_1^{\gamma\,(1)\,{\rm SFSG}} & = & 
-h_1^{\gamma\,(1)\,{\rm SFSG}}  \;,\\
\makebox[-0.7cm]{}
h_1^{Z\,(1)\,{\rm SFSG}} & = & \frac{1}{16\pi^2m_W\hatgz\hatssq}\bigg\{
\Big( g^{Z W}_{{\scup_j}{\scdown_i}} \Big)^\ast
g^{\chi}_{{\scup_j}{\scdown_i}}
B_0(m_W^2; m_{\scdown_i},m_{\scup_j})
+ \Big(g^{Z W}_{{\scnul}{\sclepton_i}}\Big)^\ast
g^{\chi}_{{\scnul}{\sclepton_i}}
B_0(m_W^2; m_{\sclepton_i},m_{\scnul}) \bigg\} , \\
\makebox[-0.7cm]{}
\overline{h}_1^{Z\,(1)\,{\rm SFSG}} & = & 
-h_1^{Z\,(1)\,{\rm SFSG}}  \;,
\end{eqnarray}
\end{subequations}
with $i,j = 1,2$.
As mentioned in the previous section, our BRS sum rules effectively test
the form factors except for the wavefunction renormalization corrections, 
so we write  
\begin{eqnarray}
\hhbar\mbox{}_i^{V\,(1)}(s) & = & 
\hhbar\mbox{}_i^{V\,(1)\,{\rm SFT}}(s)+\hhbar\mbox{}_i^{V\,(1)\,{\rm
SFSG}}(s)\;.
\end{eqnarray}

\subsection{$e^-e^+ \rightarrow \chi^- \chi^+$}
\label{app-eexx-ff}

\hspace*{12pt}
The vertex corrections for $V \chi^-\chi^+$ ($V = \gamma$, $Z$) 
are rather simple; by taking into account the electron current conservation, 
we see that the sfermion sector contribute to the triangle type diagrams.  
The sfermion effects on the $V W^-W^+$ form factor coefficients are 
calculated as 
\begin{eqnarray}
r^{\gamma(1)} &=& \frac{1}{16 \pi^2 \hate}  
             \left\{ 3 g^{\gamma}_{\tilde{u}_i \tilde{u}_i}
                     g^{\chi}_{\tilde{u}_i \tilde{d}_j}
                    \left( g^{\chi}_{\tilde{u}_i \tilde{d}_j}
                                                   \right)^\ast 
             \left( C_{12} - C_{11} \right) 
             (p_1^2,p_2^2,s; 
                  m_{\tilde{u}_i}, m_{\tilde{d}_j}, m_{\tilde{u}_i})
\right. 
\nonumber \\
&&   -     3 g^{\gamma}_{\tilde{d}_i \tilde{d}_i}
                     g^{\chi}_{\tilde{d}_i \tilde{u}_j}
                    \left( g^{\chi}_{\tilde{d}_i \tilde{u}_j}
                                                   \right)^\ast
             \left( C_{12} - C_{11} \right) 
             (p_1^2,p_2^2,s; 
                  m_{\tilde{d}_i}, m_{\tilde{u}_j}, m_{\tilde{d}_i})
                \nonumber \\
&&\left.     -     g^{\gamma}_{\tilde{e}_i \tilde{e}_i}
                     g^{\chi}_{\tilde{e}_i \tilde{\nu}_L^{}}
              \left( g^{\chi}_{\tilde{e}_i \tilde{\nu}_L^{}}
                                                    \right)^\ast
             \left( C_{12} - C_{11} \right) 
             (p_1^2,p_2^2,s; 
                  m_{\tilde{e}_i}, m_{\tilde{\nu}_L^{}}, m_{\tilde{e}_i})
               \right\} , \\
r^{Z(1)} &=& \frac{1}{16 \pi^2 \hatgz}  
             \left\{ 3 g^{\gamma}_{\tilde{u}_k \tilde{u}_i}
                     g^{\chi}_{\tilde{u}_i \tilde{d}_j}
                    \left( g^{\chi}_{\tilde{u}_i \tilde{d}_k}
                                                   \right)^\ast 
             \left( C_{12} - C_{11} \right) 
             (p_1^2,p_2^2,s; 
                  m_{\tilde{u}_i}, m_{\tilde{d}_j}, m_{\tilde{u}_k})
\right. 
\nonumber \\
&&   -     3 g^{\gamma}_{\tilde{d}_k \tilde{d}_i}
                     g^{\chi}_{\tilde{d}_i \tilde{u}_j}
                    \left( g^{\chi}_{\tilde{d}_i \tilde{u}_k}
                                                   \right)^\ast
             \left( C_{12} - C_{11} \right) 
             (p_1^2,p_2^2,s; 
                  m_{\tilde{d}_i}, m_{\tilde{u}_j}, m_{\tilde{d}_k})
                \nonumber \\
&&     +     g^{\gamma}_{\tilde{\nu}_L^{} \tilde{\nu}_L^{}}
                     g^{\chi}_{\tilde{\nu}_L^{} \tilde{e}_i}
              \left( g^{\chi}_{\tilde{\nu}_L^{} \tilde{e}_i}
                                                   \right)^\ast
             \left( C_{12} - C_{11} \right) 
             (p_1^2,p_2^2,s; 
                  m_{\tilde{\nu}_L^{}}, m_{\tilde{e}_i}, m_{\tilde{\nu}_L^{}})
\nonumber \\
&&\left.     -     g^{\gamma}_{\tilde{e}_i \tilde{e}_i}
                     g^{\chi}_{\tilde{e}_i \tilde{\nu}_L^{}}
              \left( g^{\chi}_{\tilde{e}_i \tilde{\nu}_L^{}}
                                                   \right)^\ast
             \left( C_{12} - C_{11} \right) 
             (p_1^2,p_2^2,s; 
                  m_{\tilde{e}_i}, m_{\tilde{\nu}_L^{}}, m_{\tilde{e}_i})
               \right\} , 
\end{eqnarray}
where summation for $i, j, k = 1, 2$ is taken. The tensor coefficient 
functions $C_{ij}$ follow the notation in Ref~\cite{hhkm94}.

\section{The analytic formulas of the integral functions}
\label{app-lowenergy}
\cleqn

\hspace*{12pt}
We present convenient analytic formulas of the integral functions in  
the low-  and high- energy limit. The formulas for the  
Passarino and Veltman's $A$, $B_0$ and $C_0$ 
functions\cite{pv79} are given in the $\ov{\rm MS}$ scheme 
in the notation in Ref~\cite{hhkm94}.    

\subsection{The low energy limit (heavy mass limit)}

\hspace*{12pt}
The $A$ function does not depend on the momentum,   
\begin{eqnarray}
  A (m) = m^2 \left( 1 -  \ln \frac{m^2}{\mu^2} \right).  
\end{eqnarray}

The $B_0$ function and its first and second derivative are given 
for $m_1^2, m_2^2  \gg q^2$  ($m_1 \neq m_2$) by  
\begin{eqnarray}
  B_0(q^2; m_1, m_2) &=& 1 - \frac{m_1^2}{m_1^2 - m_2^2} 
                               \ln \frac{m_1^2}{\mu^2} 
                           + \frac{m_2^2}{m_1^2 - m_2^2} 
                               \ln \frac{m_2^2}{\mu^2} 
               + {\cal O} \left( \frac{m_i^2}{q^2}\right), \\
  B_0'(q^2; m_1, m_2) &=& \frac{1}{(m_1^2 - m_2^2)^2} 
                   \left\{ \frac{1}{2} (m_1^2 + m_2^2) 
                         - \frac{m_1^2 m_2^2}{m_1^2 - m_2^2} 
                           \ln \frac{m_1^2}{m_2^2} \right\} 
               + {\cal O} \left( \frac{m_i^2}{q^4}\right), \\
  B_0''(q^2; m_1, m_2) &=& \frac{1}{(m_1^2 - m_2^2)^2} 
                  \left\{ \frac{1}{3} 
                          + 4 \frac{m_1^2 m_2^2}{(m_1^2 - m_2^2)^2} 
                          - 2 \frac{m_1^2 m_2^2 (m_1^2 + m_2^2)}
                                   {(m_1^2 - m_2^2)^3} 
                           \ln \frac{m_1^2}{m_2^2} \right\} 
               + {\cal O} \left( \frac{m_i^2}{q^6}\right),    
\end{eqnarray} 
where $m_i$ symbolizes $m_1$ or $m_2$.
For the case of ($m_1 = m_2 = m$), the above expressions become 
\begin{eqnarray}
  B_0(q^2; m, m) &=& - \ln \frac{m^2}{\mu^2} 
               + {\cal O} \left( \frac{m^2}{q^2}\right), \\ 
  B_0'(q^2; m, m) &=& \frac{1}{6} \frac{1}{m^2} 
               + {\cal O} \left( \frac{m^2}{q^4}\right),\\
  B_0''(q^2; m, m) &=& \frac{1}{30} \frac{1}{m^4} 
               + {\cal O} \left( \frac{m^2}{q^6}\right).
\end{eqnarray}

The expressions of the $C_0$ function and its derivative are given 
for $p_1^2, p_2^2, q^2=(p_1 + p_2)^2 \ll m_1^2, m_2^2, m_3^2$ by
\begin{eqnarray}
  C_0[123] 
      &=& \frac{-1}{m_1^2 - m_3^2} \left\{ 
       \frac{1}{m_1^2 - m_2^2} (m_1^2 \ln m_1^2 - m_2^2 \ln
      m_2^2) -  \frac{1}{m_3^2 - m_2^2} (m_3^2 \ln m_3^2 - m_2^2 \ln
   m_2^2)
       \right\} 
               + {\cal O} \left( \frac{m_i^2}{q^4}\right),\nonumber \\ \\
   C_0'[123] 
      &=&  \frac{1}{(m_1^2 - m_3^2)^2} 
           \left\{ 1 + \frac{m_2^2}{2} 
         \frac{m_1^2 + m_3^2 - 2 m_2^2}{(m_1^2 - m_2^2)(m_3^2 - m_2^2)}   
       - \frac{1}{2} \frac{m_1^2 + m_3^2 + 2 m_2^2}{m_1^2 - m_3^2} 
              \ln \frac{m_1^2}{m_3^2} \right. \nonumber\\
 && \!\!\!\!\!\!\!\!\!\!\!\!\!\!\!\!\!\!
 \left.+ \frac{1}{2} \frac{m_3^2 - 3 m_1^2 + 2 m_2^2}{m_1^2 - m_3^2} 
        \left( \frac{m_2^2}{m_1^2 - m_2^2}  \right)^2 
       \ln \frac{m_1^2}{m_2^2}
 - \frac{1}{2} \frac{m_1^2 - 3 m_3^2 + 2 m_2^2}{m_1^2 - m_3^2} 
        \left( \frac{m_2^2}{m_3^2 - m_2^2}  \right)^2 
       \ln \frac{m_3^2}{m_2^2} \right\} 
               + {\cal O} \left( \frac{m_i^2}{q^6}\right), \nonumber \\
\end{eqnarray}
where $C_0(p_1,p_2,q^2; m_i, m_j, m_k)$ is written as $C_0[ijk]$ shortly and 
$m_i$ symbolizes $m_1$ or $m_2$.    
The formulas for $m_1^2 = m_3^2$ are given by 
\begin{eqnarray}
\!\!\!\!\!\!   C_0[121] 
      &=& \frac{-1}{m_1^2 - m_2^2} \left\{ 
       1 -  \frac{m_2^2}{m_1^2 - m_2^2} \ln \frac{m_1^2}{m_2^2}
       \right\} 
               + {\cal O} \left( \frac{m_i^2}{q^4}\right), \\
\!\!\!\!\!\!   C_0'[121] 
      &=&  \frac{1}{m_1^4} 
           \left\{  
       - \frac{1}{12} \frac{m_1^6}{(m_1^2 - m_2^2)^3}    
       + \frac{5}{12} \frac{m_1^4 m_2^2}{(m_1^2 - m_2^2)^3}
  + \frac{1}{6}  \frac{m_1^2 m_2^4}{(m_1^2 - m_2^2)^3} 
         - \frac{1}{2}  \frac{m_1^4 m_2^4}{(m_1^2 - m_2^2)^4} 
              \ln \frac{m_1^2}{m_2^2} \right\} 
               + {\cal O} \left( \frac{m_i^2}{q^6}\right). \nonumber\\
\end{eqnarray}
Finally, for the complete degenerating case $m_1 = m_2 = m_3$, 
we have
\begin{eqnarray}
  C_0[111] 
      &=& - \frac{1}{2} \frac{1}{m_1^2}
+ {\cal O} \left( \frac{m^2}{q^4}\right), \\
   C_0'[111] 
      &=&  - \frac{1}{24} \frac{1}{m_1^4}
+ {\cal O} \left( \frac{m^2}{q^6}\right). 
\end{eqnarray}

\subsection{The high-energy limit}
\label{app-highenergy}

We will list formulas needed to reproduce the analytic high-energy 
expressions of $M^{00}_\tau$ amplitudes, Eqs.~(\ref{mx001}) and (\ref{my001}). 
The leading contribution of $B_i$-function in such cases is obtained as 
\begin{eqnarray}
  B_0(q^2; m_1, m_2) &=&  \ln \mu^2 - \ln |q^2| 
                                     + i \pi \theta (q^2) 
                                     + 2 
           + {\cal O} \left( \frac{m_i^2}{q^2}\right),\label{bf0}\\
  B_1(q^2; m_1, m_2) &=& - \frac{1}{2} 
                                 \left\{ \ln \mu^2 - \ln |q^2| 
                                     +  i \pi \theta (q^2) 
                                     + 2 
           + {\cal O} \left( \frac{m_i^2}{q^2}\right) \right\},  \label{bf1}\\
  B_2(q^2; m_1, m_2) &=& \frac{1}{3} 
                                 \left\{ \ln \mu^2 - \ln |q^2| 
                                     +   i \pi \theta (q^2) 
                                     + \frac{13}{6} 
           + {\cal O} \left( \frac{m_i^2}{q^2}\right) \right\},
\label{bf2}\\
   B_5(q^2; m_1, m_2) &=& \frac{q^2}{3}      
                      \left\{ \ln \mu^2 -  \ln |q^2| 
                               +   i \pi \theta (q^2) 
                               + \frac{8}{3} 
           + {\cal O} \left( \frac{m_i^2}{q^2}\right) \right\}, \label{bf3}\\  
  B_5'(q^2; m_1, m_2) &\rightarrow& \frac{1}{3}      
                      \left\{ \ln \mu^2 -  \ln |q^2| 
                               +   i \pi \theta (q^2) 
                               +  \frac{5}{3} 
           + {\cal O} \left( \frac{m_i^2}{q^2}\right) \right\},  \label{bf4}
\end{eqnarray}
where $m_i$ symbolizes $m_1$ or $m_2$.

\end{document}